 \documentclass[preprint2]{aastex61}

\usepackage{natbib,amsmath}
\usepackage{hyperref}

\shorttitle{Chemistry of the infrared luminous merger NGC 3256}
\shortauthors{Harada et al.}

\begin{document}
\bibliographystyle{apj}

\title{ALMA Astrochemical Observations of the Infrared-Luminous Merger NGC 3256}

\author{Nanase Harada}
\affiliation{Academia Sinica Institute of Astronomy and Astrophysics, P.O. Box 23-141, Taipei 10617, Taiwan}

\author{Kazushi Sakamoto}
\affiliation{Academia Sinica Institute of Astronomy and Astrophysics, P.O. Box 23-141, Taipei 10617, Taiwan}

\author{Sergio Mart\'in}
\affiliation{European Southern Observatory, Alonso de C\'ordova 3107, Vitacura, Santiago, Chile}
\affiliation{Joint ALMA Observatory, Alonso de C\'ordova 3107, Vitacura, Santiago, Chile}

\author{Susanne Aalto}
\affiliation{Department of Space, Earth and Environment, Chalmers University of Technology, Onsala Space Observatory, SE- 43992 Onsala, Sweden}

\author{Rebeca Aladro}
\affiliation{Max Planck Institute for Radio Astronomy, Auf dem H\"ugel 69, D-53121 Bonn, Germany}

\author{Kazimierz Sliwa}
\affiliation{Max Planck Institute for Astronomy, K\"onigstuhl 17, D-69117 Heidelberg, Germany}

\correspondingauthor{Nanase Harada}
\email{harada@asiaa.sinica.edu.tw}

\begin{abstract}

%
%
%
In external galaxies, molecular composition may be influenced by extreme environments such as starbursts and galaxy mergers.
To study such molecular chemistry, we observed the luminous-infrared galaxy and merger NGC 3256 using the Atacama Large Millimeter/sub-millimeter Array. 
We covered most of the 3-mm and 1.3-mm bands for a multi-species, multi-transition analysis.
We first analyzed intensity ratio maps of selected lines such as HCN/HCO$^+$, which shows no enhancement at an AGN.
We then compared the chemical compositions within NGC 3256 at the two nuclei, tidal arms, and positions with influence from galactic outflows. We found the largest variation in SiO and CH$_3$OH, species that are likely to be enhanced by shocks.
Next, we compared the chemical compositions in the nuclei of NGC 3256, NGC 253, and Arp 220; these galactic nuclei have varying star formation efficiencies. 
Arp 220 shows higher abundances of SiO and HC$_3$N than NGC 3256 and NGC 253. Abundances of most species do not show strong correlation with the star formation efficiencies, although the CH$_3$CCH abundance seems to have a weak positive correlation with the star formation efficiency.
Lastly, the chemistry of spiral arm positions in NGC 3256 is compared with that of W 51, a Galactic molecular cloud complex in a spiral arm. We found higher fractional abundances of shock tracers, and possibly also higher dense gas fraction in NGC 3256 compared with W 51.
\end{abstract}


\section{Introduction}
Molecular composition (or simply ``chemistry") of various species in the interstellar medium (ISM) varies with the environment surrounding the molecular clouds.
Although the dominant species in molecular regions is H$_2$, and the second most abundant molecule CO is most commonly observed, 
more minor species are more sensitive to physical conditions.
For example, the compositions of those minor species vary with stages of star formation or external radiation such as UV-photons, X-rays, cosmic-rays, or shocks
 \citep[see ][ and references therein]{2017arXiv170807269H}.
 Molecular composition also changes as a molecular cloud evolves from diffuse clouds to form denser cores before the star formation.

Despite faint emission due to large distance, about 60 species have been previously detected in external galaxies \footnote{https://www.astro.uni-koeln.de/cdms/molecules}.
In starburst galaxies, the chemistry is likely to show features of photon-dominated regions (PDRs) due to the high UV-photon flux.
At the same time, starburst galaxies may have higher fractions of dense, star-forming gas compared to galaxies with lower star formation rates.
If so, this difference in the ISM properties can also affect the chemistry. 

\sloppypar{Recent development of Atacama Large Millimeter/sub-millimeter Array (ALMA) and some pre-ALMA radio interferometers
has opened up the possibility of spatially-resolved astrochemistry in external galaxies 
\citep[e.g., ][]{2005ApJ...618..259M,2014PASJ...66...75T,2014A&A...570A..28V,2015PASJ...67....8N,2015ApJ...801...63M,2015A&A...573A.116M}.}
The chemical compositions seen in those observations vary with types of galaxies.
One of the most characteristic chemical compositions is seen in so-called compact obscured nuclei (CONs).
CONs are extremely compact ($<$ 10s of parsecs), obscured ($A_{\rm V}> 10^{3}$ mag), and luminous ($> 10^{10} L_{\odot}$) nuclei in some ultra/luminous infrared galaxies (U/LIRGs).
 For example, Arp 220 and NGC 4418 have CONs, and have very high abundances of 
 HC$_{3}$N or CH$_{3}$CN \citep{2011A&A...527A..36M,2015A&A...582A..91C}. In the Galaxy, these molecules are usually seen only towards star-forming regions, and their abundances 
 averaged in the galaxy scale are very low. Even in local starburst galaxies such as NGC 253 and M82, these molecules are much less abundant than in CONs.
 In addition, the CONs have strong emission of vibrationally-excited lines \citep{2010ApJ...725L.228S,2015A&A...584A..42A,2016A&A...590A..25M}
 with upper state energies ranging from a few hundreds to $\sim 1000$ K. Their detections suggest strong infrared radiation for the excitation.
Because of the high degree of obscuration, it is still unclear whether these chemical features are caused by extreme starbursts or active galactic nuclei (AGNs), or the compactness of the nuclei.

What the observed chemistry is most sensitive to in external galaxies is still unknown partly because of the lack of previous observations,
and partly because of the complexity contained in beams typically of giant molecular cloud sizes or larger.
 In starburst galaxies, molecular abundances species associated with dense cores could correlate with the current star formation rate,
but the correlation may not be tight because the molecular clouds are ingredients of future star formation, 
and they may be showing different stages from existing massive stars used to measure the star formation rate.
This time delay is suggested in the molecular study of NGC 253 by \citet{2017ApJ...849...81A},
where the chemistry significantly varies among clumps with similar star formation rates.
Another driving force of the chemistry is feedback from the existing stars such as irradiation from UV-photons 
 \citep[e.g., ][]{2009ApJ...694..610M}.
To understand the mechanisms behind the chemical composition, it is important to observe starburst galaxies with various star formation rates.

NGC 3256 is an ideal target for such a test. NGC 3256 ($D=35$\,Mpc, 170pc/$1''$) is a late-stage merger (by the definition of merger stage by \citet{2013ApJS..206....1S};
 two nuclei in a common envelope). 
An intense starburst caused by the merging activity generates a high infrared luminosity of $L_{IR} = 3 \times 10^{11} L_{\odot}$. 
Star formation rates in the northern and the southern nuclei (N and S nuclei hereafter) are 
$15\,M_{\odot}$ yr$^{-1}$ and $6\,M_{\odot}$ yr$^{-1}$ respectively, according to the spectral energy distribution (SED)
 fitting in the near- and mid-infrared wavelengths \citep{2008MNRAS.384..316L}.
 The total star-formation rate within this galaxy is estimated to be $\sim 50\,M_{\odot}$ yr$^{-1}$ \citep{2014ApJ...797...90S}, and it is expected that the star formation rate even
 in the off-nucleus positions should still be much higher compared with local starburst galaxies such as NGC 253 (total SFR $\sim 5\,M_{\odot}$ yr$^{-1}$).
 
 Although the main energy source of NGC 3256 is known to be the starburst event, there are other interesting features that can be studied in NGC 3256.
 One of them is the AGN in the S nucleus.
 The configuration of the southern galaxy is almost edge-on, and the very central part of the S nucleus is highly obscured. 
 Therefore, presence of an AGN in the S nucleus lacked convincing evidence \citep[e.g., ][]{2002MNRAS.330..259L}.
 Recently, \citet{2015ApJ...805..162O} have revisited this issue using IR and X-rays, and found that the fit shows an AGN-like feature in the S nucleus.
Such AGN/starburst activities are giving feedback to the interstellar medium (ISM) in NGC 3256.
From the velocity components shifted from the range expected from the galaxy rotation, 
outflows are detected in ionized gas \citep{2013ApJ...772..120L}, cold molecular gas \citep{2006ApJ...644..862S,2014ApJ...797...90S}, 
and hot molecular gas \citep{2014A&A...572A..40E}.
\citet{2014ApJ...797...90S} found that the energy in the outflow from the N nucleus can be explained by star formation alone, 
but the outflow from the S nucleus cannot be accounted for just by a starburst, and needs the presence of an AGN.
By studying the chemistry of NGC 3256, we can examine various topics such as the effects of star formation, 
merging events, and outflows on the molecular compositions.

In this paper, we present a molecular line survey to cover most of the ALMA Bands 3 and 6 to study the chemical abundances
 at locations within the central kpc of NGC 3256.
The organization of this paper is as follows. 
In Section \ref{sec:obs}, we explain our observations and analysis. 
The results obtained from those observations are described in Section \ref{sec:res} for the continuum and molecular lines.
Then, the column densities of individual molecules and their ratios are discussed in Section \ref{sec:column}.
In Section \ref{sec:disc}, we further discuss the implication of the results in Sections \ref{sec:res} and \ref{sec:column}.
Finally, we summarize our findings in Section \ref{sec:summary}.

\section{Observations}\label{sec:obs}
Our ALMA observations (project code 2015.1.00412.S and 2016.1.00965.S) span through
 Cycles 3 and 4 using the 12-meter array and Atacama Compact Array (ACA) for higher-frequency observations.
The observation parameters are summarized in Table \ref{tab:obs_param}. 
For the better signal-to-noise (S/N) ratio, we also used the ALMA data 2015.1.00993.S in the archive for frequency ranges overlapping with ours.
To get the data on the CN ($1-0$) line, we also used the ALMA data 2011.0.00525.S, whose observational parameters 
are described in \citet{2014ApJ...797...90S}.
Our observations covered most of Bands 3 and 6.
For calibration, we ran the pipeline calibration script using $CASA$ version~4.5.3 for cycle 3 data and version~4.7 for cycle 4 data.
Then, we checked the amplitude and phase stability of the pipelined data, and no major problem was found.
 We note that the calibrator of the scheduling block B6\_h taken in cycle 4 has 
fluctuation near the atmospheric absorption at 258 GHz and 271 GHz because the frequency dependence of $T_{\rm sys}$ there was not accurately 
corrected due to channel smoothing. 
Since this fluctuation only affects specific frequencies and there is no lines of interest around these frequencies, we proceed with those calibrated data. 
To improve the accuracy of flux calibration, we corrected the flux using the continuum fluxes as follows.
Continuum flux values at both nuclei were plotted as a function of frequency. 
Then, we fitted them with the power law $\nu^{\alpha}$ ($\nu$: frequency) for Band 3 and Band 6 separately. 
We determined the scaling factor for each sideband and each nucleus to match the power law fit.
By taking the average of scaling factors between both sidebands and nuclei, we determined the degree of correction for the amplitude error.
We applied this correction to scheduling blocks with more than 3\% discrepancy from the power law fit. 
Our resulting flux errors should be within $\sim 5 \%$.
Imaging and simple image analysis were done with $CASA$ version 4.6.
Missing fluxes were evaluated by comparing the images with ACA and without ACA.
Since we do not have ACA data for all the scheduling blocks, we used HCN($3-2$), HCO$^+$($3-2$),
and C$^{18}$O($2-1$) for this comparison, and the estimated missing fluxes are 4.7 \%, 0.2 \%, and 2.8 \%, respectively.
We do not have the ACA data for $^{13}$CO, and the missing flux is expected to be higher for this line because of the extended emission.

\section{Observational Results}\label{sec:res}
\begin{figure*} 
\centerline{
\includegraphics[width=.35\textwidth]{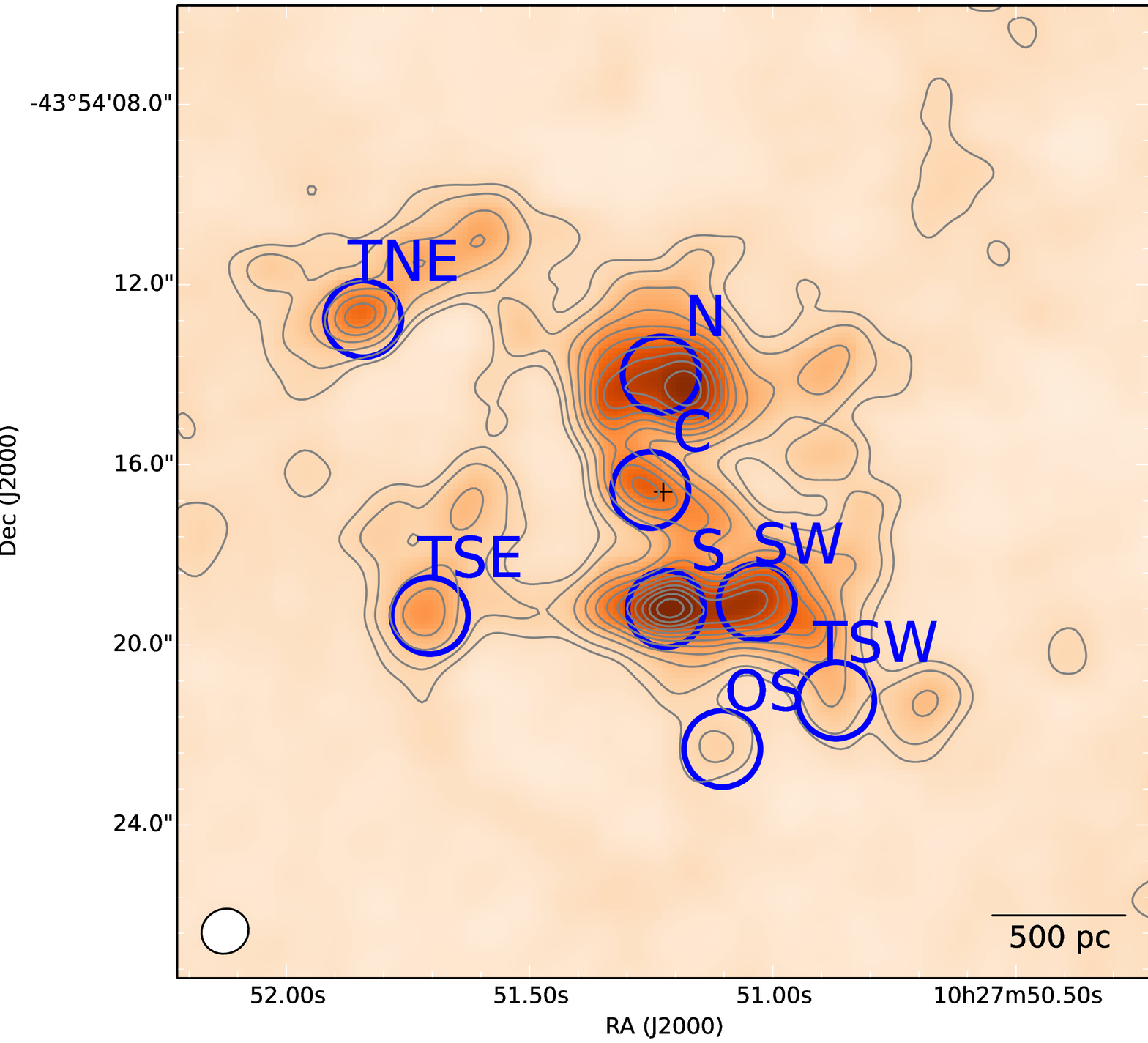}
\includegraphics[width=.35\textwidth]{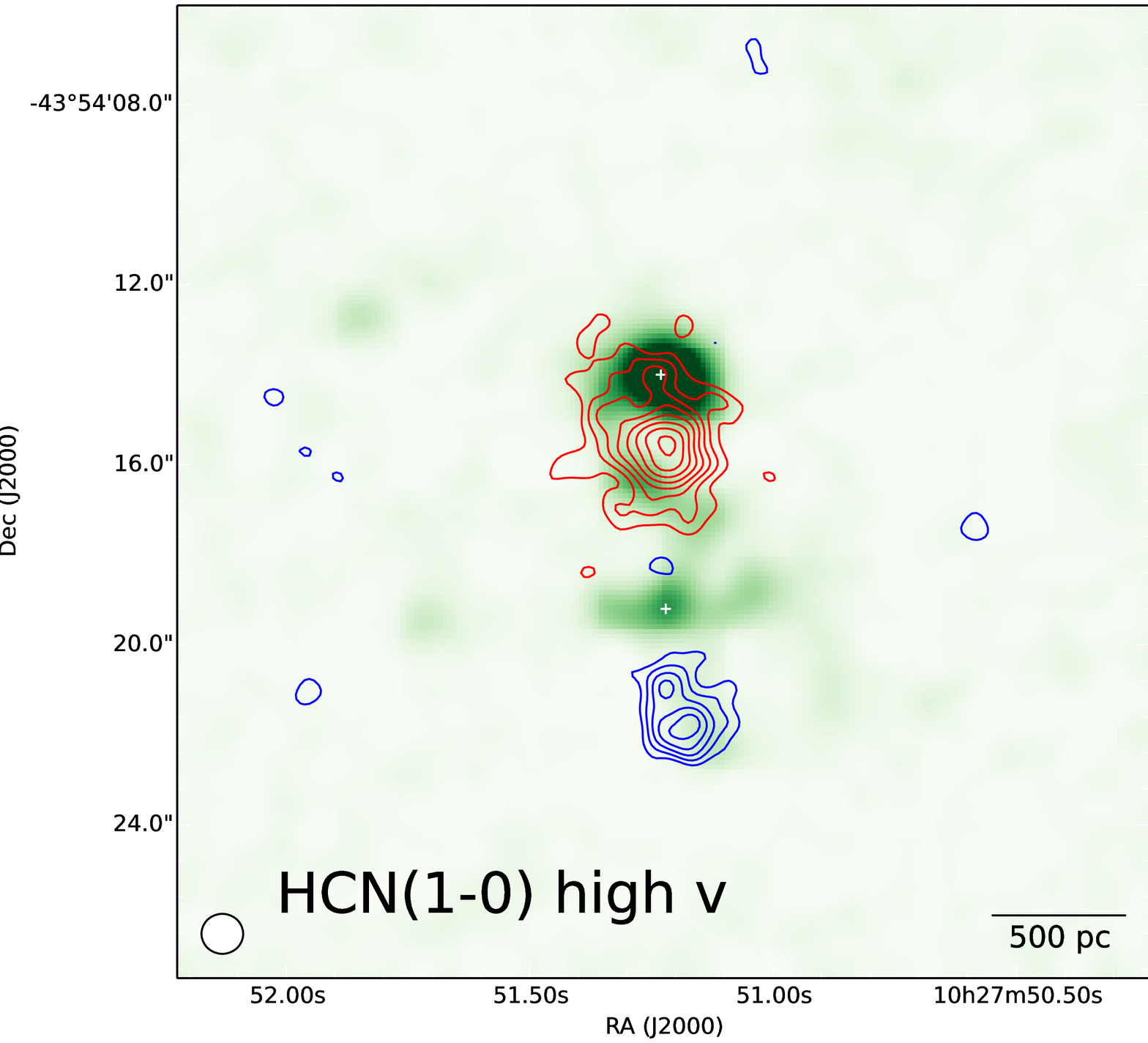}
\includegraphics[width=.20\textwidth, trim =  0 0 0 0]{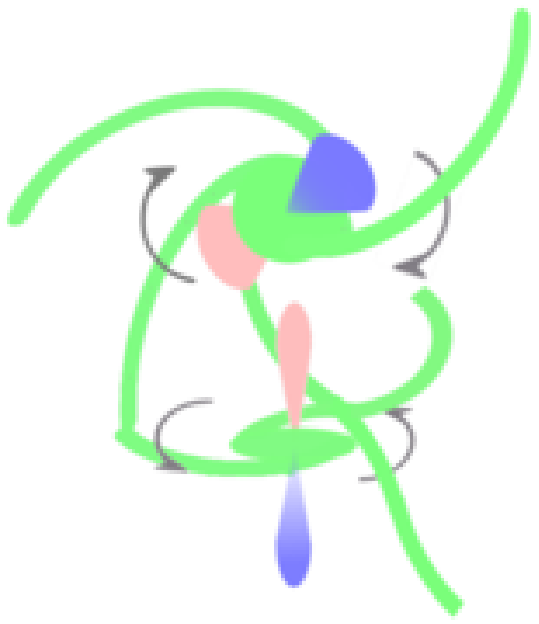}
}
\caption{(Left) Eight positions analyzed in the Section \ref{sec:column}  are shown on the velocity-integrated intensity map of $^{13}$CO ($2-1$). 
(Middle) Positions of components in HCN ($1-0$) with large velocity shifts from the systemic velocity are shown in red contours for the velocity range +225 to +375 km s$^{-1}$,
 and blue contours for the velocity range -435 to -195 km s$^{-1}$ relative to the systemic velocity.
Contour levels are starting from $3 \sigma$ for every $1 \sigma$.
(Right) A schematic image of the merger system of NGC 3256 reproduced from a similar image in \citet{2014ApJ...797...90S}.
 Two nuclear disks, spiral arms, and outflows from both galaxies are shown. \label{fig:8pos}}
\end{figure*}

Before we explain our results later in this section, we briefly explain the morphology of NGC 3256 for clarity.
In NGC 3256, the two nuclei, N and S nuclei are indicated as ``N" and ``S" in Figure \ref{fig:8pos} ($left$).
Spiral arm positions are indicated as ``C" (central peak), ``TNE" (tidal arm northeast), ``TSE" (tidal arm southeast), ``TSW" (tidal arm southwest)\footnote{ 
Although it is likely that those arms are influenced by the tidal interaction, we note that all the spiral arms may not have the tidal origin. 
It is known from simulations and observations that galaxy mergers can create such arm-like features through tidal interaction \citep[e.g., ][]{2010ApJ...725..353D,2011AJ....141..100H}.}.
The position of the outflow from the S galaxy is indicated as ``OS" (outflow south).
There is also a peak at western part of the S galaxy (south galaxy west indicated as ``SW"). 
At position OS and around position C, there are components red-shifted blue-shifted from the systemic velocity (Figure \ref{fig:8pos} $middle$), 
which are thought as outflow components. These high-velocity components were already found with CO by \citet{2014ApJ...797...90S} at similar positions.
A proposed image for this galaxy merger system similar to the one in \citet{2014ApJ...797...90S} is shown in Figure \ref{fig:8pos} ($right$).

\subsection{Continuum Emission}
\begin{figure*}
\centerline{
\includegraphics[width=.5\textwidth, trim =  0 0 0 0]{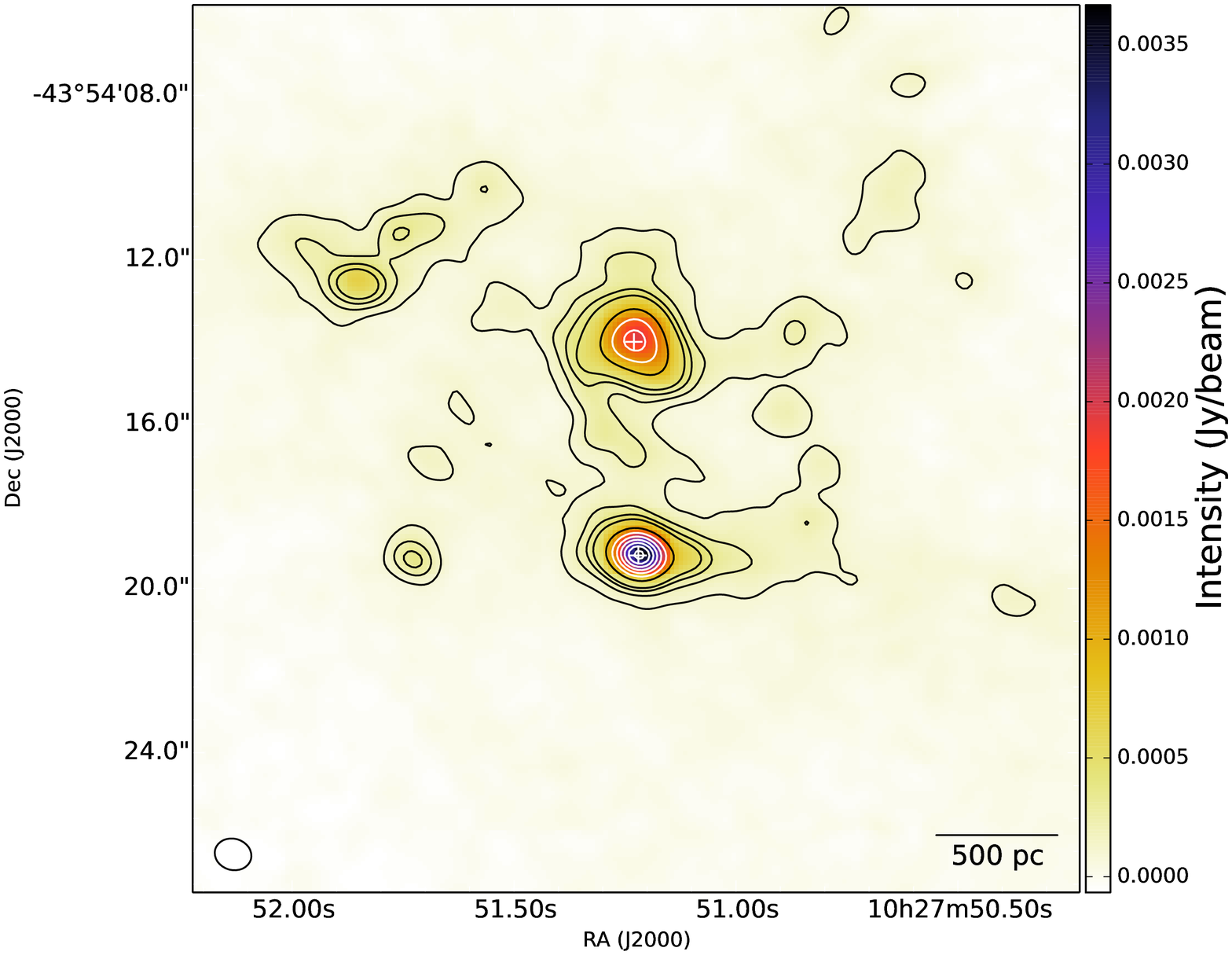}
\includegraphics[width=.5\textwidth, trim =  0 0 0 0]{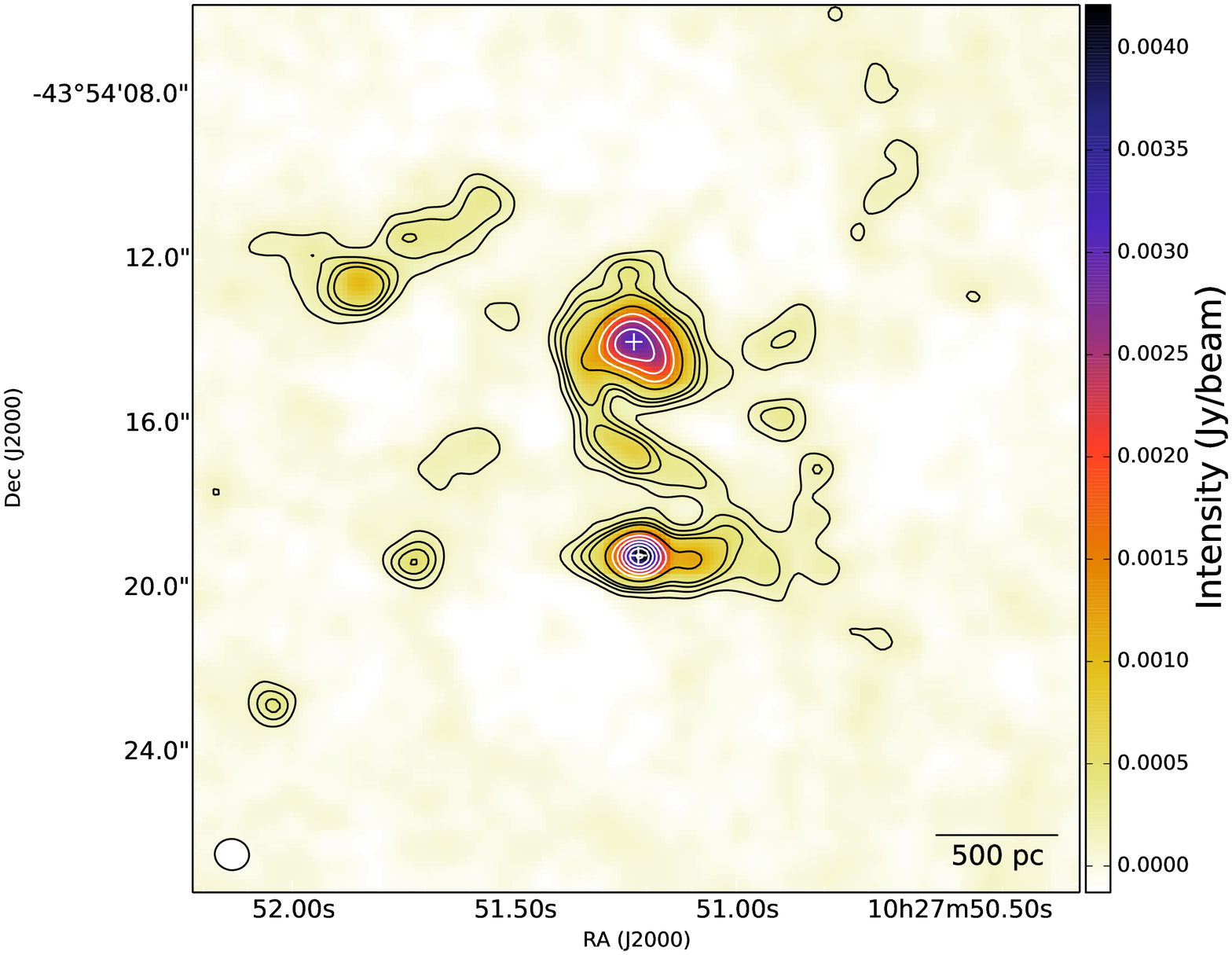}
}
\caption{Continuum maps at ($left$) $\lambda=3.0$\,mm and ($right$) $\lambda=1.2$\,mm. Contour levels are ($left$) 5 $\sigma$, 10 $\sigma$, 15 $\sigma$, 20 $\sigma$, and every 20 $\sigma$ afterwards where 1 $\sigma = 0.022$\,mJy, and 3 $\sigma$, 6 $\sigma$, 9 $\sigma$, 12 $\sigma$, and every 12 $\sigma$ afterwards where 1 $\sigma=0.043\,$mJy. \label{fig:cont}}
\end{figure*}

\begin{figure*}
\centerline{
\includegraphics[width=.5\textwidth, trim =  0 0 0 0]{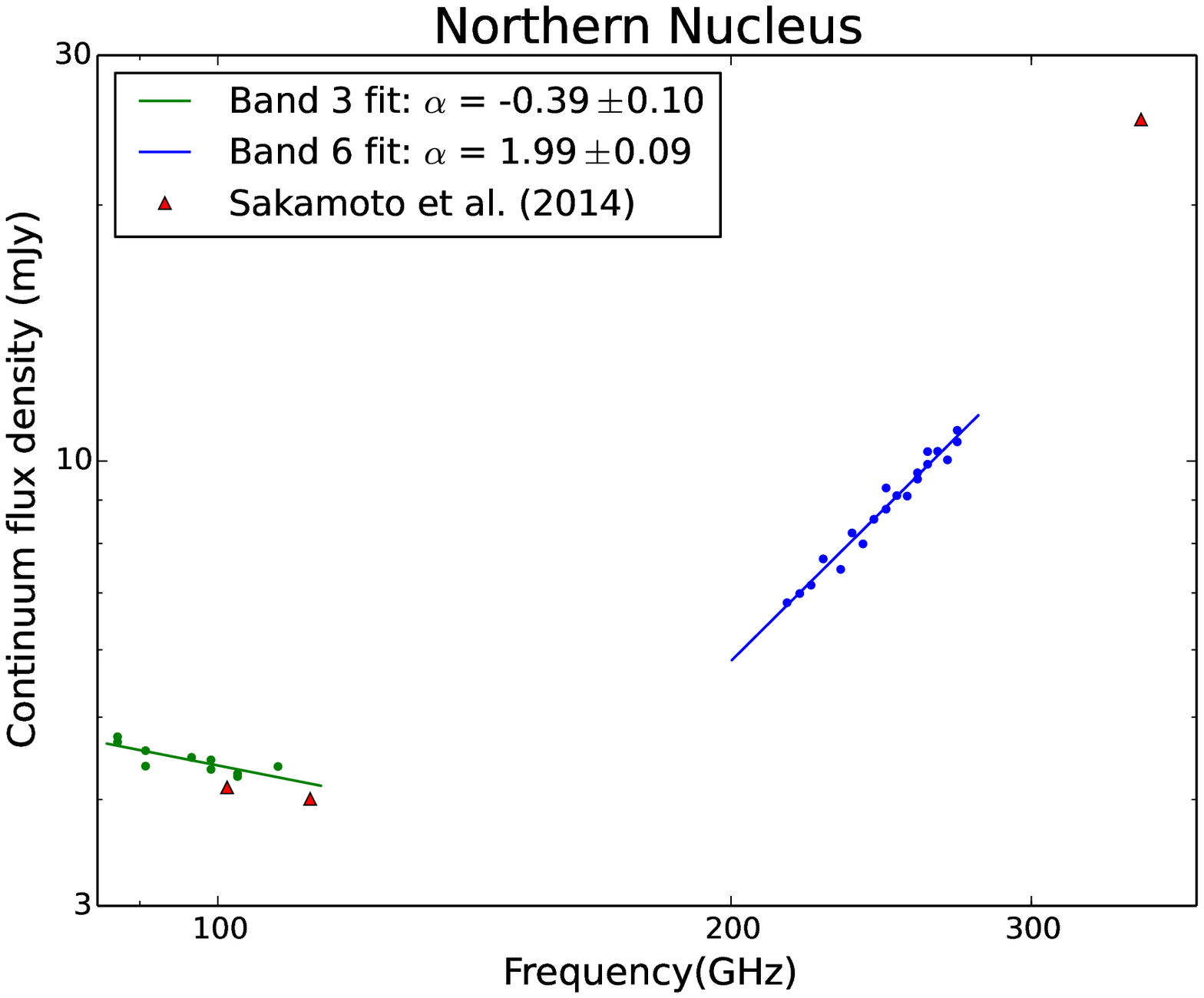}
\includegraphics[width=.5\textwidth, trim =  0 0 0 0]{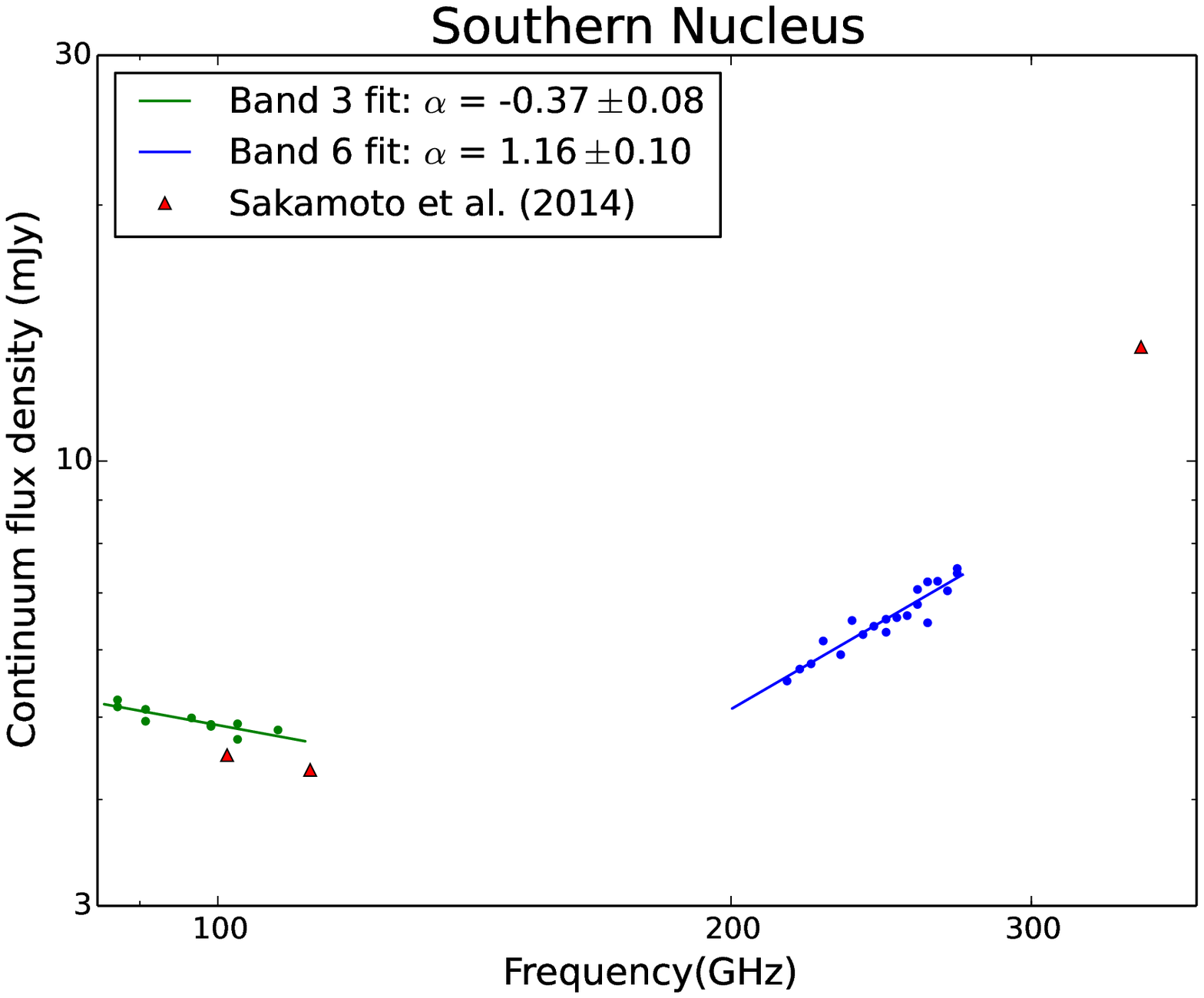}
}
\caption{Continuum flux densities within $2.0''$ from the radio peaks of (left:) the northern nucleus and (right:) the southern nucleus
as a function of frequency from our observations in the log scale on both axes. Solid line show fits of spectral index for Bands 3 and 6. 
Values from Cycle 0 by \citet{2014ApJ...797...90S} are also plotted as a reference with red triangles. 
The value in Band 7 from the Cycle 0 observation was taken by using both sidebands. \label{fig:cont_fit}}
\end{figure*}

Continuum images at $\lambda = 3.0$\,mm and 1.2\,mm are shown in Figure \ref{fig:cont}.
Images at these two wavelengths are similar to each other, and to continuum images at 2.81 mm, 0.86\,mm in \citet{2014ApJ...797...90S}.
Those images are overall similar to the radio continuum image at 3.6 cm by \citet{2003ApJ...599.1043N}, 
but only the 3.6-cm image has the feature south from the southern galaxy at the position of outflow from the S nucleus
 \citep[see Figure 16 of ][ for the 3.6-cm feature at the position of the outflow]{2014ApJ...797...90S}.
Spectral indices $\alpha$ of N and S nuclei where $S_{\nu} \propto \nu^{\alpha}$ are derived using the continuum data using 
each scheduling block in each sideband. We use only the PI data (2015.1.00412.S and 2016.1.00975.S) for the fit,
and the fits were obtained for Band 3 and Band 6 separately. 
Continuum flux densities within $2''$ are plotted in Figure \ref{fig:cont_fit}.
Each sideband has a 1.875-GHz width, and the frequency ranges 
of each sideband can be found in Table \ref{tab:obs_param}.
Those values are obtained after convolving all the images to 2.0$''$.
For the northern nucleus, $\alpha = -0.39 \pm 0.10$ for Band 3, and $\alpha = 1.99 \pm 0.09$ for Band 6. The value in the southern nucleus is
equivalent to the northern nucleus in Band 3 ($\alpha = -0.37 \pm 0.08$), but the one for Band 6 is lower than the northern nucleus ($\alpha=1.16 \pm 0.10$).
Values obtained from $f_{\rm obs} = 85.5 - 110.3$ GHz are lower than the ones obtained from $f_{\rm obs} = 99.6 - 115.0$ GHz 
by \citet{2014ApJ...797...90S}, which are -0.12 for northern nucleus and -0.19 for southern nucleus.
This is reasonable because there should be more contribution from Synchrotron radiation ($\alpha \sim -1$) than the contribution
 from free-free emission ($\alpha \sim -0.1$) at the lower frequency. 
 At the same time, this level of change in $\alpha$ can be easily caused by the flux error,
 which seems to be a more likely cause (see Figure \ref{fig:cont_fit}).
Similarly, values of $\alpha$ at $\lambda=0.86\,$mm in \citet{2014ApJ...797...90S} are higher than the ones at $\lambda=1.2\,$mm
because the dust emission ($\alpha \sim 3-4$) has more contribution at the higher frequency.

\subsection{Molecular line images}\label{sec:molimage}
\begin{figure*} 
\centerline{
\includegraphics[width=.33\textwidth, trim =  0 0 0 0]{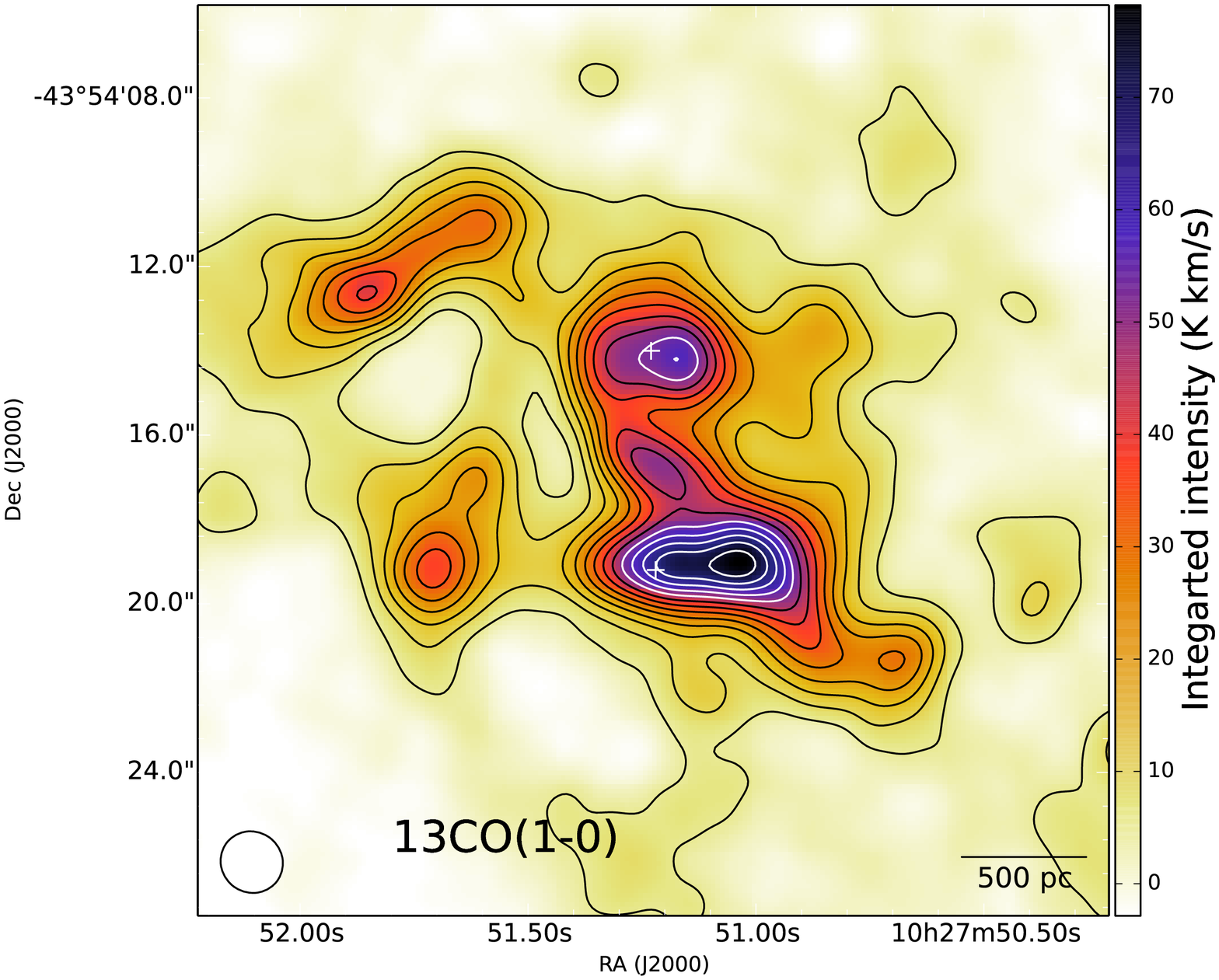} 
\includegraphics[width=.33\textwidth, trim =  0 0 0 0]{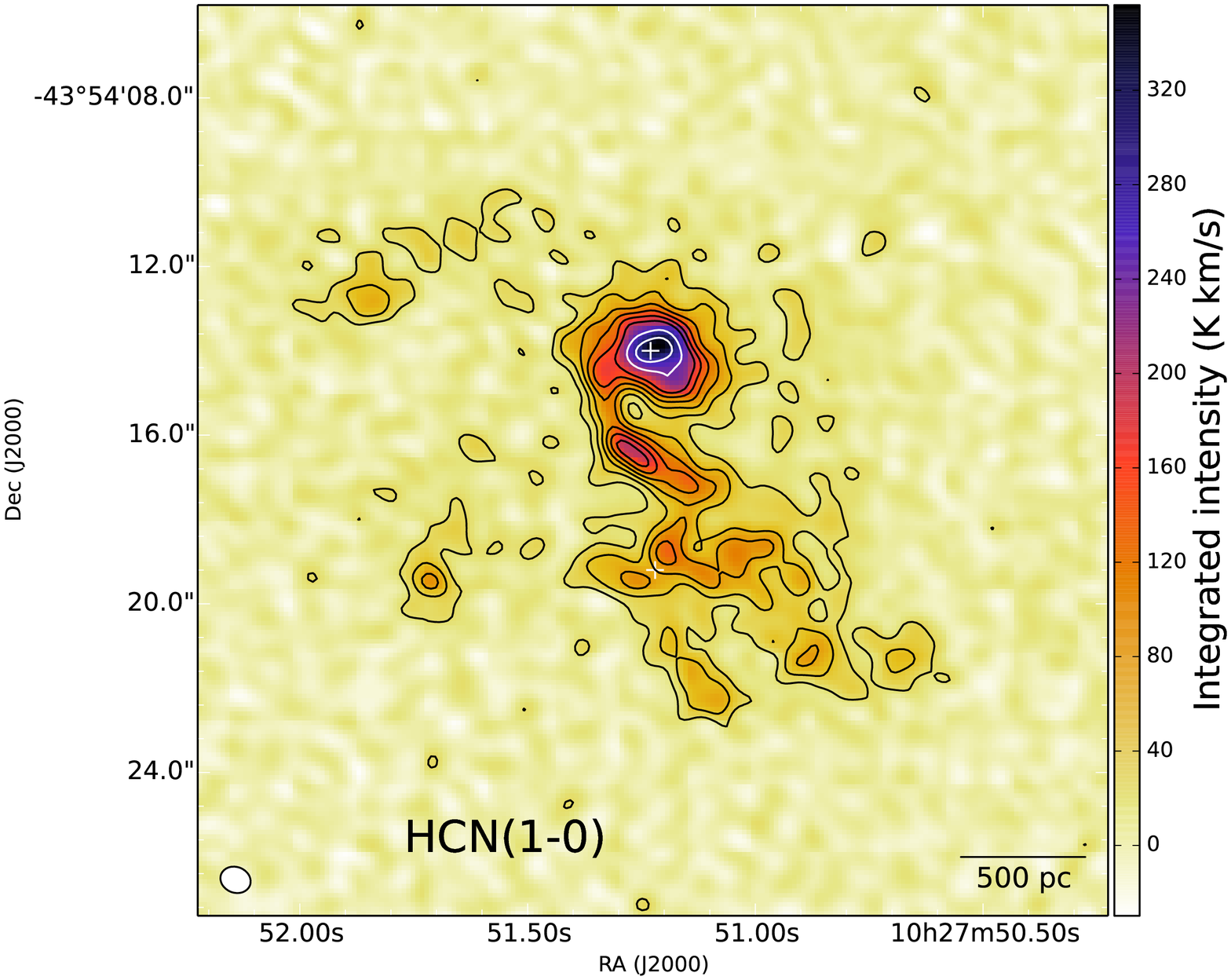} 
\includegraphics[width=.33\textwidth, trim =  0 0 0 0]{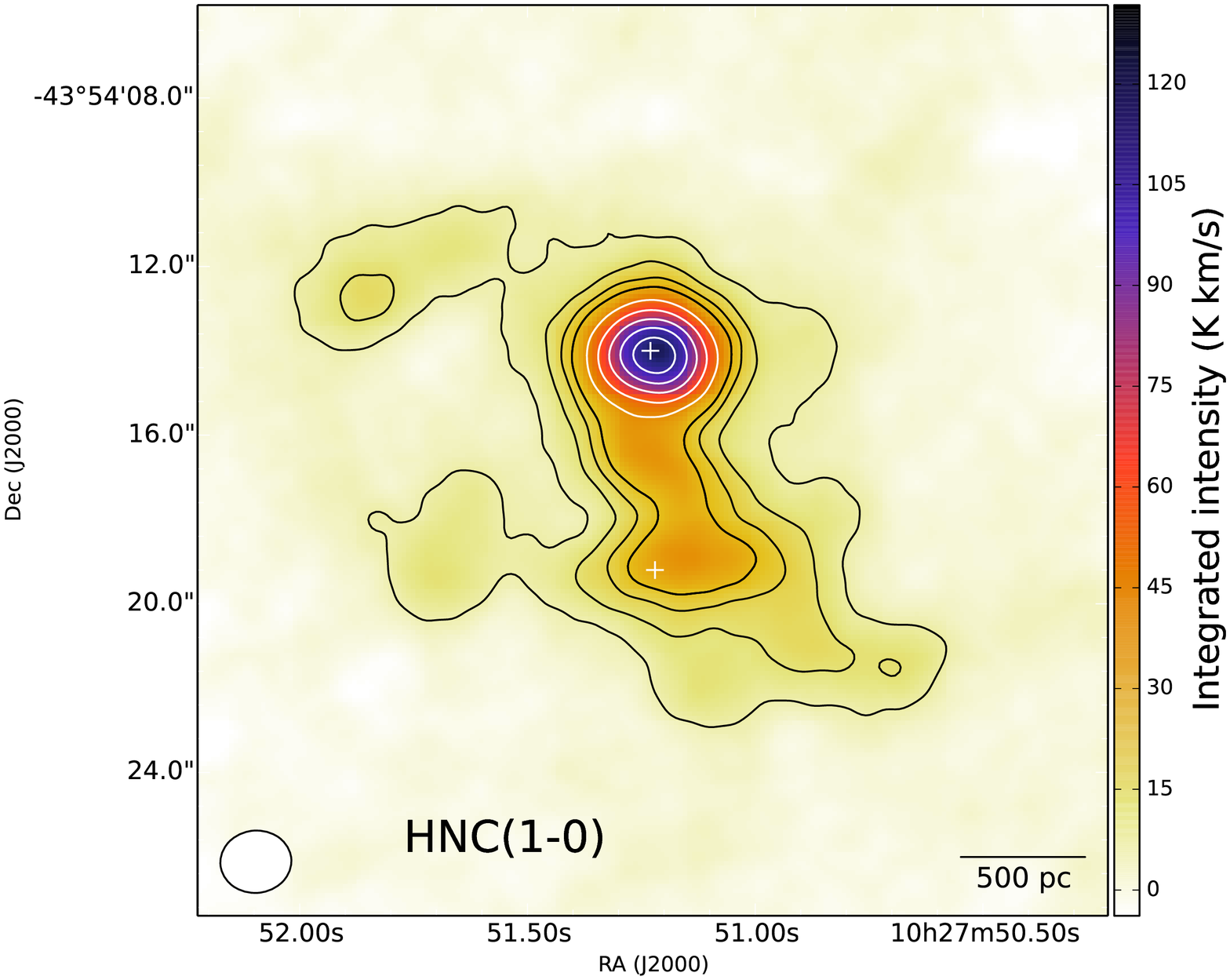} 
}

\centerline{
\includegraphics[width=.33\textwidth, trim =  0 0 0 0]{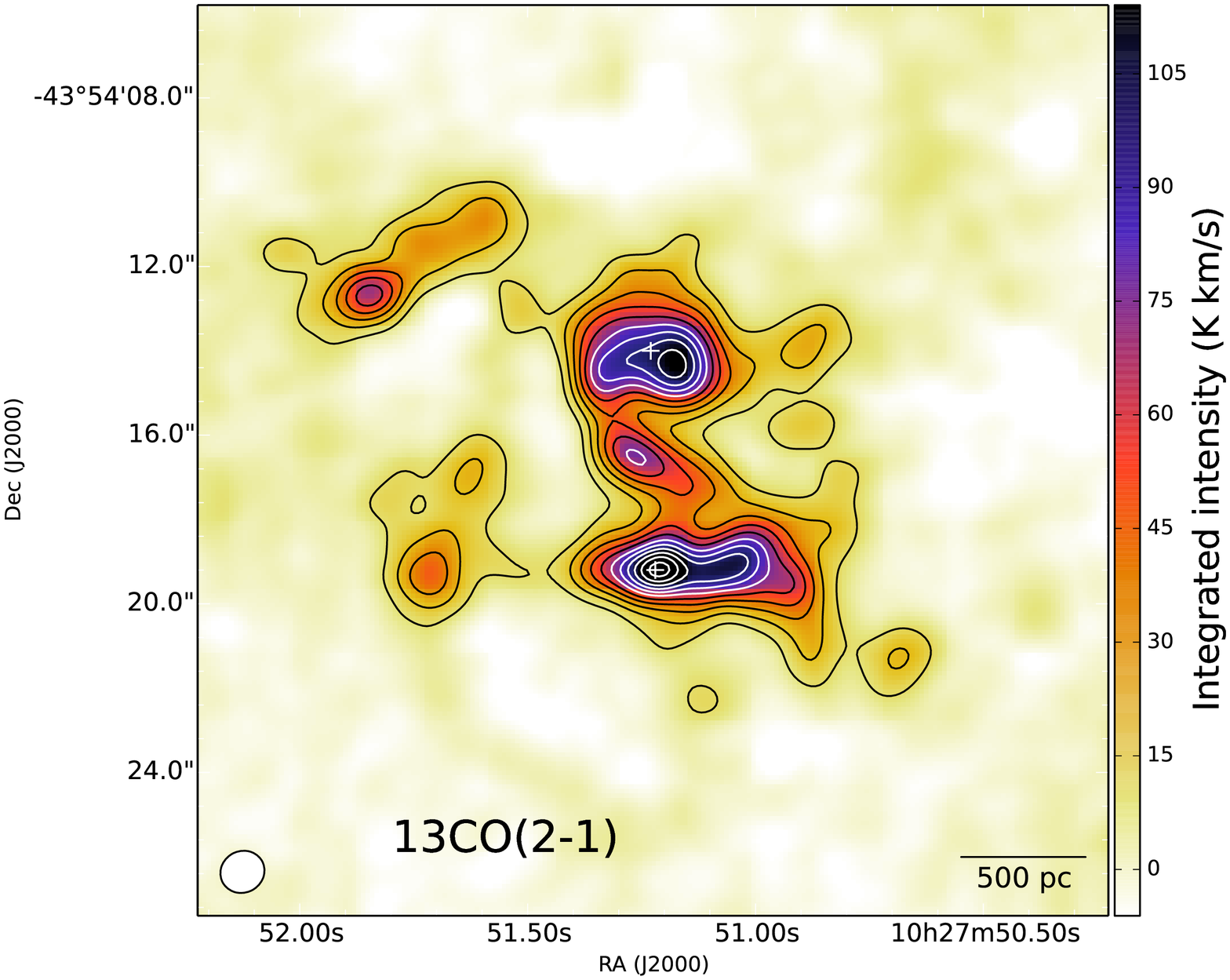} 
\includegraphics[width=.33\textwidth, trim =  0 0 0 0]{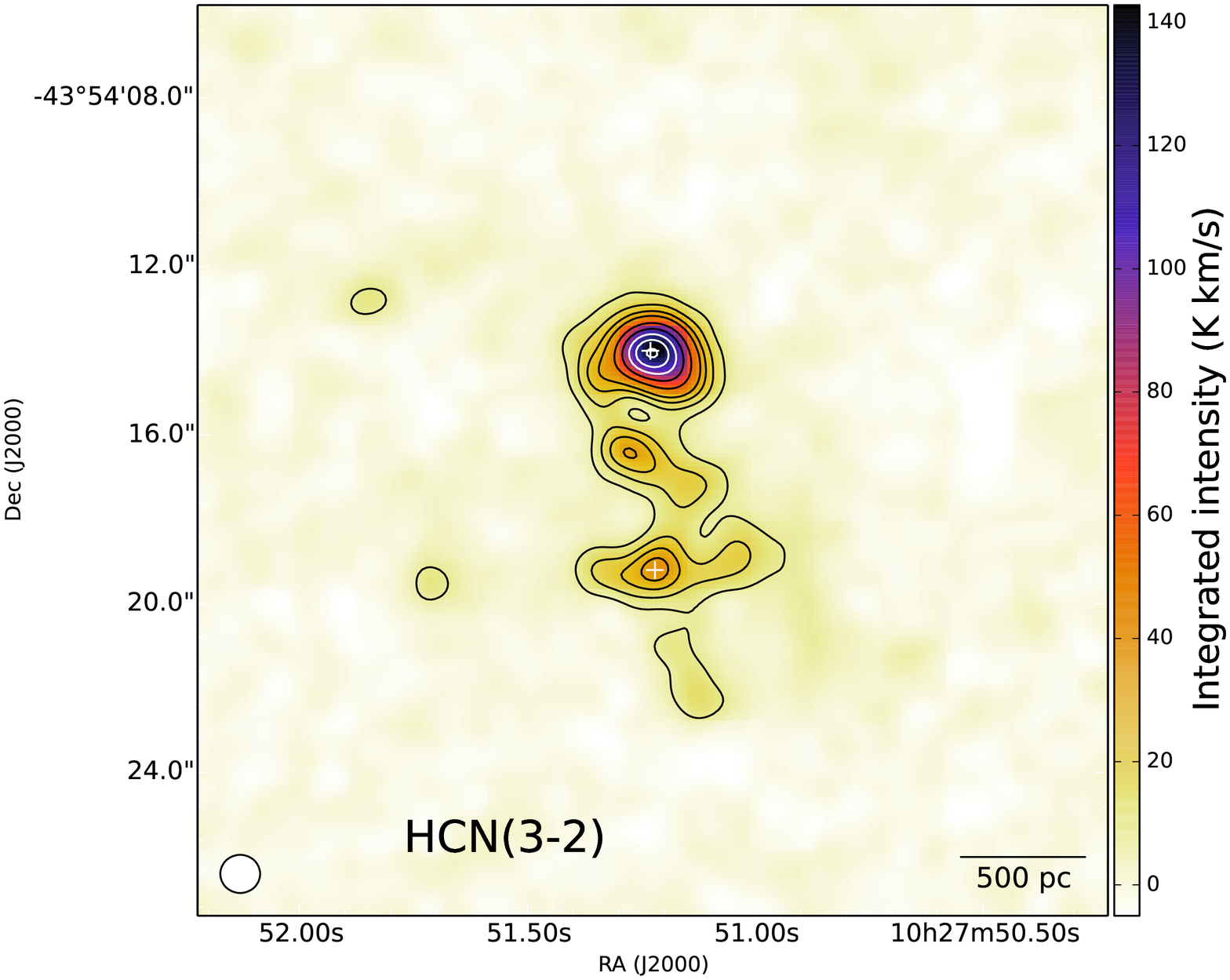} 
\includegraphics[width=.33\textwidth, trim =  0 0 0 0]{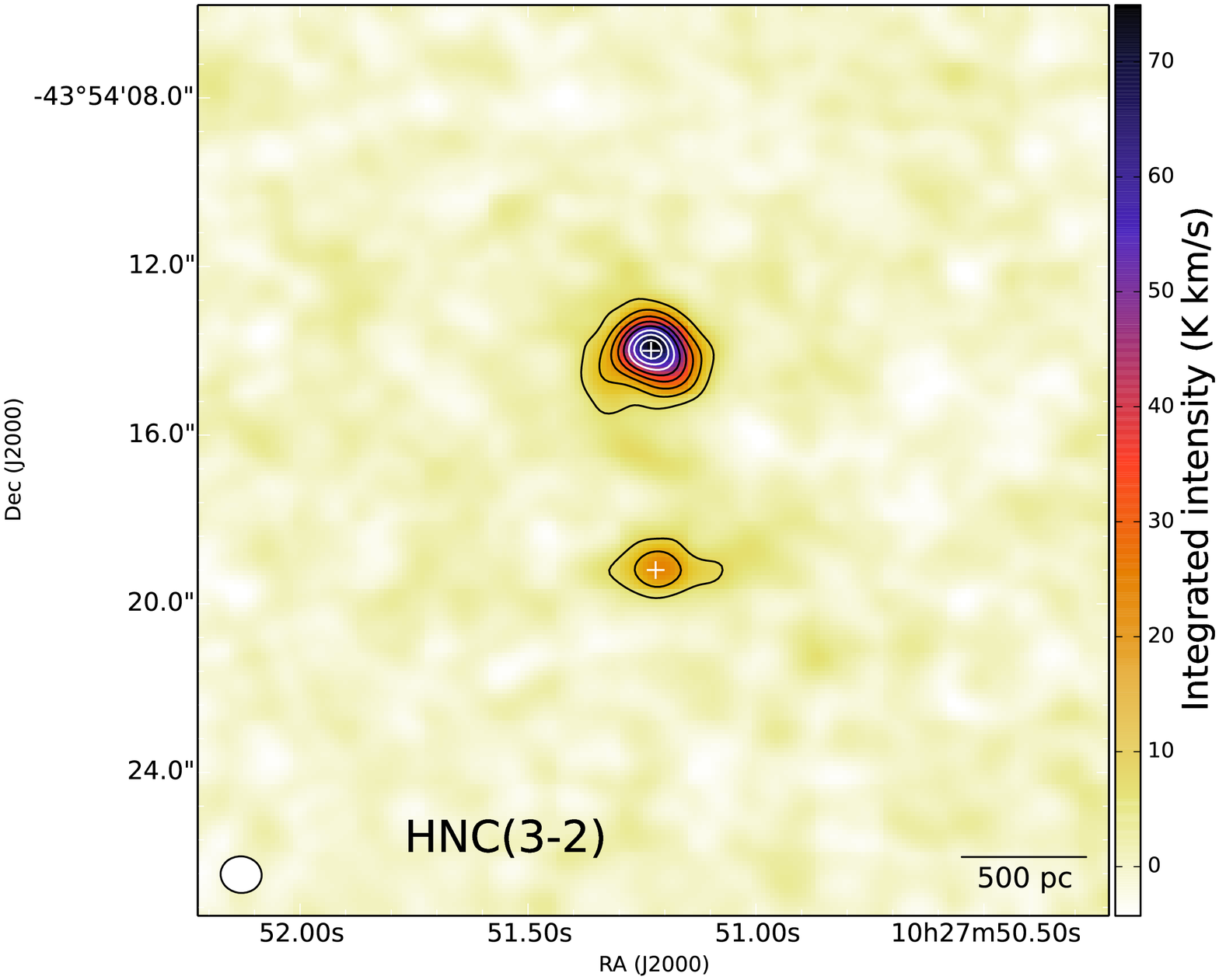}
}
\centerline{
\includegraphics[width=.33\textwidth, trim =  0 0 0 0]{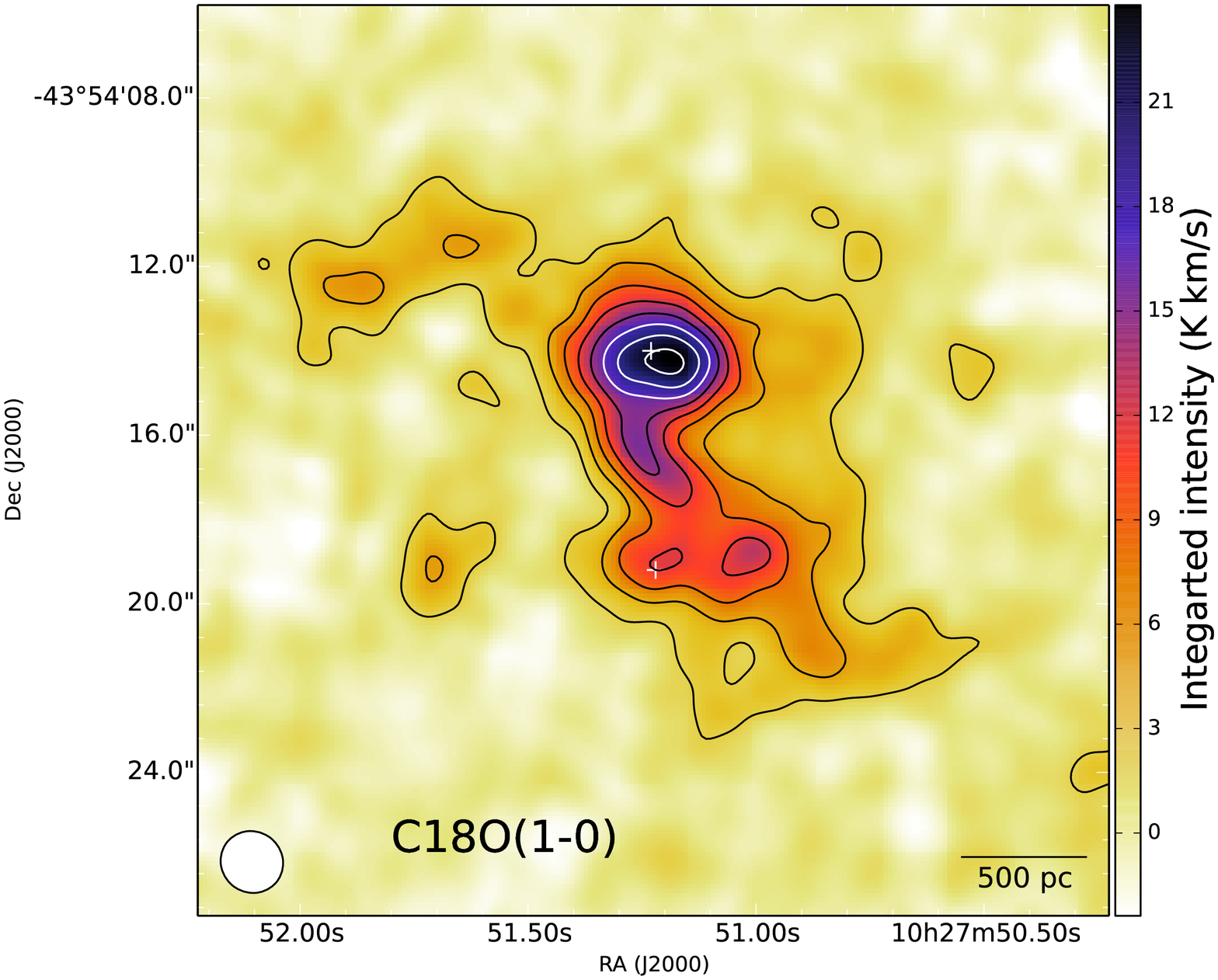} 
\includegraphics[width=.33\textwidth, trim =  0 0 0 0]{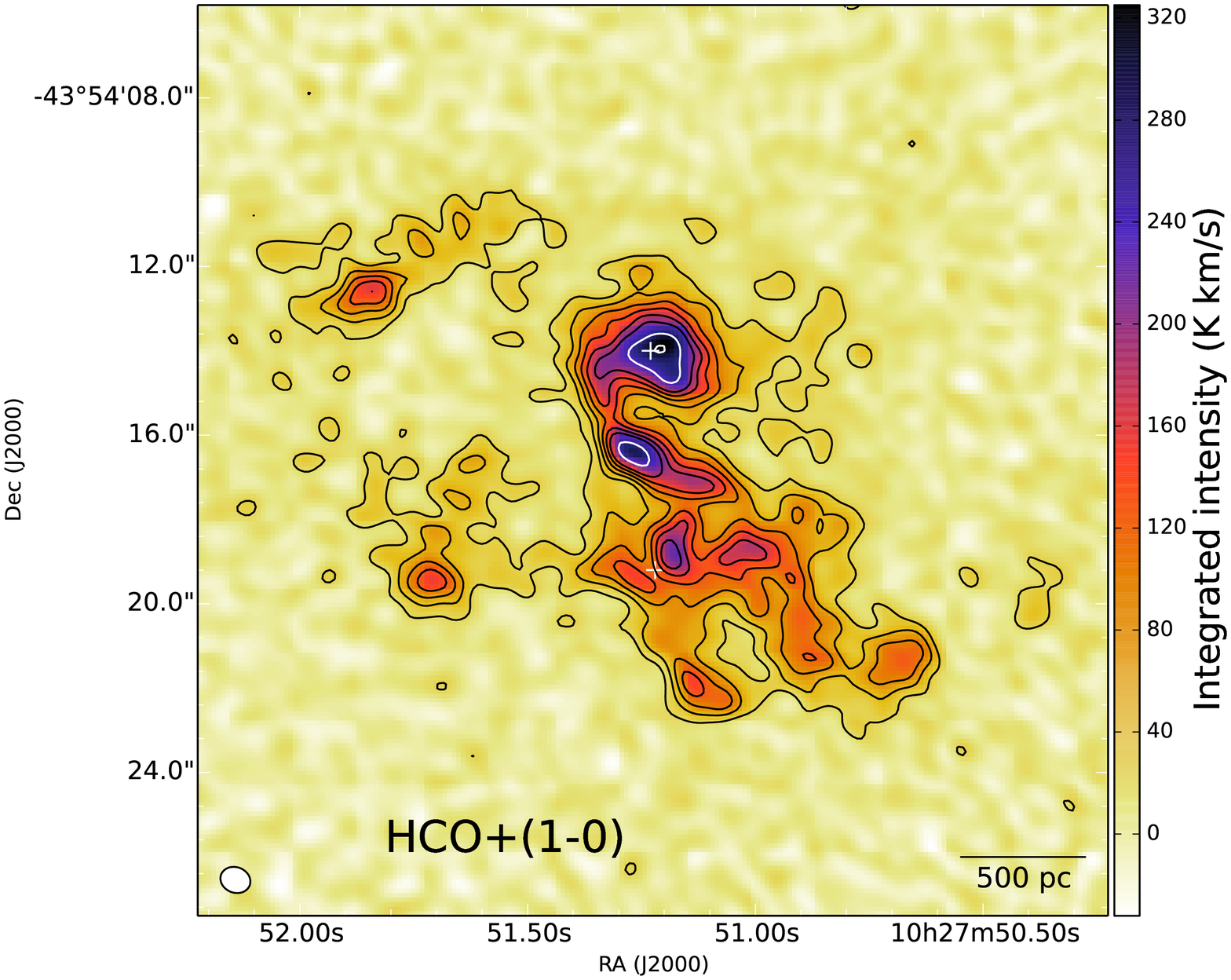} 
\includegraphics[width=.33\textwidth, trim =  0 0 0 0]{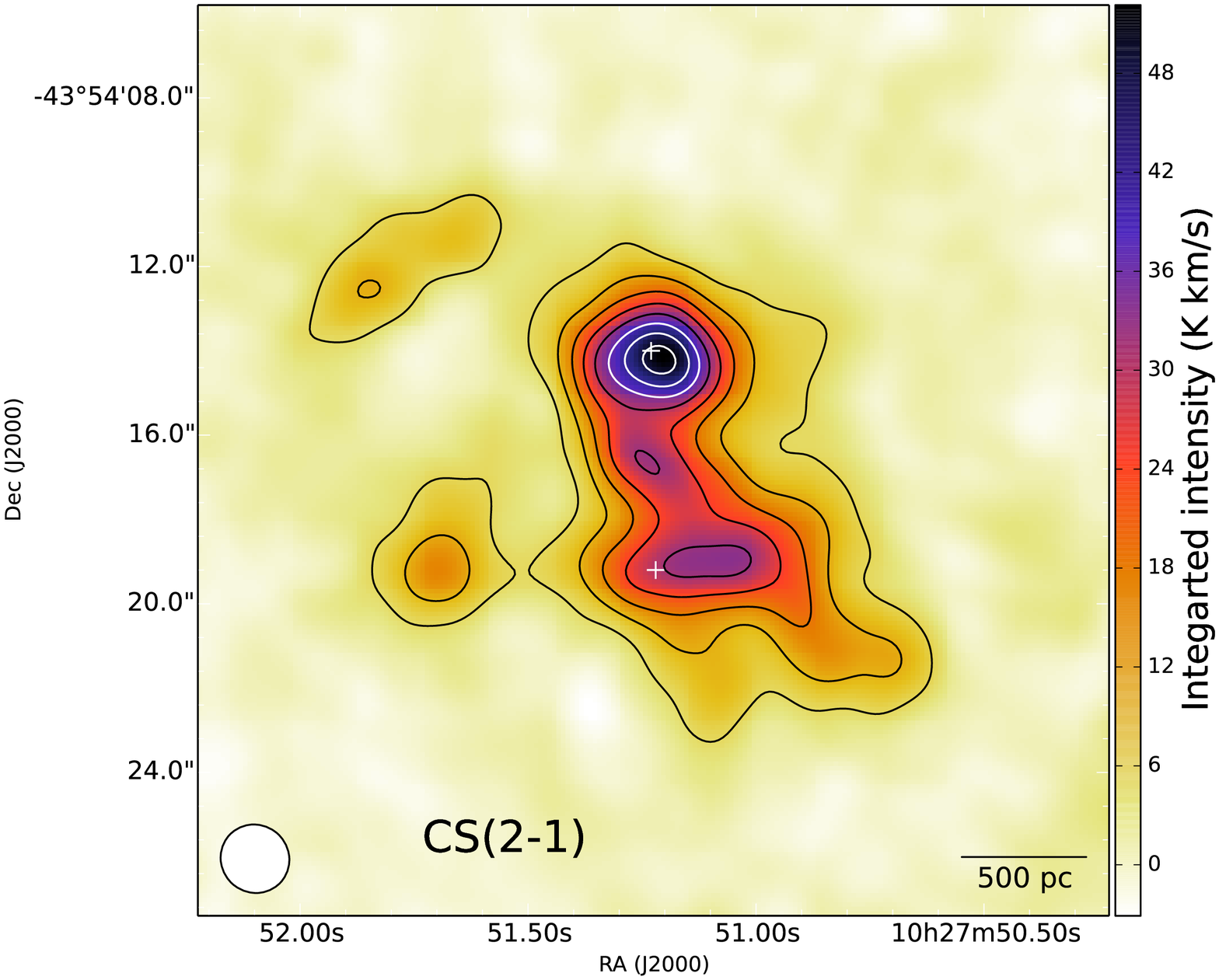} 
}
\centerline{
\includegraphics[width=.33\textwidth, trim =  0 0 0 0]{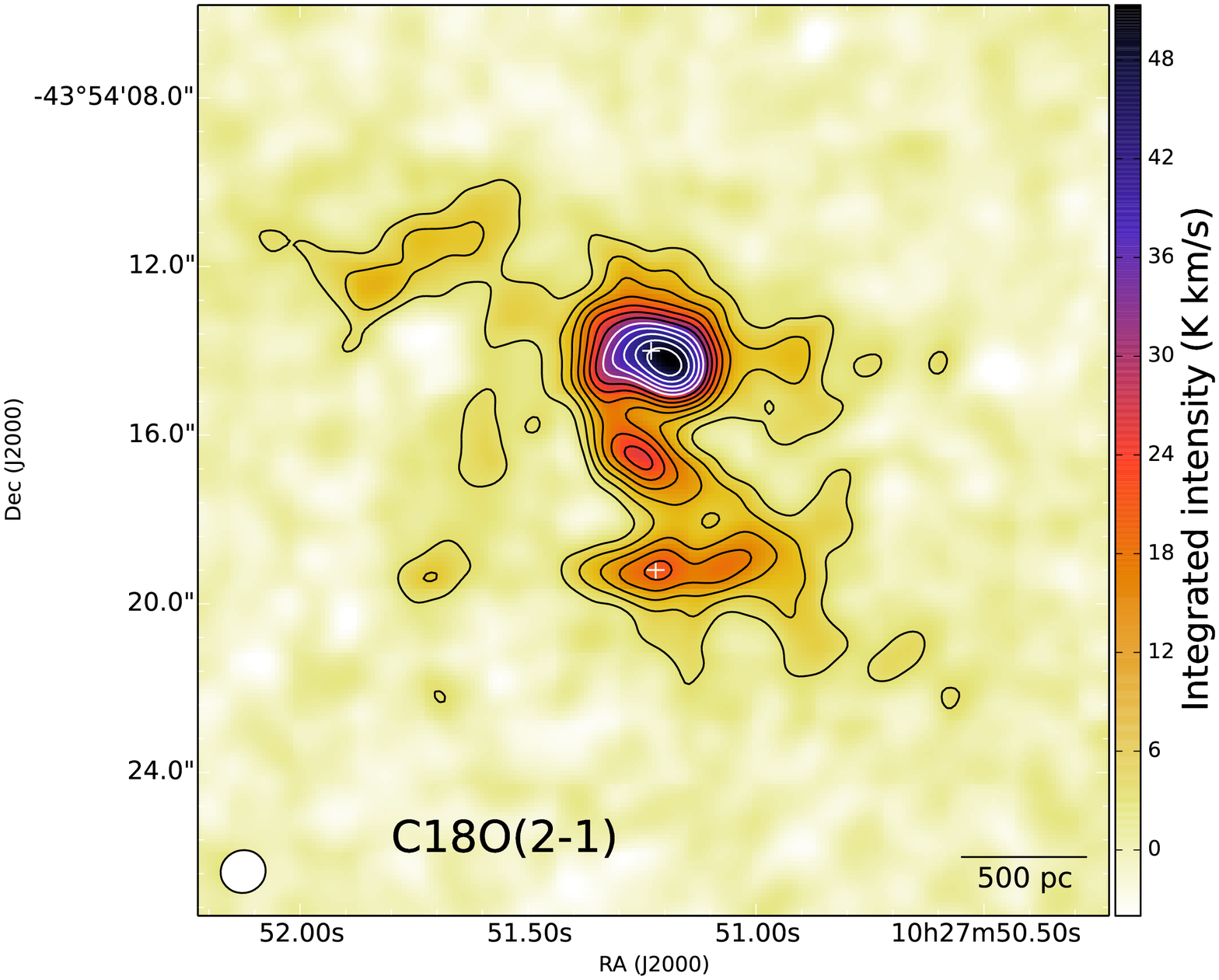}
\includegraphics[width=.33\textwidth, trim =  0 0 0 0]{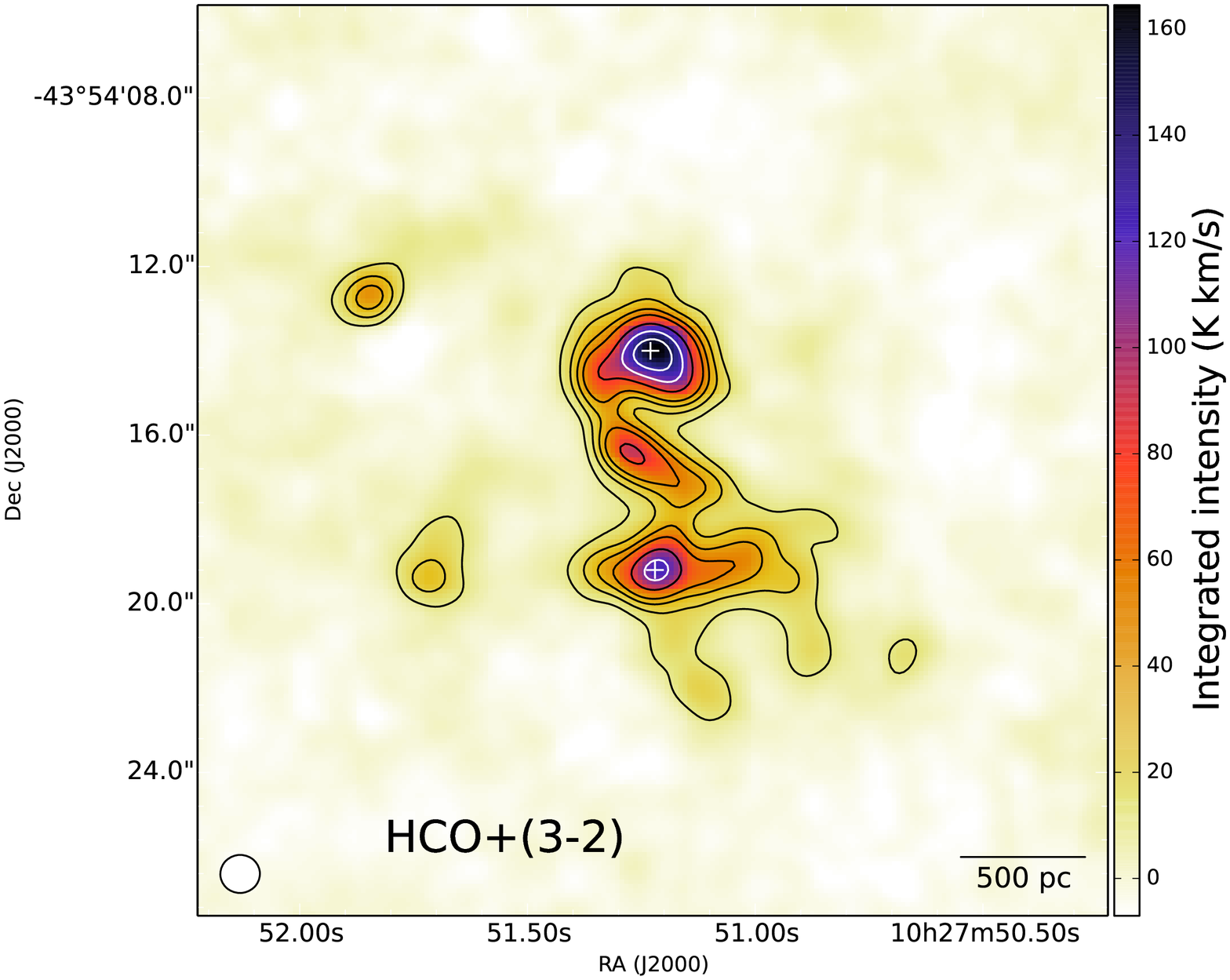} 
\includegraphics[width=.33\textwidth, trim =  0 0 0 0]{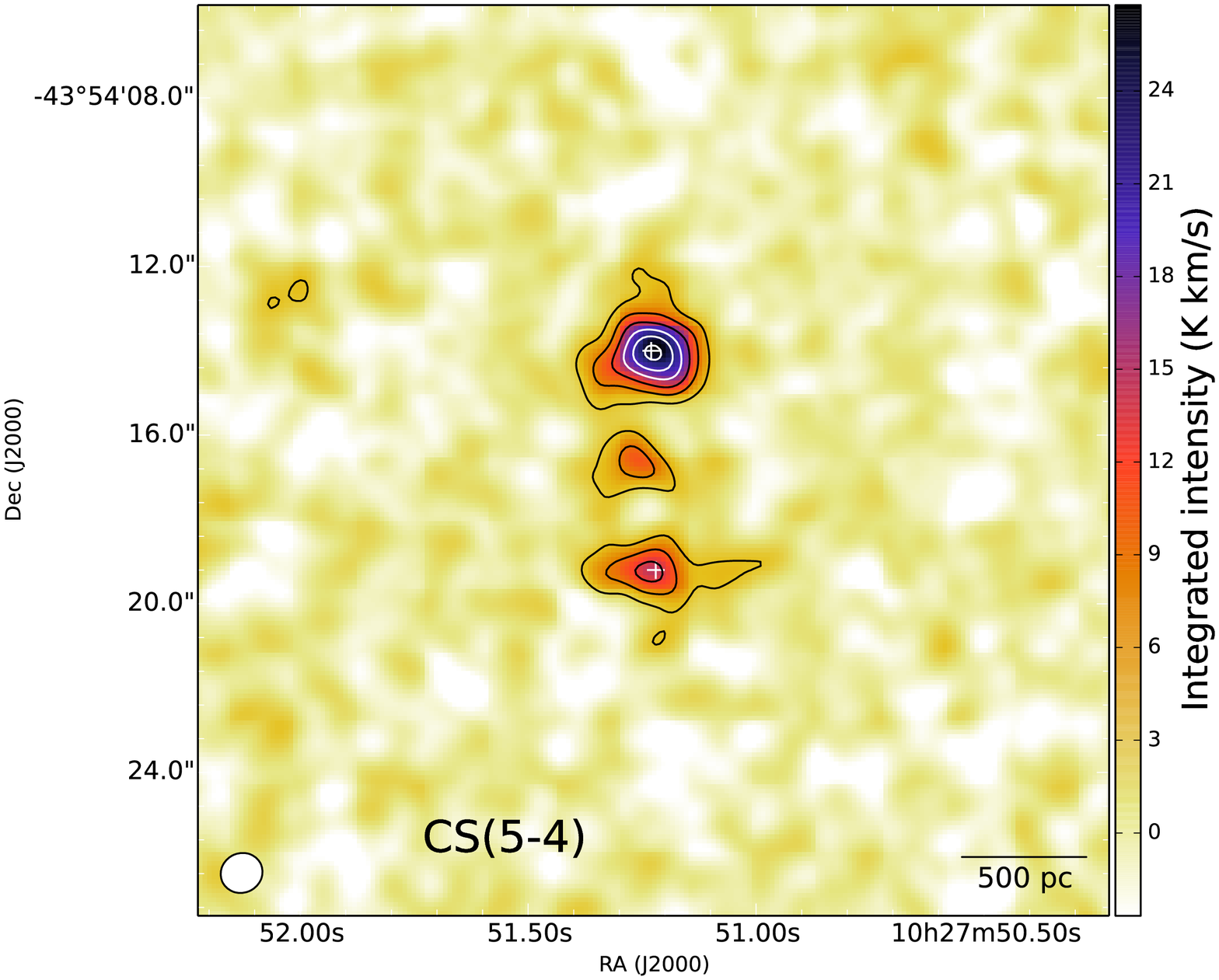} 
}
\caption{Velocity-integrated intensity maps of selected observed lines. Contour levels are every $6\, \sigma$ for $^{13}$CO ($1-0$), $^{13}$CO ($2-1$),
HNC($3-2$), every $3\, \sigma$ for HCN($1-0$), C$^{18}$O($1-0$), C$^{18}$O($2-1$), HCO$^+$($1-0$), CS($5-4$), $6\, \sigma$, $12\, \sigma$, $18\, \sigma$, $24\, \sigma$,  
and every $12\, \sigma$ for HCN($3-2$), HCO$^+$($3-2$), and HNC($1-0$). The rms values are listed in Table \ref{tab:imparline}. \label{fig:mom0}} 
\end{figure*}

\begin{figure*} 
\centerline{
\includegraphics[width=.33\textwidth, trim =  0 0 0 0]{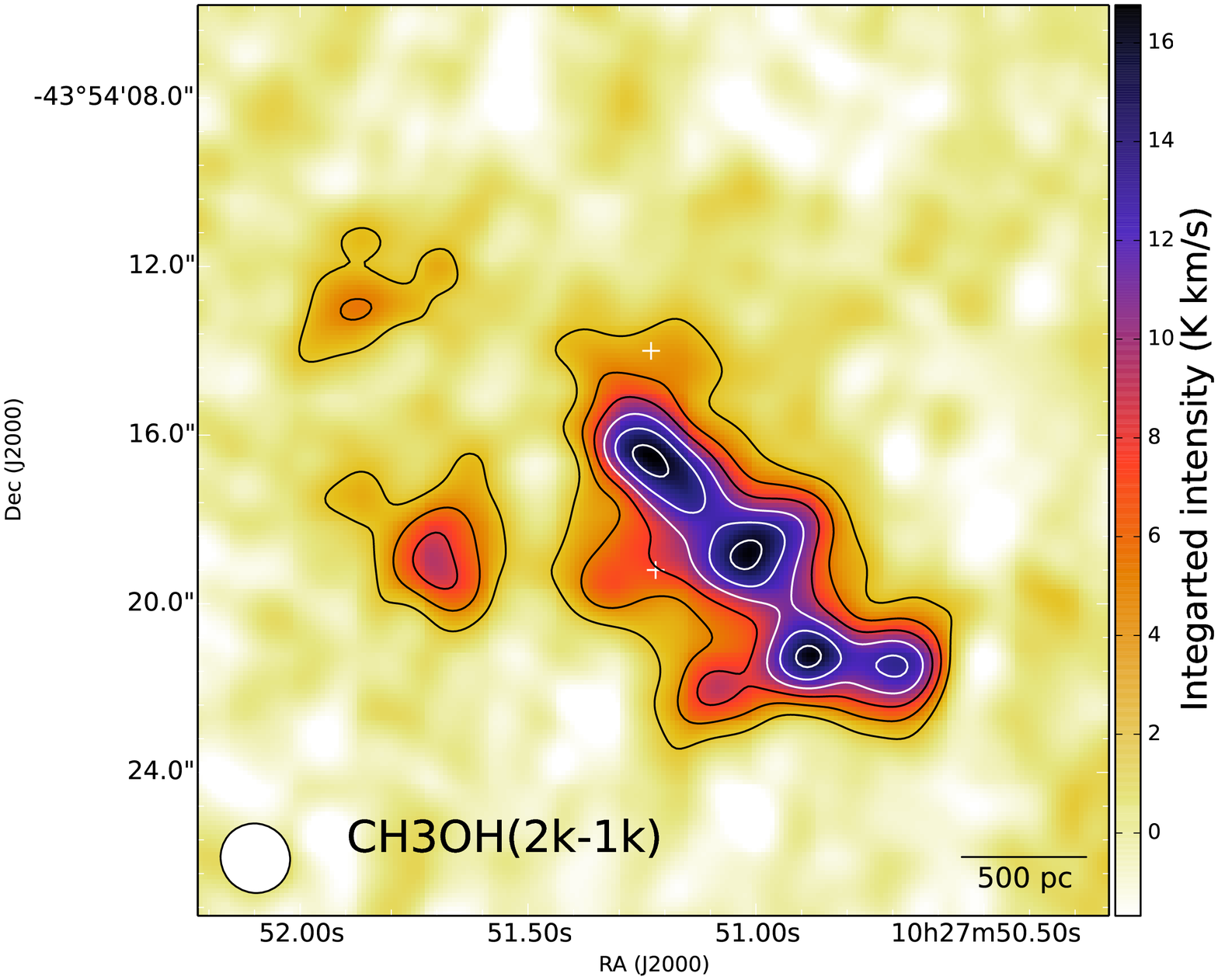}
\includegraphics[width=.33\textwidth, trim =  0 0 0 0]{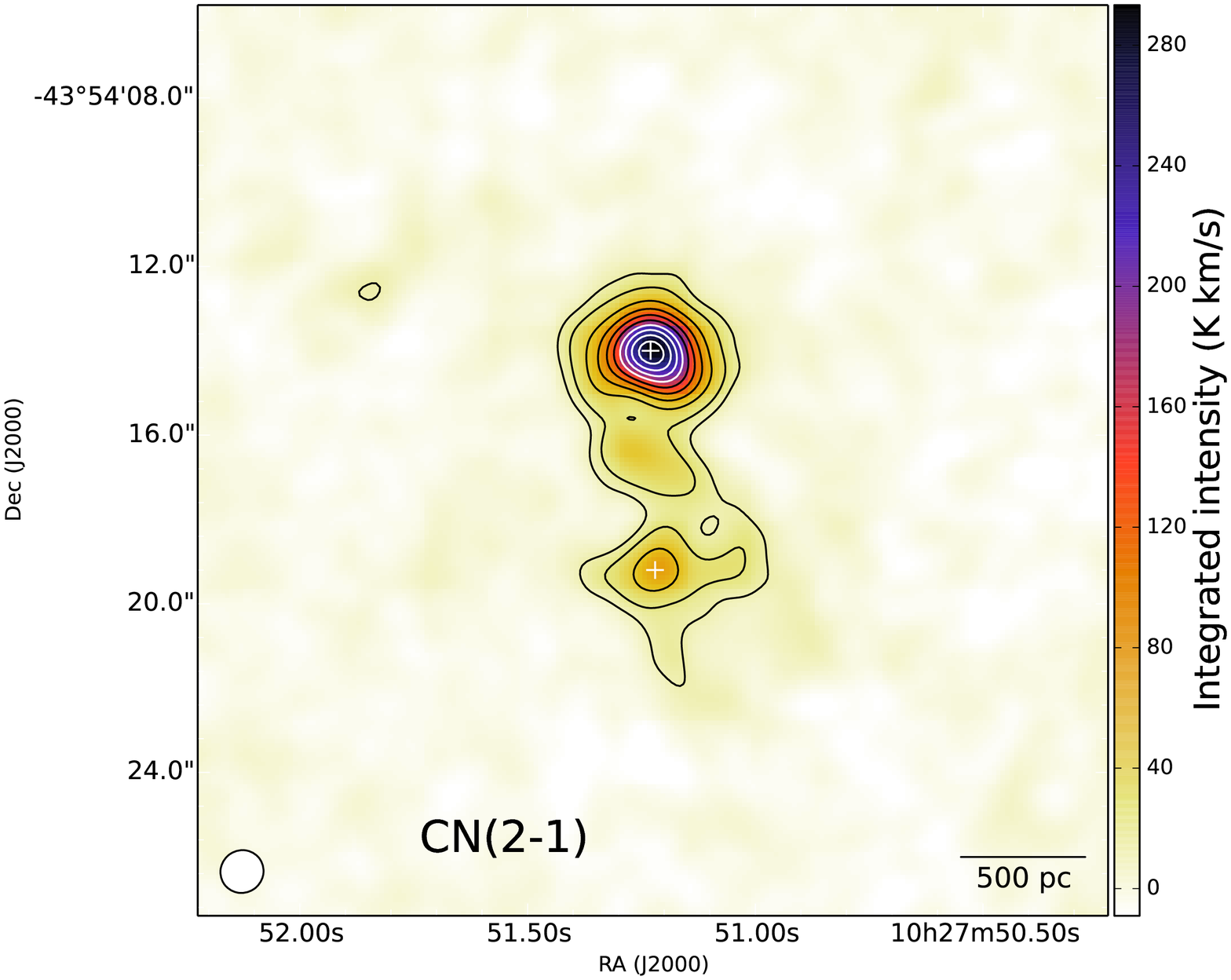} 
\includegraphics[width=.33\textwidth, trim =  0 0 0 0]{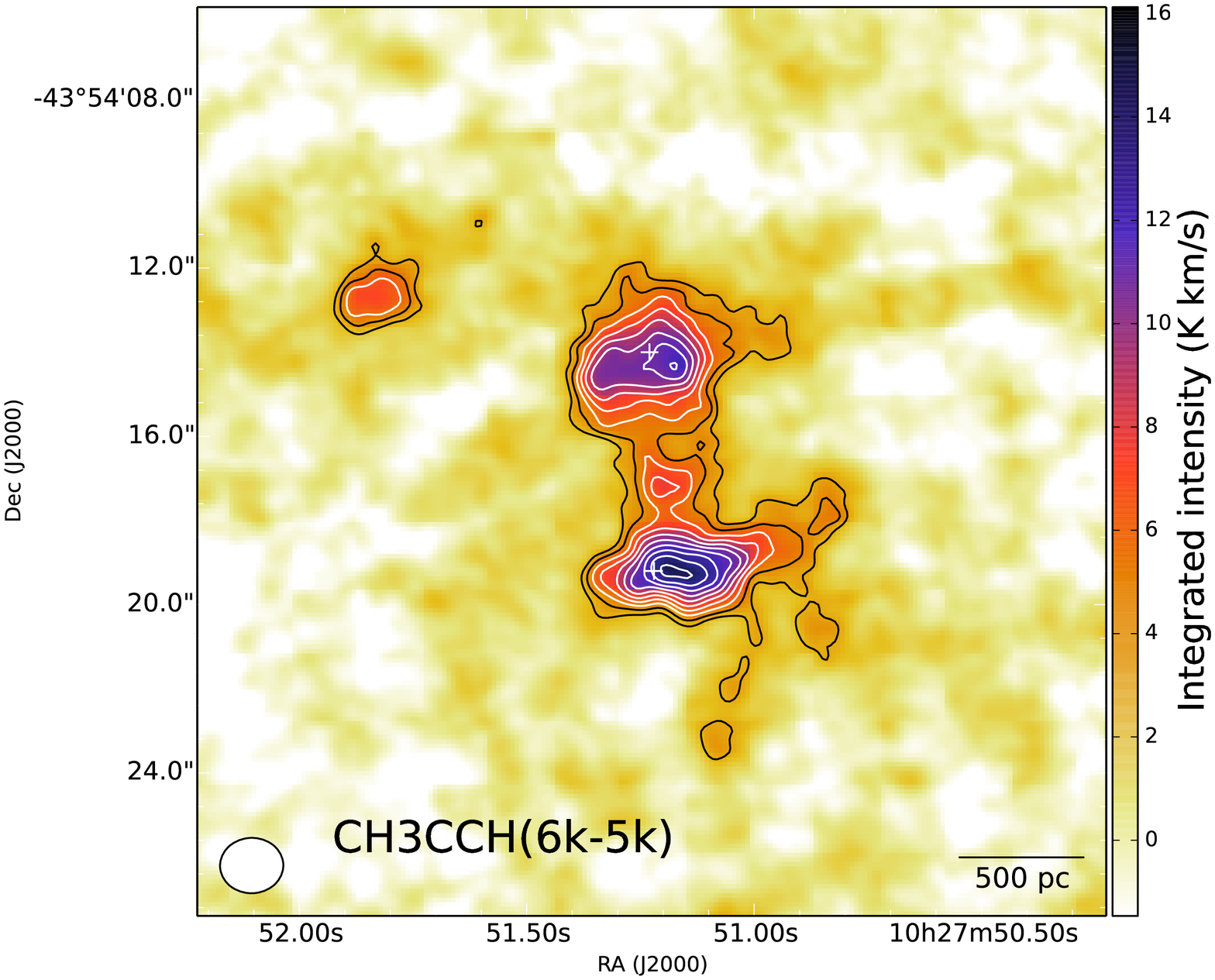} 
}
\centerline{
\includegraphics[width=.33\textwidth, trim =  0 0 0 0]{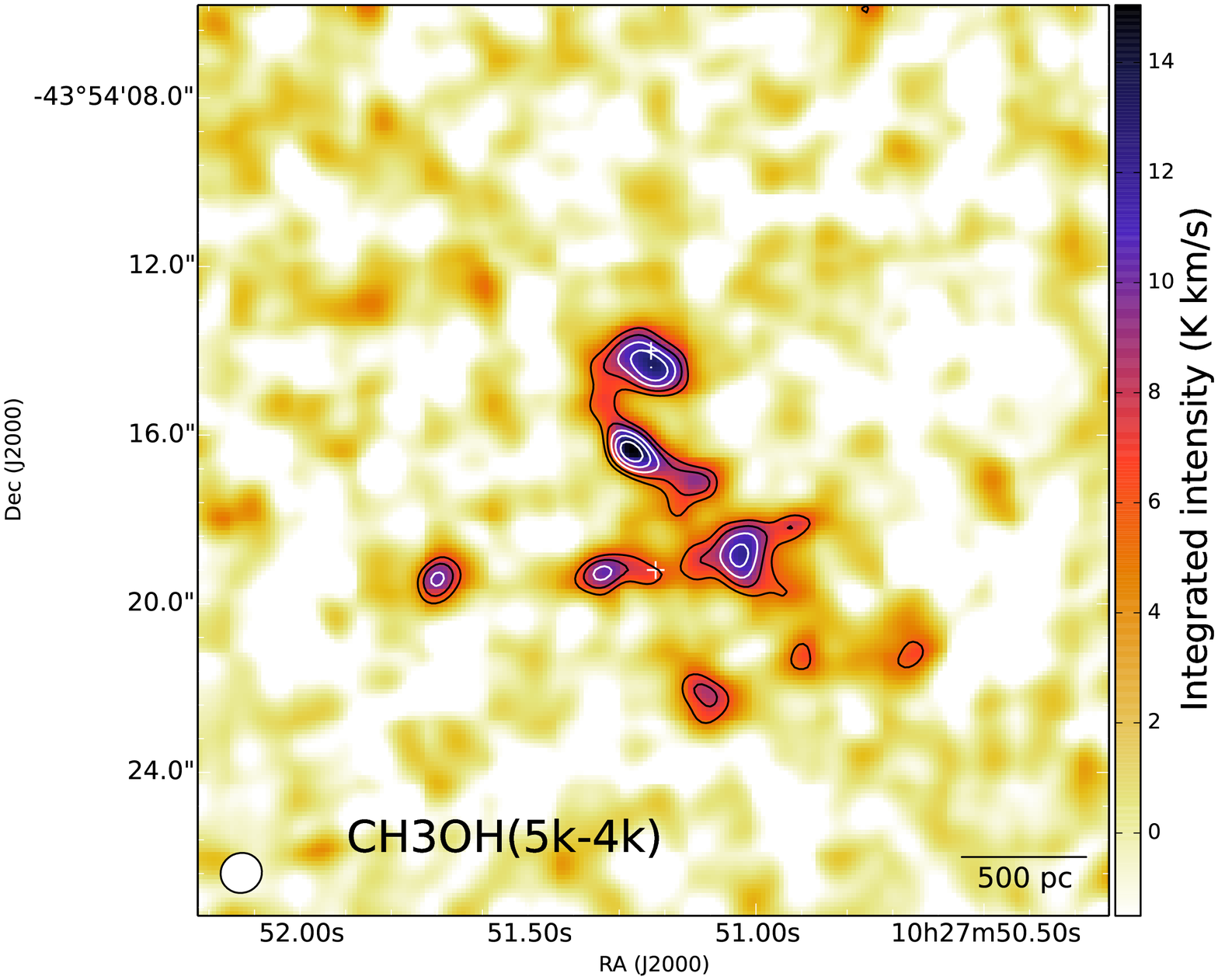} 
\includegraphics[width=.33\textwidth, trim =  0 0 0 0]{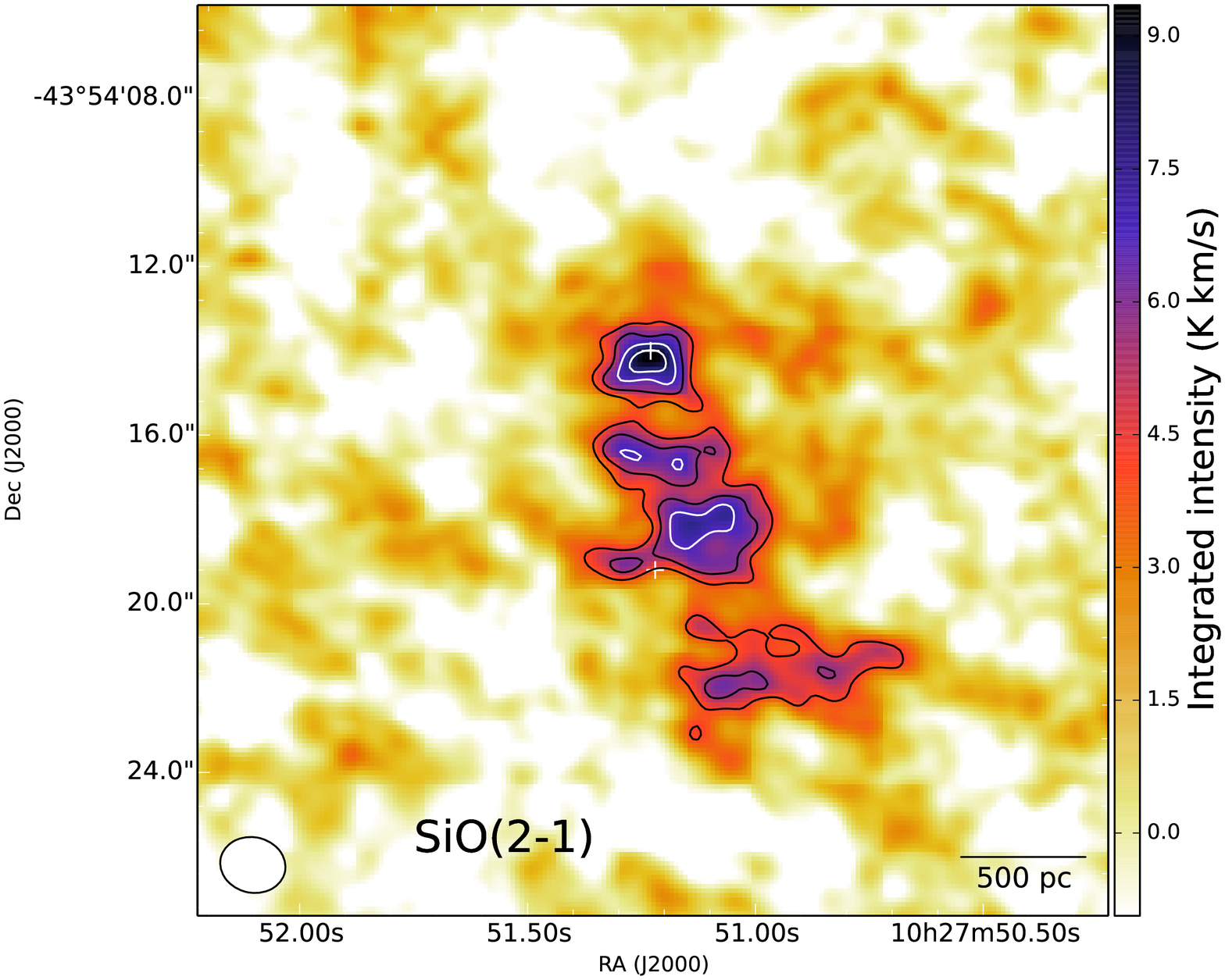} 
\includegraphics[width=.33\textwidth, trim =  0 0 0 0]{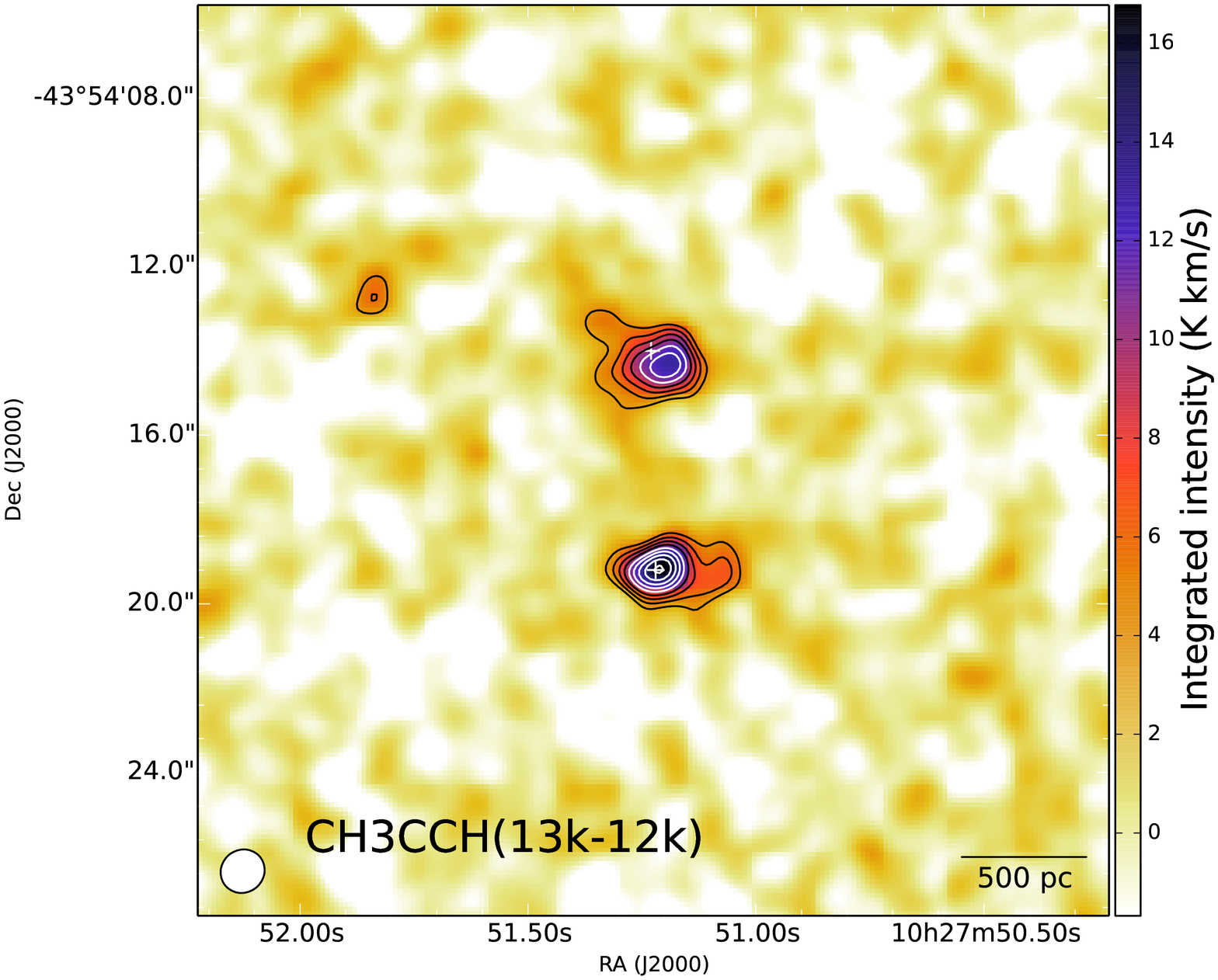} 
}
\centerline{
\includegraphics[width=.33\textwidth, trim =  0 0 0 0]{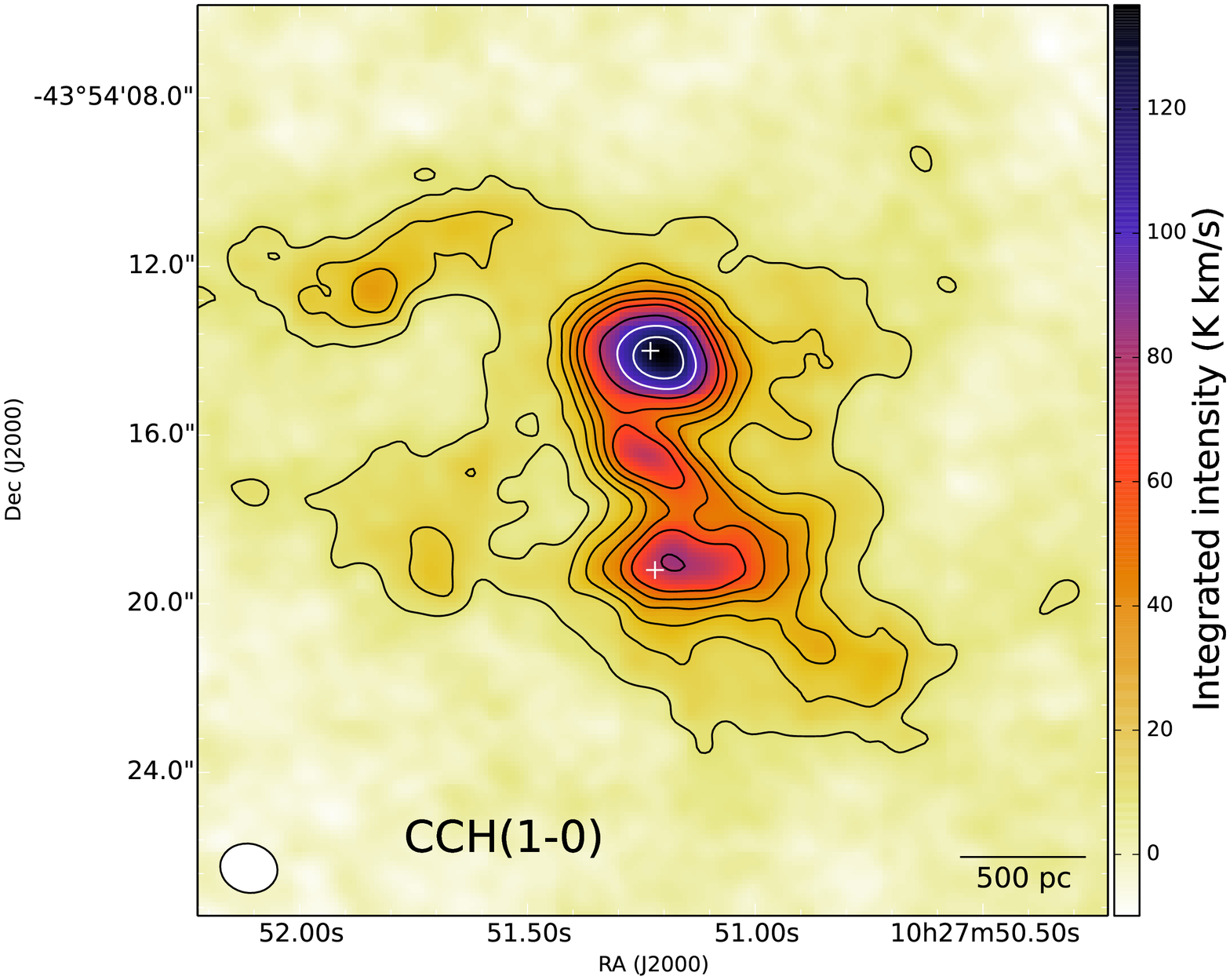} 
\includegraphics[width=.33\textwidth, trim =  0 0 0 0]{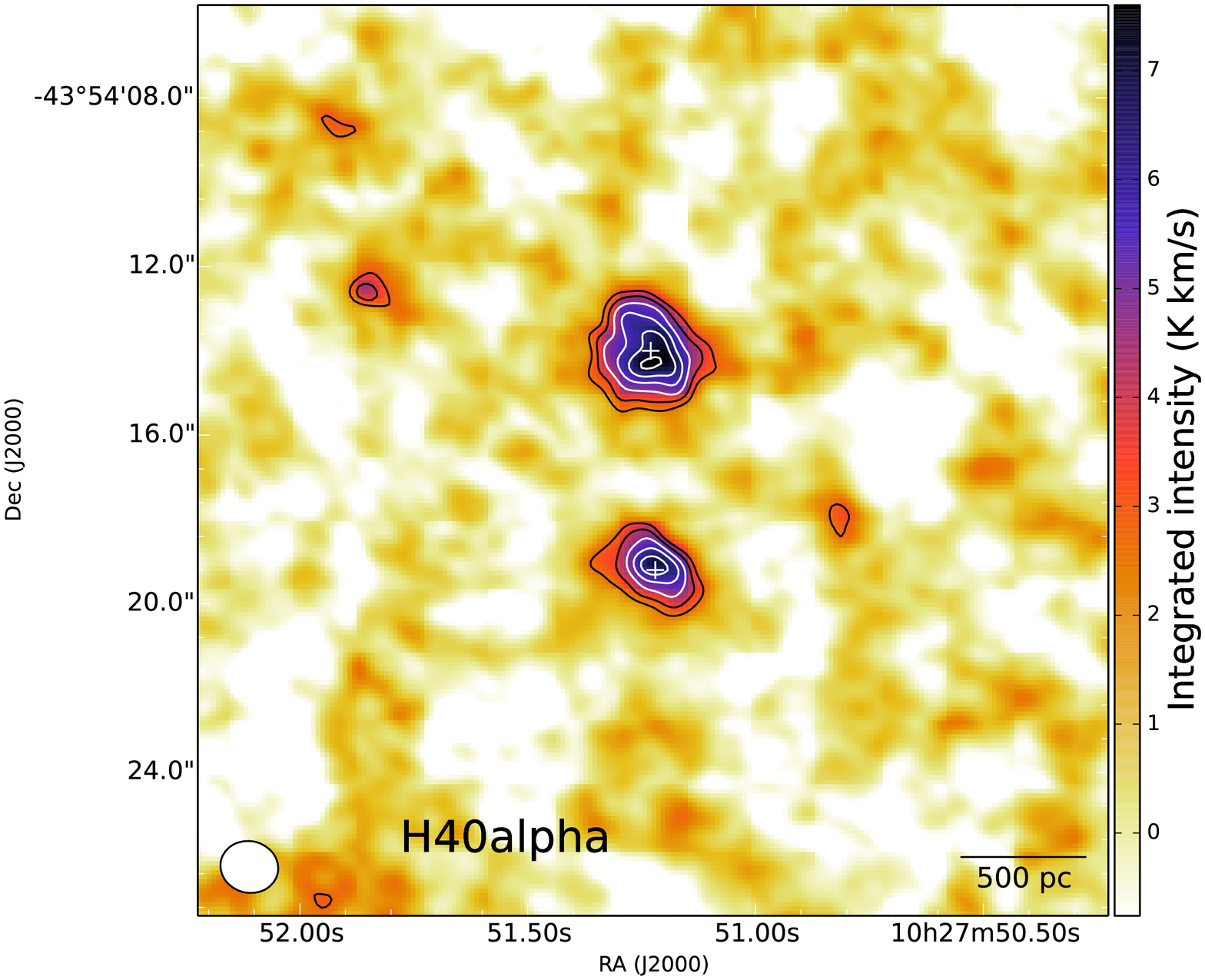} 
\includegraphics[width=.33\textwidth, trim =  0 0 0 0]{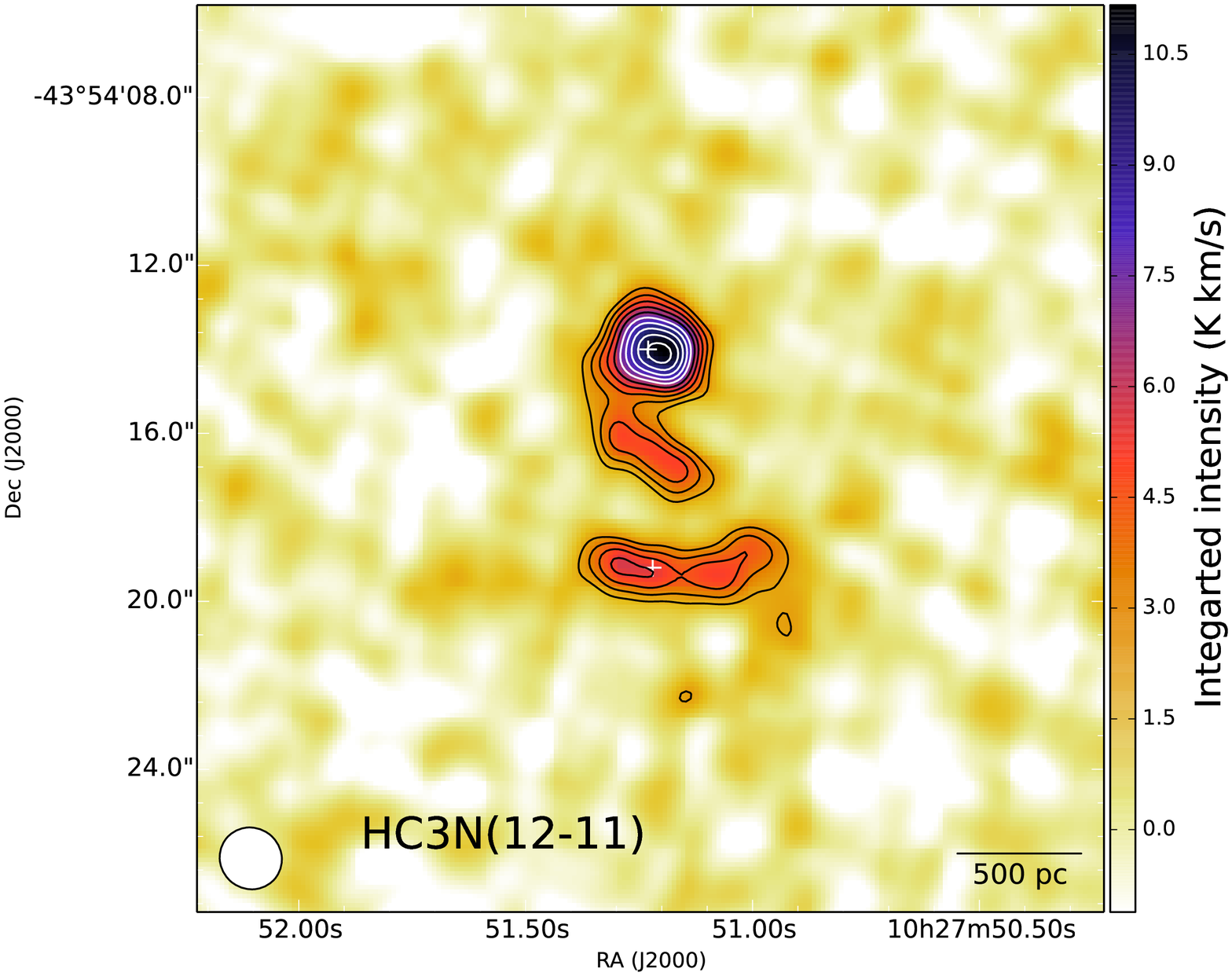} 
}
\centerline{
\includegraphics[width=.33\textwidth, trim =  0 0 0 0]{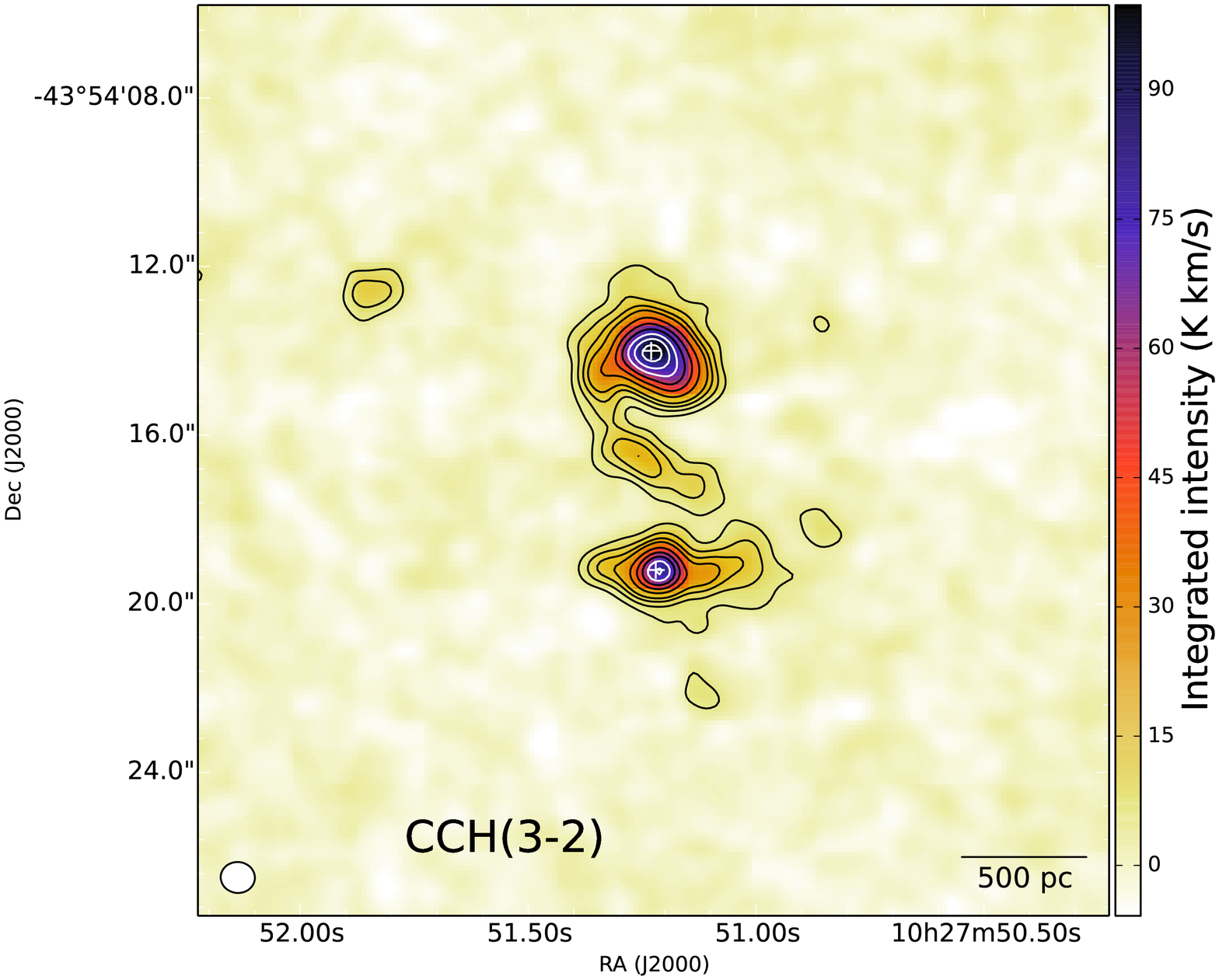} 
\includegraphics[width=.33\textwidth, trim =  0 0 0 0]{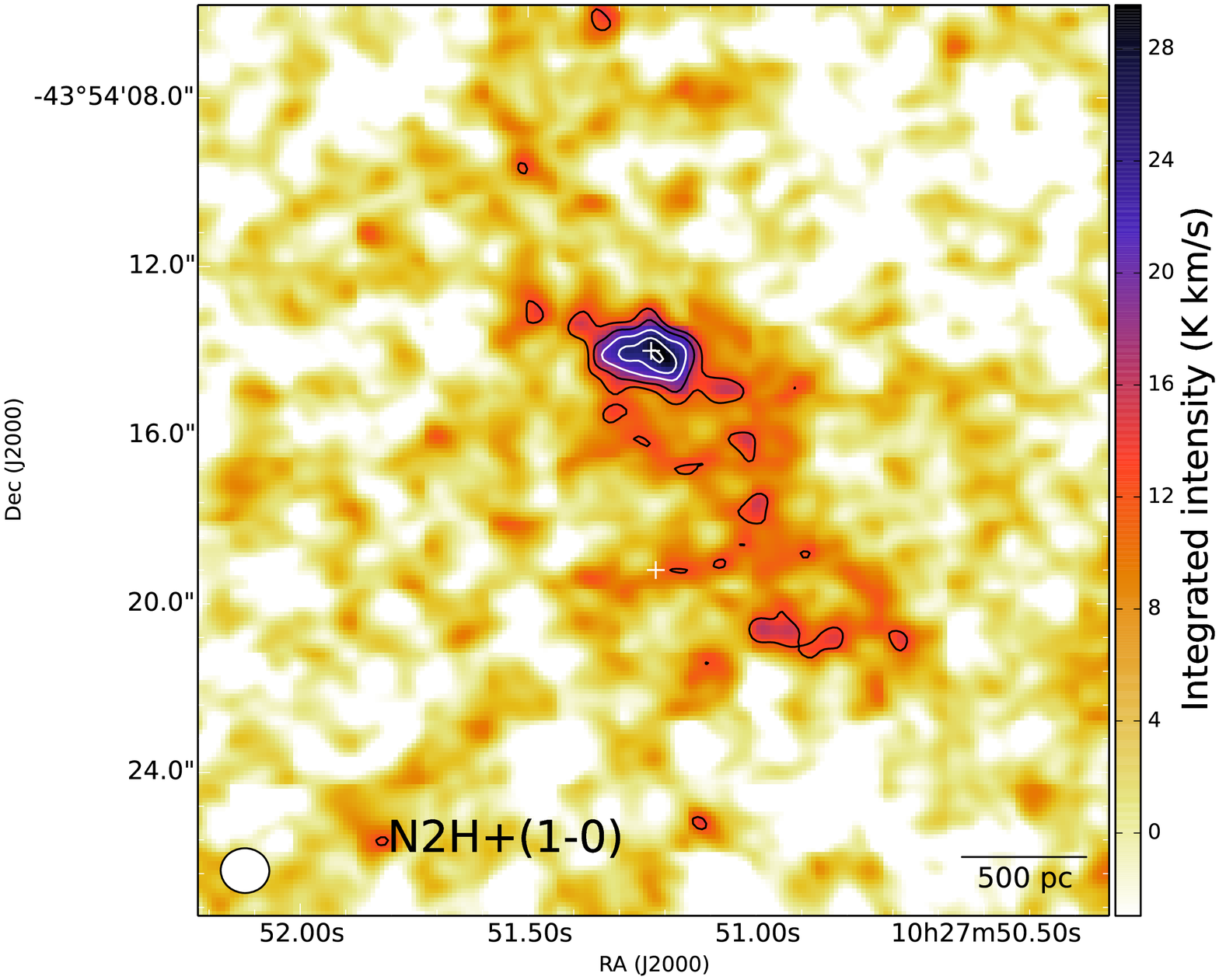}
\includegraphics[width=.33\textwidth, trim =  0 0 0 0]{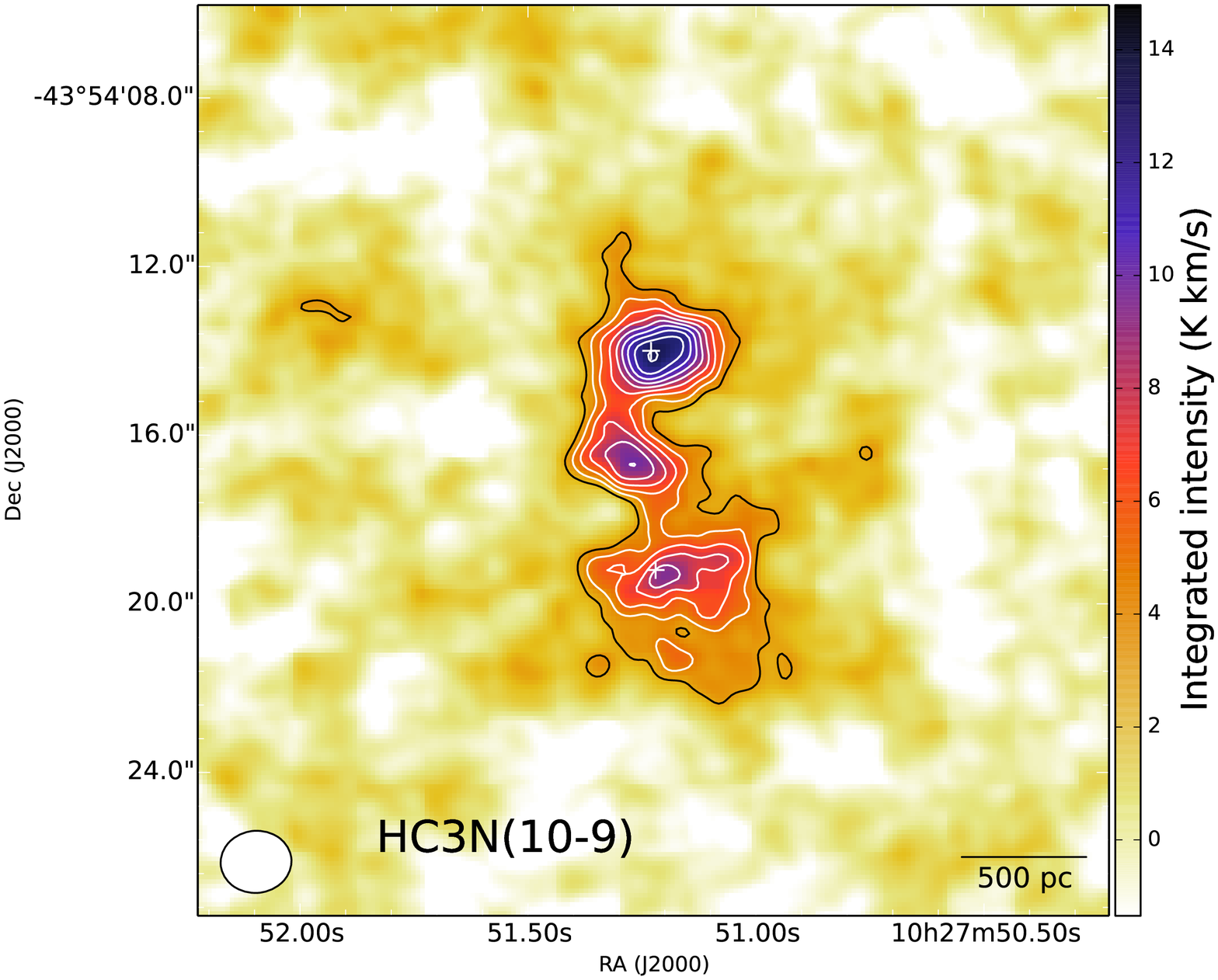}
}

\caption{Same as Figure \ref{fig:mom0}, but for other transitions. Contour levels are chosen as every $3\, \sigma$ for CH$_3$OH($2_k - 1_k$), CCH($1-0$), 
every $1\, \sigma$ starting from $3\, \sigma$ for CH$_3$OH ($5_k - 4_k$), SiO($2-1$), H40$\alpha$, N$_2$H$^+$($1-0$), 
CH$_3$CCH ($6_k - 5_k$), CH$_3$CCH ($13_k - 12_k$), HC$_3$N ($12-11$), HC$_3$N ($10-9$), and every $5\, \sigma$ for CN($2-1$),
and $3\, \sigma$, $6\, \sigma$, $9\, \sigma$, $12\, \sigma$, and every $6\, \sigma$ for CCH($2-1$).
For CCH and CN, we integrated the velocity ranges to include all the doublets / triplets. 
The rms values are listed in Table \ref{tab:imparline}. \label{fig:mom0-2}} 
\end{figure*}

Velocity-integrated moment maps of selected molecular lines and a radio recombination line (H40$\alpha$) are shown in Figures \ref{fig:mom0} and \ref{fig:mom0-2}. 
In those images, the primary beam is not corrected. The sizes of primary beams are $\frac{51.5''}{100 (GHz)}$,
and most of the emission is within $10''$.
For each transition and without taking line blending into account, we integrated the velocity range $\pm 165$ km s$^{-1}$ around the transition to produce the moment 0 maps.
Image parameters and rms values of those images are listed in Table \ref{tab:imparline}.
For molecules with doublet or triplet transitions such as CN or CCH, we integrated velocity ranges of all transitions.
Main structures such as the two nuclei, tidal arms, and the outflow (see Figure \ref{fig:8pos} right for a schematic image)
 are well traced by major lines such as $^{13}$CO, C$^{18}$O, HCN, and HCO$^{+}$,
while weaker molecular lines are detected only in selected positions.
Spatial distribution of emission intensities of some molecular lines are obviously different from that of CO isotopologues.
For example, the distribution of SiO(2-1) is enhanced in the position between the two nuclei and the southwest tidal arm position (Figure \ref{fig:mom0-2}). 
The integrated intensity maps of methanol (CH$_3$OH) also have different distributions from that of CO isotopologues
although transitions at 96.7 GHz and around 241 GHz have different distribution to each other as well (Figure \ref{fig:mom0-2}).
For both transitions of methanol, emission at the two nuclei is not dominant, and emission from tidal arms or the outflow is more visible.
For the transition at 96.7 GHz, the emission from the northern nucleus is not obvious while it is visible in 241 GHz.
Another molecule with significantly different distribution from CO isotopologues is N$_{2}$H$^+$.
While other molecules have some contribution from the both nuclei, N$_2$H$^+$ is very weak in the S nucleus.
Those features mentioned here will be quantitatively analyzed and discussed in Section \ref{sec:column}.

\subsubsection{Ratio maps}\label{sec:ratio_maps}
\begin{figure*}
\centerline{
\includegraphics[width=.45\textwidth, trim =  0 0 0 0]{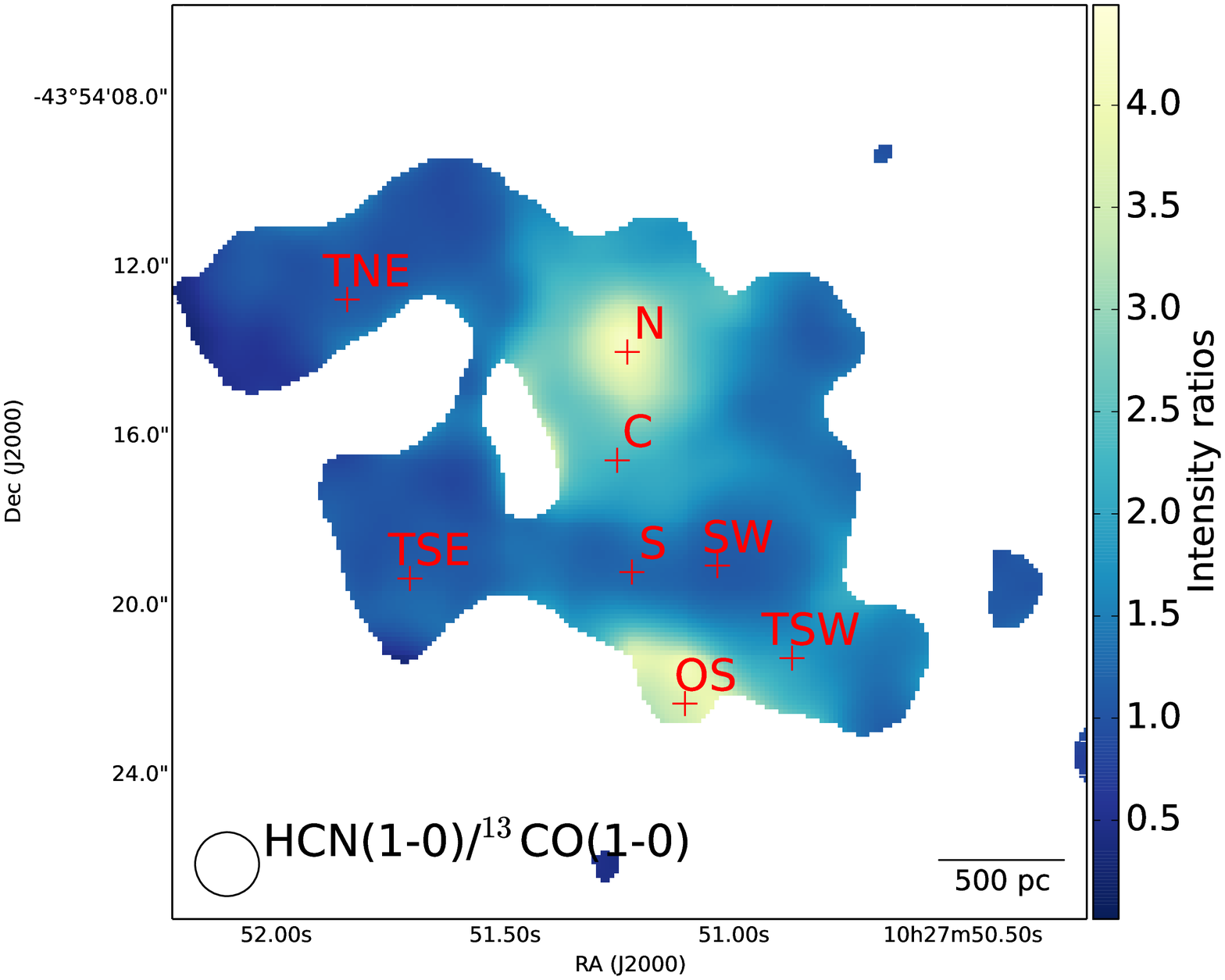} 
\includegraphics[width=.45\textwidth, trim =  0 0 0 0]{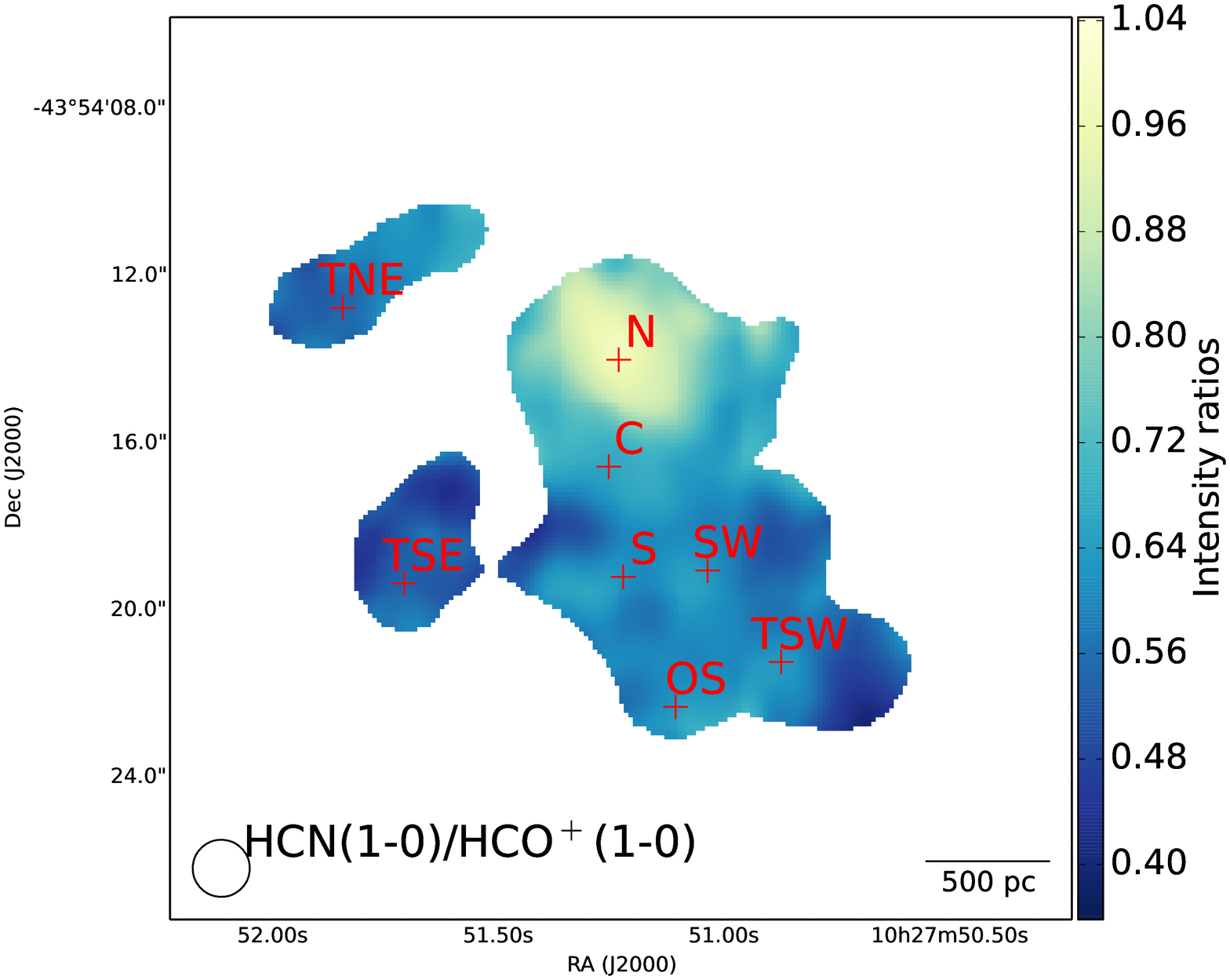} }
\centerline{
\includegraphics[width=.45\textwidth, trim =  0 0 0 0]{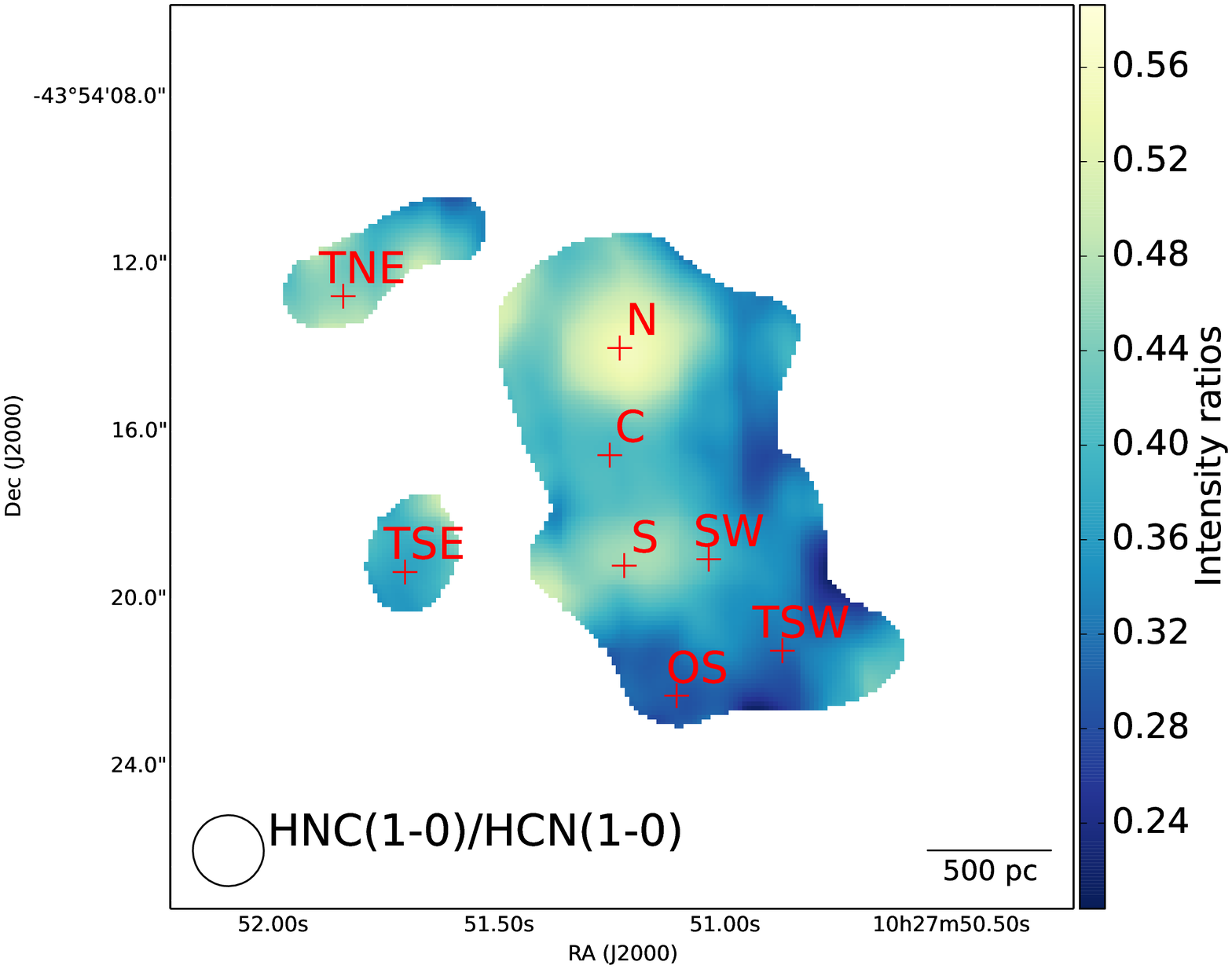} 
\includegraphics[width=.45\textwidth, trim =  0 0 0 0]{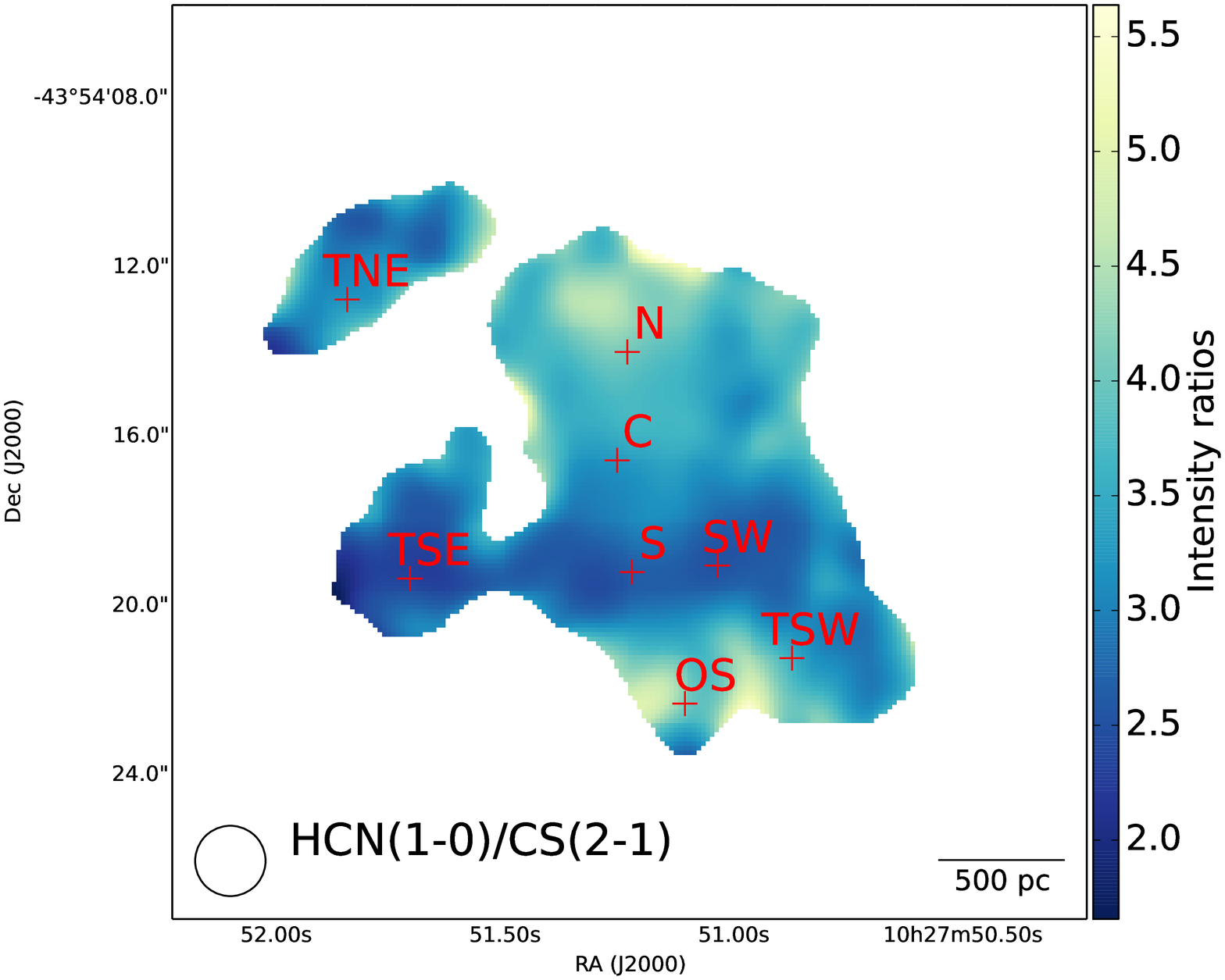} 
}
\caption{Maps of intensity ratios of (upper left) HCN($1-0$)/$^{13}$CO (1-0), (upper right)  HCN ($1-0$)/HCO$^+$($1-0$),  
(lower left) HNC ($1-0$)/HCN ($1-0$), and (lower right) HCN ($1-0$)/CS($2-1$). The ratios are calculated for the intensity units in Kelvin scale. 
Positions shown in Figure \ref{fig:8pos} are marked with red crosses. \label{fig:ratio}}
\end{figure*}

Spatial variations of intensity ratios are more clearly shown in the ratio maps. 
In Figure \ref{fig:ratio}, ratios of molecular emission intensities in Kelvin units are shown.
Primary beams are corrected for each moment 0 maps before the ratios are taken to produce these images.
The HCN(1-0)/$^{13}$CO($1-0$) ratio is higher in the N nucleus than in the S nucleus (Figure \ref{fig:ratio}, left). 
This ratio is also high at the outflow position of the southern galaxy. 
The HCN($1-0$)/HCO$^+$($1-0$) ratio is $\sim 1.0$ in the N nucleus while $\sim 0.6$ in the S nucleus, and it is
lower in the S nucleus than in the N nucleus. 
This HCN($1-0$)/HCO$^+$($1-0$) ratio has been suggested as the AGN/starburst diagnostics first by \citet{2001ASPC..249..672K},
and later followed up by \citet{2016ApJ...818...42I} and \citet{2017ApJ...835..213P}.
Results by \citet{2001ASPC..249..672K} (HCN/HCO$^+$ for $J=1-0$) and \citet{2016ApJ...818...42I} (HCN/HCO$^+$ for $J=4-3$ and HCN($4-3$)/CS($7-6$)) 
show the values of  HCN/HCO$^+$ or HCN/CS are lower in the starburst galaxies (HCN/HCO$^+ < 1$ for both transitions)
than in AGN-containing galaxies (HCN/HCO$^+ \sim 1-3$). 
At the same time, \citet{2016ApJ...818...42I} pointed out that the high-resolution data of AGN-containing 
galaxies do not show enhanced HCN($1-0$)/HCO$^+$($1-0$) at the location of AGNs \citep[see also ][for the data in the individual galaxies]{2014A&A...567A.125G,2015A&A...573A.116M}.
On the other hand, \citet{2017ApJ...835..213P} found that there is no significant difference between starburst and AGN-containing galaxies
for the sample of LIRGs, possibly due to the high opacity.
Even for LIRGs, \citet{2014AJ....148....9I} argued that AGNs show elevated HCN($4-3$)/HCO$^+$($4-3$) ratios in their ALMA observations.
In NGC 3256, S nucleus is the one that shows the past activity of AGN activity although it has the lower HCN($1-0$)/HCO$^+$($1-0$) ratio
showing the opposite trend from the AGN/starburst diagnostics mentioned above. 
The ratios in the both nuclei are within the range of values in typical starburst galaxies.
Even in AGN-containing galaxies, the line ratios may become similar to those in starburst galaxies
if the beam sizes are too large. However, it is unlikely that the observed ratios in NGC 3256 from our observations 
show starburst-like line ratios of HCN($1-0$)/HCO$^+$($1-0$) simply because of the beam smearing.
The beam size of our ratio map is 230 pc, which is equivalent to or smaller than beam sizes in \citet{2001ASPC..249..672K}
and ``high-resolution samples" in \citet{2016ApJ...818...42I}.
Although we cannot make direct comparison with line ratios for different transitions, 
both HCN($1-0$)/HCO$^+$($1-0$) and HCN($1-0$)/CS($2-1$) show similar ratio to values of 
HCN($4-3$)/HCO$^+$($4-3$) and HCN($4-3$)/CS($7-6$) for samples of starburst galaxies shown in \citet{2016ApJ...818...42I}.
The intensity ratios of HNC($1-0$)/HCN($1-0$) do not differ significantly between the N and S nuclei.
Variation of these ratios reflect either the change in molecular abundances or excitation.

Some features can be attributed to the change in the excitation and critical densities.
For example, if the mean density in the N nucleus is higher than the S nucleus, it can explain the higher HCN(1-0)/$^{13}$CO($1-0$)
and the higher HCN($1-0$)/HCO$^+$($1-0$) ratios in the N nucleus. The critical densities of $^{13}$CO, HCO$^+$, HCN for $J=1-0$ transitions are
$n_{\rm crit} \sim 2 \times 10^3\,$cm$^{-3}$,  $2 \times 10^5\,$cm$^{-3}$, and $1 \times 10^6\,$cm$^{-3}$, respectively
 \footnote{Critical densities listed here are estimated by simply assuming the two level approximation without considering the photon trapping 
 and take $n_{\rm crit} = A_{ul}/\gamma_{ul}$
 where $A_{ul}$ is the Einstein A coefficient for a transition from the level $u$ and to the level $l$, $\gamma_{ul}$ is a collisional coefficient 
 for a transition from the level $u$ and to the level $l$. Values of $A_{ul}$ and $\gamma_{ul}$ were obtained from Leiden Atomic and Molecular Database
 \citep{2005A&A...432..369S}. The temperature of 10 K was assumed for $\gamma_{ul}$.}.

On the other hand, the HNC($1-0$)/HCN($1-0$) ratios cannot be explained solely by the excitation.
The critical density of HNC($1-0$) is $n_{\rm crit} \sim 3 \times 10^5\,$cm$^{-3}$, and 
the HNC($1-0$)/HCN($1-0$) ratio should be lower in the N nucleus if the critical density is the only contributing factor.
It has been pointed out that molecules emit at densities 1-2 orders of magnitudes lower than critical densities \citep{2015PASP..127..299S},
and also that the distribution of molecular emission within GMCs do not vary as expected from critical density alone
due to both excitation and chemical abundances
 \citep{2017A&A...599A..98P,2017A&A...605L...5K,2017ApJ...845..116W,2017ApJ...848...17N}.
\citet{2017ApJ...845..116W} and \citet{2017ApJ...848...17N} examined the fractions of molecular line intensities coming from extended diffuse components 
and denser star forming regions in Galactic GMCs. 
Among species mentioned above, $^{13}$CO, HCN, HCO$^+$, and HNC, 
their results suggest that $^{13}$CO emission has more contribution from more extended regions, 
HCN and HCO$^+$ emission from more compact regions,
and HNC has the highest contribution from the compact star-forming regions probably from dense, yet still cold gas.
If distribution of molecular emission in the nuclei of NGC 3256  is similar to that of Galactic GMCs, 
the observed line ratios can be explained if the N nucleus has more fraction of dense and compact clouds than in the S nucleus.
However, we should also note that \citet{2005ApJ...618..259M} did not see any temperature dependence of HNC/HCN intensity ratio
in a starburst galaxy IC 342 possibly because of the high opacity. Another explanation of the lack of temperature dependence is 
from the high ionization rate, which can create higher abundance of HCNH$^+$ to form HNC through the recombination with an electron \citep{2002A&A...381..783A}.

Another feature to note is the higher HCN(1-0)/$^{13}$CO($1-0$) ratio in the outflow position.
It is surprising if it is due to the higher 
density in the outflow because outflows are usually considered to be hot and tenuous.
This feature might be a reflection of higher fractional abundance\footnote{Fractional abundances refer to abundances of certain species over total hydrogen 
or molecular hydrogen abundances.
In this paper, we discuss the variation of fractional abundances from abundances of certain species over some reference species such as 
$^{13}$CO or CS because of the difficulty of obtaining the total hydrogen abundances.} of HCN due to the higher gas temperature
 \citep{2008A&A...488L...5L,2010ApJ...721.1570H} possibly due to shocks \citep{2012A&A...537A..44A,2015A&A...573A.116M,2016ApJ...818...42I}.
 Higher HCN($1-0$)/CS($2-1$) and lower HNC($1-0$)/HCN($1-0$) also supports this scenario.
 On the other hand, the HCN(1-0)/HCO$^+$($1-0$) do not show strong enhancement at the outflow position.
 It is possible that HCO$^+$ is also enhanced due to ionization in the outflow,
 but the precise density structure and molecular abundances need to be further explored.
 However, this enhancement may also be due to the higher density, similar to the case of the outflow in Mrk 231 \citep{2015A&A...574A..85A}.

\subsection{Molecular spectra}\label{sec:spectra}
To analyze individual positions of interest, we extracted spectra from our image cubes.
For this analysis, we use the primary-beam corrected cubes.
Since we did not have the ACA data for all the Band 6 scheduling blocks, we only used the 12-meter data for consistency.
The spectra used to calculate column densities in the next section are produced by convolving the image cubes to a common 1.7$''$ resolution.
The rms values of individual spectral windows are listed in Table \ref{tab:imparam}.
To include the CN($1-0$) line in the analysis, we also used the data produced for \citet{2014ApJ...797...90S}.
The positions we analyzed are shown in Figure \ref{fig:8pos} ($left$) as discussed at the beginning of Section \ref{sec:res}
(N: northern nucleus, S: southern nucleus, C: central peak, TNE: tidal arm northeast, TSE: tidal arm southeast, TSW: tidal arm southwest,
OS: outflow south, SW: southern galaxy west).
Note that in the analysis of column densities in the later sections, we use the entire velocity range around the systemic velocity 
for the position OS, and it may have some contamination from non-outflow components is possible. 
Plots of spectra are shown in Figures \ref{fig:N_b3} - \ref{fig:N_b6_3} for the position N
(see the online journal of the published version for all positions). 
Observed peak intensities and integrated intensities are listed in Table \ref{tab:intens} for 
detected species and tentatively detected species with $> 2\sigma$ at the peak.
 The integrated intensities are calculated by integrating a velocity range of $\pm 150\,$km s$^{-1}$
from the systemic velocity of $2775\,$km s$^{-1}$. Errors for the integrated intensities are estimated as $\Delta I \Delta v \sqrt{N}$
where $\Delta I$ is the rms of the original cube corrected according to the position in the primary beam, $\Delta v$ is the velocity resolution of the cube, and $N = v_{range}/\Delta v$
for $v_{range}=300\,$km s$^{-1}$.
 If lines are narrower than $v_{range}=300\,$km s$^{-1}$, the actual errors may be smaller. That is why some of weak lines have low values of $I \Delta V$.

\begin{figure}
\includegraphics[width=.5\textwidth, trim =  0 -0 0 0 0]{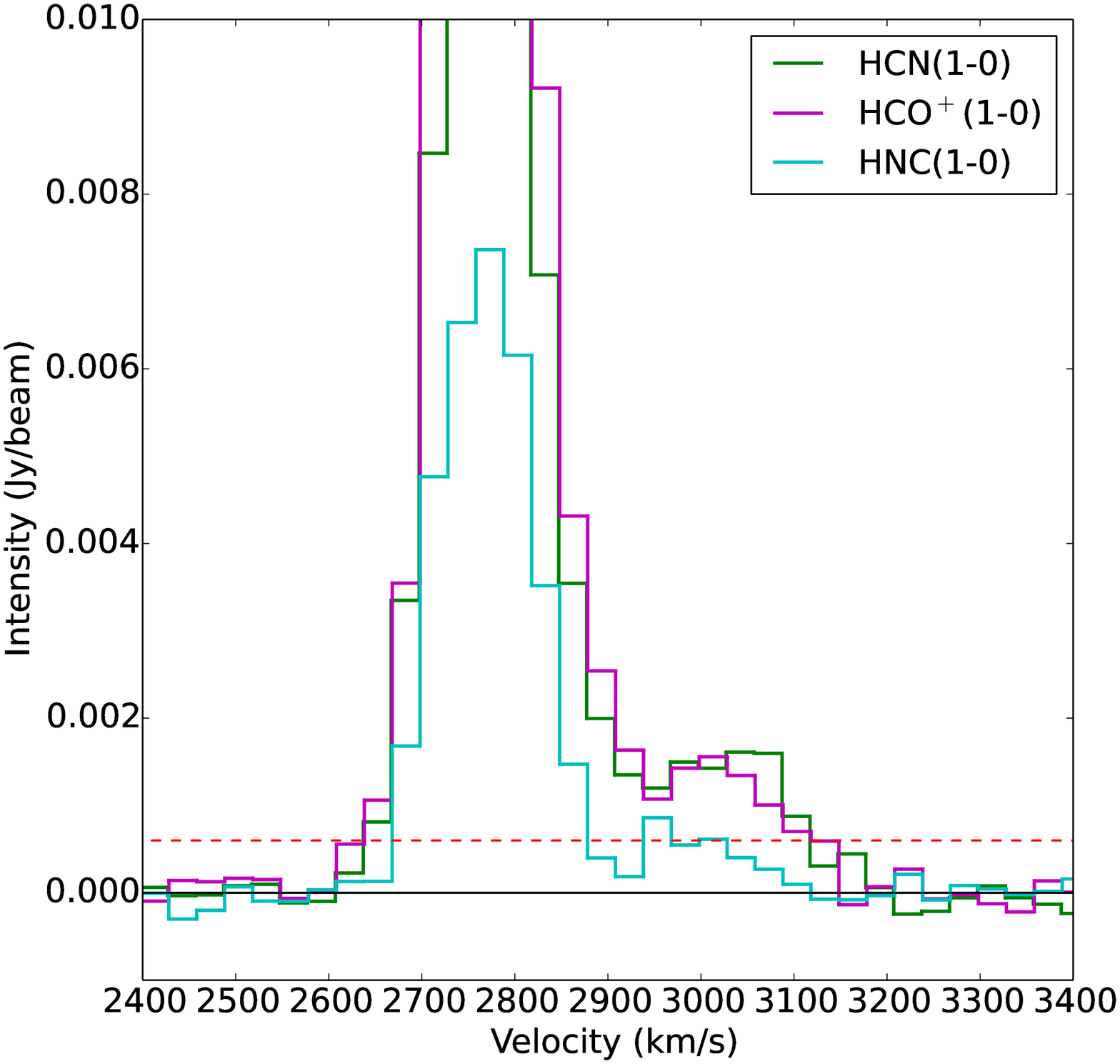} 
\caption{Spectra of HCN($1-0$), HCO$^+$($1-0$), and HNC($1-0$) at the peak of red-shifted outflow components shown in Figure \ref{fig:8pos} ($middle$)
are plotted with green, magenta, and cyan, respectively. Spectra are extracted from cubes convolved to $1.7''$. 
The value of $3\sigma$ is plotted with a red dotted line. 
The systemic velocity is 2775 km s$^{-1}$, and the emission from $v > 2975$ km s$^{-1}$ is likely to come from the outflow.  \label{fig:specwing}}
\end{figure}

While the molecular line spectra can be fit with a single Gaussian profile in most positions, there are exceptions.
For example, spectrum shapes of the N and S nuclei have double peaks for most of the species.
It is likely that these double peaks come from the rotation of the galactic nuclei, and not from self absorption
because even molecules that are thought to be optically thin have double peaks.
 Another position where spectra cannot be fit well with a single gaussian is the position OS,
 which can be explained by the complicated velocity structure of the outflow.
 Line wings from the outflow are detected clearly at the peak position of the red-shifted component seen in Figure \ref{fig:8pos} ($middle$)
 for HCN($1-0$), HCO$^+$($1-0$), and HNC($1-0$) shown in Figure \ref{fig:specwing}.

\section{Column densities}\label{sec:column}
To quantify the spatial variation of the chemical composition, we present the column densities of detected species in this section
using the spectra presented in Section \ref{sec:spectra}.
To obtain column densities, we used a spectral fitting feature of MADCUBA (Mart\'in et al., in preparation) \footnote{http://cab.inta-csic.es/madcuba/Portada.html}.
In this program, the fit between the simulated spectra and the observed one is calculated,
and column densities and excitation temperatures with the best fit can be obtained under the LTE approximation.
In the simulated spectra, optical depth is also taken into account.
The list of obtained column densities and excitation temperatures are listed in Tables \ref{tab:colN} - \ref{tab:colSW}.
Errors shown in those tables are errors from the fitting.
The estimated $T_{\rm ex}$ are mostly $\lesssim 10\,$K except for HC$_3$N, CH$_3$CCH, CH$_3$OH, and H$_2$CO.
For those molecules, $T_{\rm ex}$ can be as high as 20-40 K.
For a molecule with only one line with enough S/N, we used $T_{\rm ex}=10$ K.
Even with the detection of multiple transitions, there are cases where $T_{\rm ex}$ cannot be precisely determined.
For those cases, $T_{\rm ex}$ was fixed at the best-fit value to obtain column densities.
If the molecular species are undetected at certain positions, the upper limits are obtained by deriving column densities 
for cases of 2-$\sigma$ detections at the lowest (or strongest) transition with $T_{\rm ex}=10$ K.

In the following sections, we plot the ratios of the derived column densities. 
We use two different species with different critical densities ($^{13}$CO and CS) as denominator of the ratios for the following reasons.
Since we have detections of multiple transitions for most of the major species, the derived column densities should result
from differences in fractional abundances. However, this is not always the case if the density distribution within the beam is different.
If the beam contains higher fraction of relatively dense material, then molecules with higher critical densities have a larger area/volume 
where they can emit. This higher fraction of dense gas can result in higher apparent abundances even when the abundances are derived 
using multiple transitions. If this excitation factor plays a role in the ratio of column densities, discussion of variation of fractional abundances are 
only valid when the column density ratios are obtained using molecules with similar critical densities for both numerator and denominator. 
$^{13}$CO is a good denominator for molecules with low critical densities while CS is suitable for high $n_{\rm crit}$ molecules.
We have already mentioned in Section \ref{sec:ratio_maps} that molecules may emit at lower densities than the critical densities,
but molecules such as CS still emit more effectively at higher densities than $^{13}$CO \citep{2017ApJ...845..116W,2017ApJ...848...17N}.


\subsection{Comparison within NGC 3256}\label{sec:comp_n3256}
\begin{figure}
\includegraphics[width=.5\textwidth, trim =  0 0 0 0]{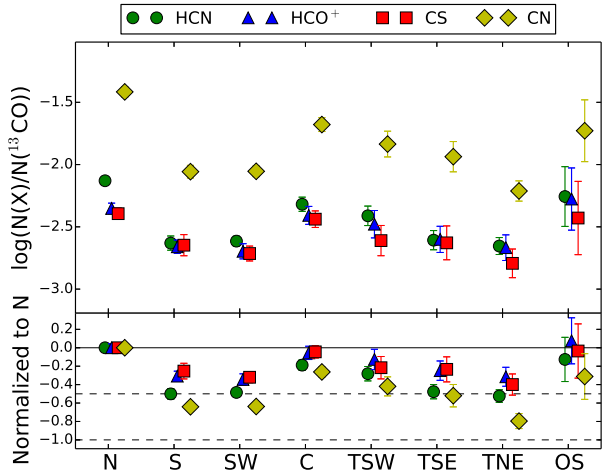} 
\caption{({\it Top}) Column density ratios of HCN, HCO$^+$, CS, and CN over $^{13}$CO at positions N, S, C, TNE, TSE, TSW, OS, and SW in NGC 3256. 
Errors of individual column densities are automatically calculated from the Gaussian fitting by MADCUBA.
({\it Bottom})The same figure as the top figure, but all the values are normalized to the position ``N". All the values are shown in a log scale. \label{fig:dense_n3256}}
\end{figure}

\begin{figure*}
\centerline{
\includegraphics[width=.5\textwidth, trim =  0 0 0 0]{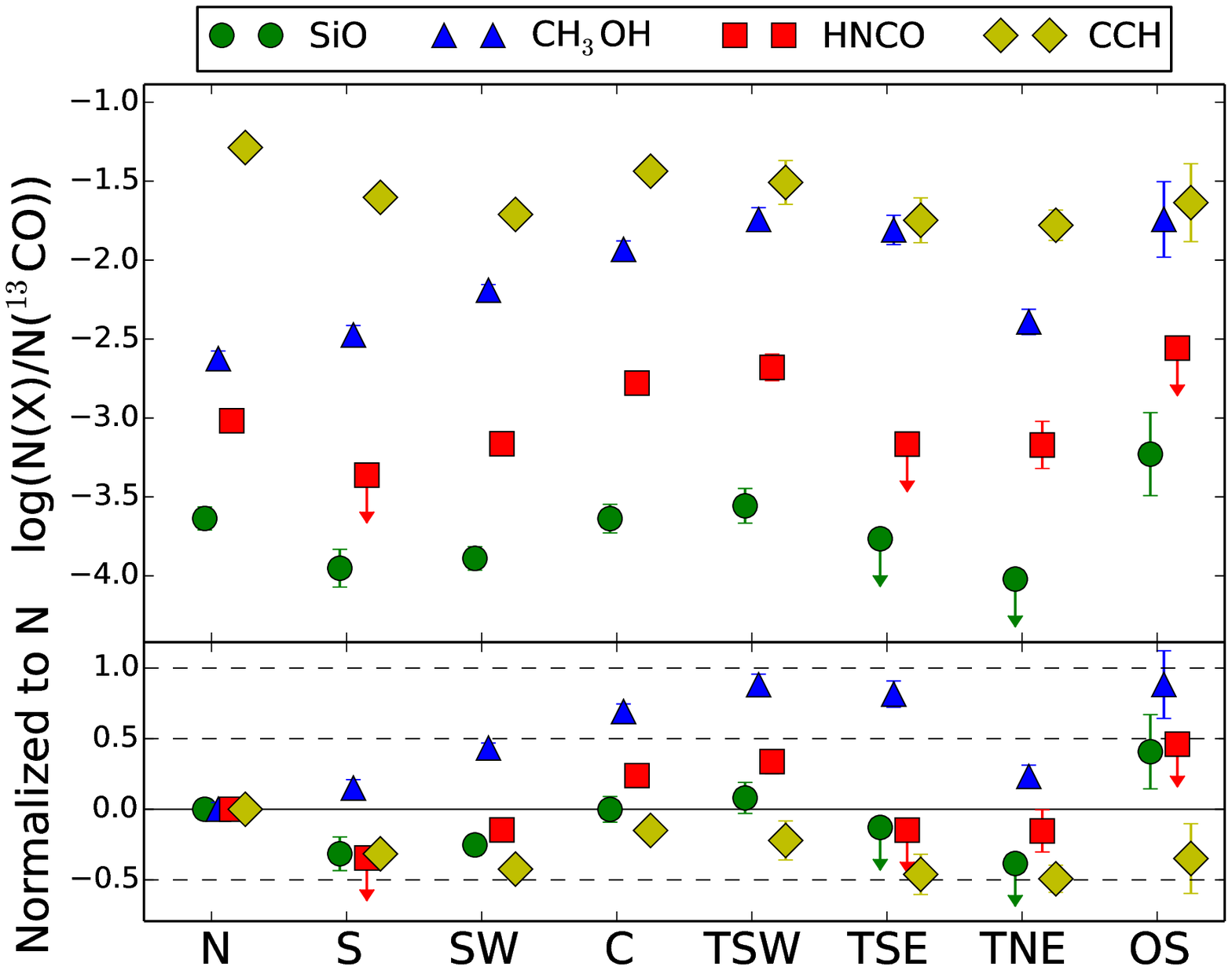} 
\includegraphics[width=.5\textwidth, trim =  0 0 0 0]{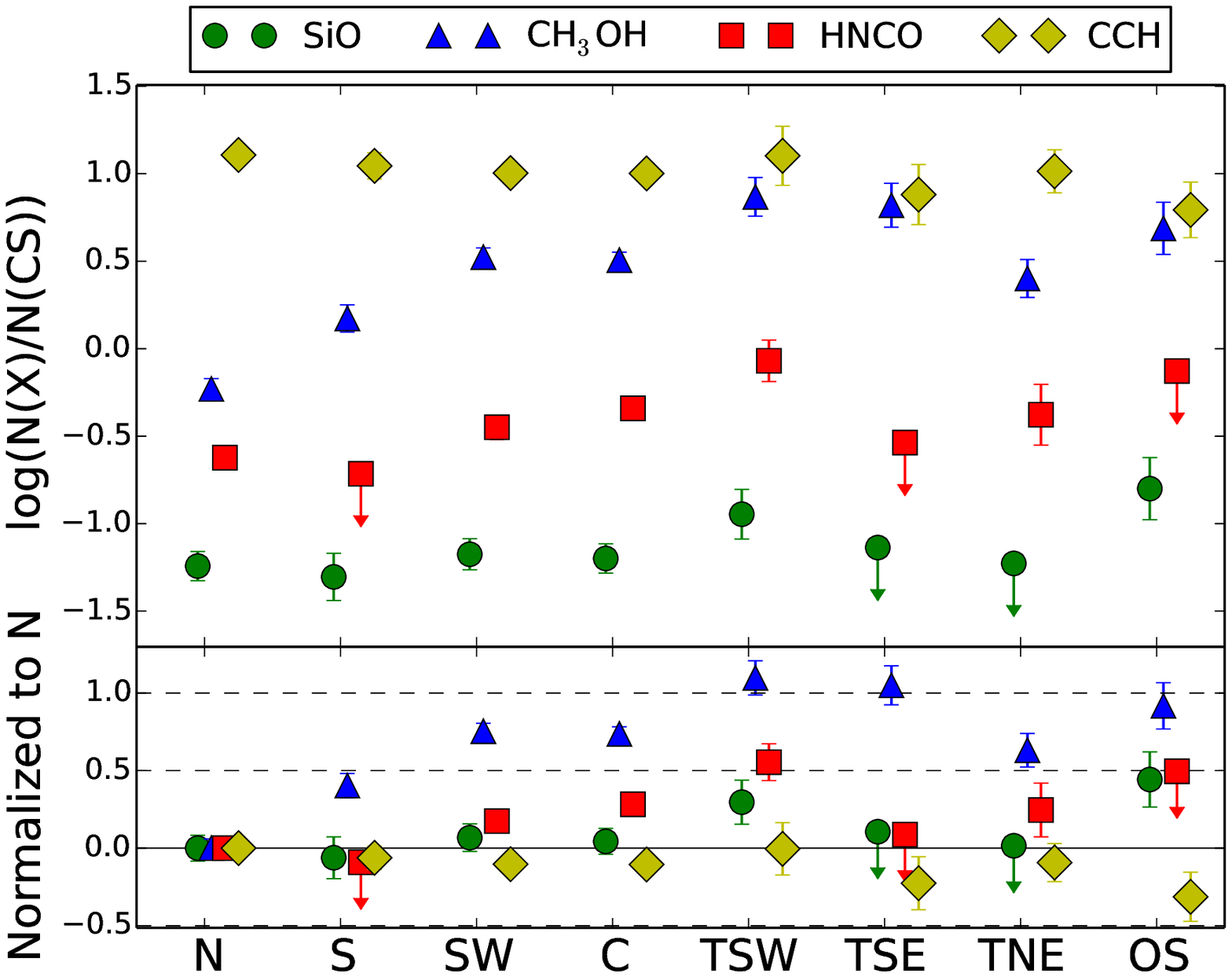} 
}
\centerline{
\includegraphics[width=.5\textwidth, trim =  0 0 0 0]{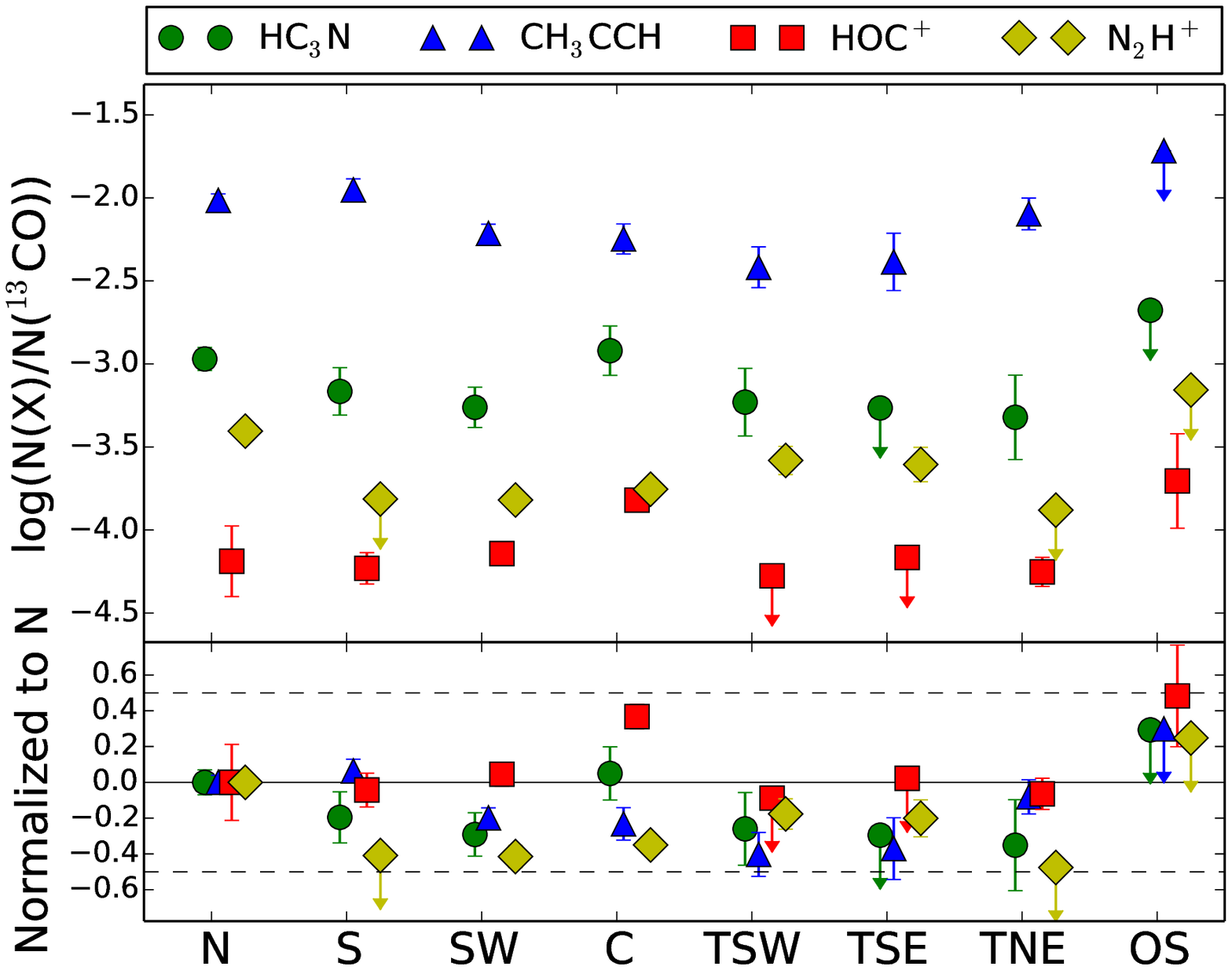} 
\includegraphics[width=.5\textwidth, trim =  0 0 0 0]{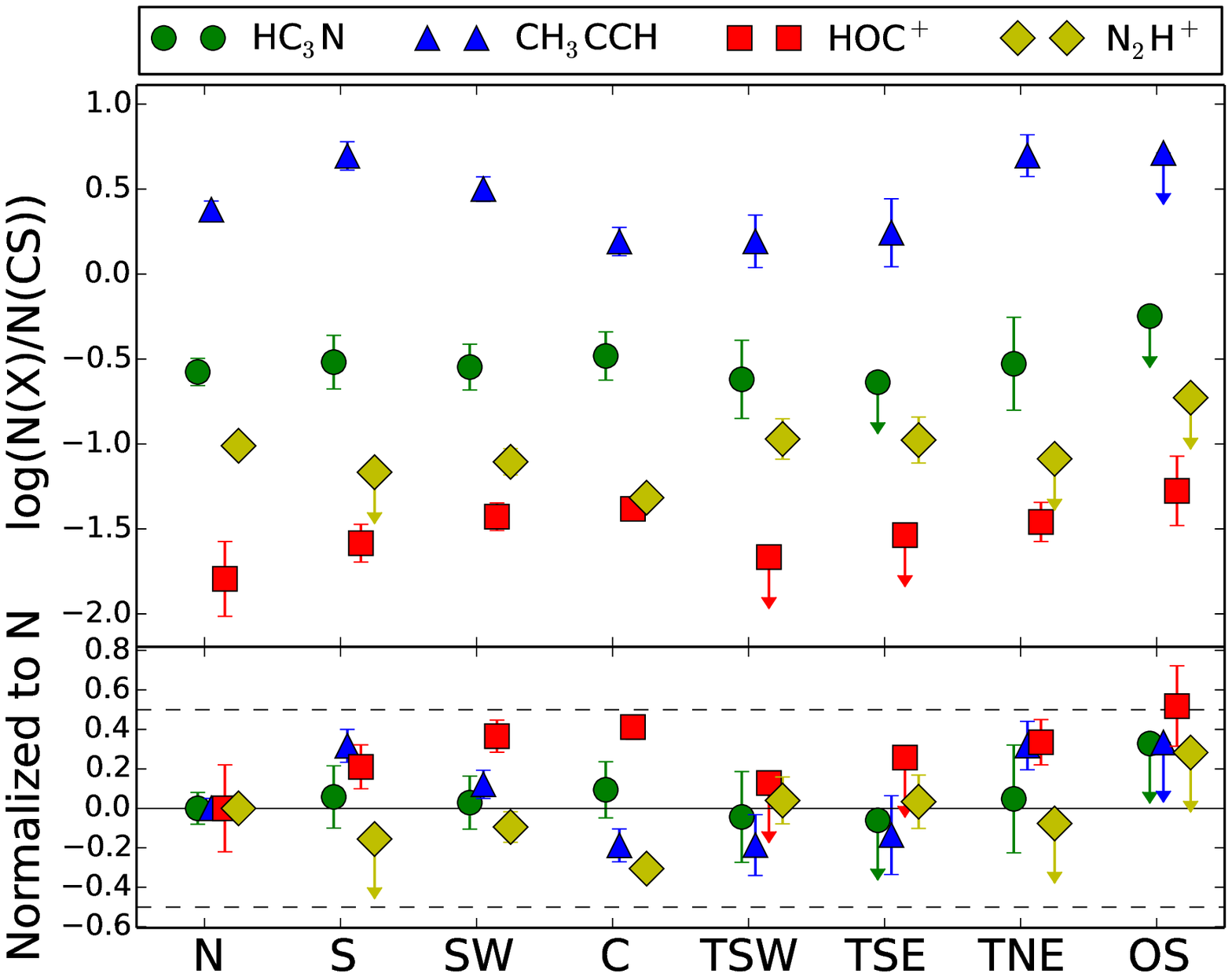} 
}
\caption{({\it Top}) Column density ratios of selected species over $^{13}$CO (left panels) and over CS (right panels)
 at positions N, S, C, TNE, TSE, TSW, OS, and SW in NGC 3256. 
({\it Bottom})The same figure as the top figure, but all the values are normalized to the position ``N". All the values are shown in a log scale.\label{fig:ratios_n3256}}
\end{figure*}

We first compare the chemistry among the 8 positions we analyzed in NGC 3256. 
Here we focus our discussion on selected species such as species with higher critical densities
(HCN, HCO$^+$, CS, and CN), species enhanced with shocks (SiO, CH$_3$OH, HNCO), species enhanced in PDRs (CCH, HOC$^+$),
and species seen in cold dense cores (N$_2$H$^+$). We also discuss CH$_3$CCH because it was proposed to be AGN/starburst 
diagnostics by \citet{2015A&A...579A.101A}. We omit discussion of species detected in 3 or less positions (c-C$_3$H$_2$, CH$_3$CN, and CO$^+$),
some isotopologues (H$^{13}$CO$^+$ and C$^{34}$S), and species with little variation among positions (SO, NO, and H$_2$CO).
We arrange the column density ratios by CS column densities in the descending order in Figures \ref{fig:dense_n3256} - \ref{fig:ratios_n3256}.

In Figure \ref{fig:dense_n3256} (top), column density ratios of molecules with relatively high critical densities over $^{13}$CO are plotted in the log scale.
To highlight the variation among the locations, the column density ratios over $^{13}$CO normalized to the value at the location ``N" are shown in 
Figure \ref{fig:dense_n3256} (bottom) also in the log scale.
Variation of column density ratios seem similar among molecules. 
Although it is possible that the fractional abundances of those molecules vary among those positions in a similar way,
it is likely that the variation of excitation condition within the beam discussed earlier in Section \ref{sec:column} is changing those apparent fractional abundances.

On the other hand, there is variation of derived column density ratios that is likely caused by differences in fractional abundances of certain molecules.
Behavior of such molecules are described below, and column densities are plotted in Figure \ref{fig:ratios_n3256}.

{\bf SiO:} The column density ratios of SiO over $^{13}$CO and over CS show both enhancement at the position OS. 
Silicon monoxide is expected to be enhanced in strong shocks of velocity $v_{\rm shock} > 25\,$km s$^{-1}$ \citep{2008A&A...482..809G}
due to sputtering from the dust core. Although it was claimed that there is a correlation between SiO intensity and X-ray strength \citep{2009ApJ...694..943A}, 
this correlation is a weak one. It is indeed theoretically possible that X-ray heats up the dust enough to make it sublimate, SiO as an XDR tracer should be 
taken with a caution. If this outflow is an AGN-induced outflow, both shocks and X-rays are naturally expected.

{\bf CH$_3$OH:} Column density ratios of CH$_3$OH vary in a similar way to that of SiO although the enhancement
 at one of the tidal arm positions (TSW) is also prominent for CH$_{3}$OH. 
CH$_{3}$OH is known to be enhanced in weak shocks. 
For methanol, there are other mechanisms to increase its abundance in the gas-phase
such as cosmic-ray or UV-induced photo-desorption, or chemical desorption.
However, chemical desorption should affect the larger galactic scale, and cannot explain variation within NGC 3256.
For the case of cosmic-ray or UV-induced photo-desorption, if it is increasing the abundances of methanol, it should be abundant in the two nuclei.
The methanol abundances in the two nuclei are rather low. Therefore, if UV-photons or cosmic rays affect the methanol abundances, it is likely to decrease methanol abundances 
by photodissociation rather than to increase by photo-desorption.
Therefore, we regard it most likely that methanol is enhanced by the shock due to the merger interaction or an outflow.
Similar enhancement of methanol in the region of merger interaction was also reported by \citet{2017ApJ...834....6S} in VV114,
a merger with a larger nuclear separation.

{\bf HNCO:} Observation in other galaxies show that the intensities of HNCO vary in a similar way as CH$_3$OH when statistically analized position by position \citep{2005ApJ...618..259M,2012ApJ...755..104M,2015ApJ...801...63M}. 
When seen in a large fractional abundance, the major formation route of HNCO is likely from grain surface reaction \citep{2010ApJ...725.2101Q} similarly to CH$_3$OH. In positions we analyze, positional variation of column densities of CH$_3$OH and HNCO are similar to each other expect for the N nucleus.
A possible reason for the difference in the N nucleus might be that HNCO and CH$_3$OH abundances may act differently to the dust temperature.
Methanol abundances decrease at high dust temperature because the hydrogen atom evaporate faster at the high dust temperatures, 
and the hydrogen atom is less likely to react \citep{2016ApJ...822..105A}. HNCO may be able to sustain its abundance at higher dust temperatures than
CH$_3$OH because there is less number of hydrogen necessary for HNCO. This temperature dependence of HNCO needs to be tested with chemical models.
For the case of SiO column density ratios, there is a good correlation between HNCO column density ratios.
Such correlation between HNCO and SiO was also proposed in Galactic high-mass star forming regions  \citep{2000A&A...361.1079Z}.
A survey of several galaxies by \citet{2009ApJ...694..610M} also suggested that a higher ratio of HNCO is a characteristic for an early stage of starburst,
dominated by shocks, instead of UV-photons.

{\bf N$_2$H$^+$:} In Galactic star-forming regions, N$_2$H$^+$ emission is usually associated with dense cores \citep[e.g.,][]{2014ApJ...794..165S,2017A&A...599A..98P}.
In our observations, column density ratios of N$_2$H$^+$ over CS do not show clear variation, but the ratios over $^{13}$CO show a slightly higher value in 
the N nucleus than the S nucleus. It can be explained if the N nucleus contains more dense clouds than the S nucleus as argued in Section \ref{sec:molimage}.

{\bf HC$_3$N:} In Galactic molecular clouds, higher abundances of cyanoacetylene (HC$_3$N) are usually detected in dense cold clouds or hot protostellar cores.
For extragalactic sources, it has been found that the compact obscured nuclei such as NGC 4418 and Arp 220 have high fractional abundance of HC$_3$N
\citep{2007A&A...475..479A,2010A&A...515A..71C,2011A&A...527A..36M,2011A&A...527A.150L}. Since HC$_3$N is enhanced in star-forming regions, one can expect that 
HC$_3$N is more enhanced in regions of higher star formation efficiency. Although only the N and S nuclei are known to have star formation \citep{2008MNRAS.384..316L}
(and possibly a position TNE with a tentative detection of H40$\alpha$ in this paper), $N({\rm HC_3N})/N({\rm ^{13}CO})$ is also enhanced in position C,
a tidal arm with possible interaction with components. 

{\bf CH$_3$CCH:} Propyne (CH$_3$CCH) has been proposed as a diagnostics of starburst versus AGN by \citet{2015A&A...579A.101A};
starburst galaxies have high abundances of CH$_3$CCH while AGN-containing galaxies do not have detections of CH$_3$CCH.
However, the reason behind this observational trend is still unknown.
In NGC 3256, although the S nucleus is the likely AGN host, CH$_3$CCH column density ratios in the S nucleus are equivalent to or higher than in the N nucleus. 

{\bf HOC$^+$:} HOC$^+$ is a metastable isomer of HCO$^+$, and it needs irradiation from strong UV field \citep{2017A&A...601L...9G}
 or X-rays \citep{2004A&A...419..897U} to keep its abundance high.
 In NGC 3256, positions where HOC$^+$ is enhanced are C and OS
 although the detection in OS is also tentative, and the error bars are large. 
  The position OS contains the high-velocity gas coming from the outflow,
   and the position C also contains gas from the red-shifted outflow component (Figure \ref{fig:8pos} $middle$).
  The connection with the HOC$^+$ abundance and the outflow can be suspected.
 However, the connection between the HOC$^+$ enhancement and the outflow cannot be directly confirmed in the high-velocity gas alone because 
 the high-velocity components are too weak to be detected for HOC$^+$.

{\bf CCH:} Ethnyl radical (CCH) is known to be enhanced in PDRs \citep[e.g., ][]{2015A&A...575A..82C},
but there are also abundant in the early-time chemistry, and has been observed in the quiescent starless core such as TMC-1(CP) \citep{2004MNRAS.350..323S}.
Although CCH was also recently found in the outflow from an AGN in NGC 1068 \citep{2017A&A...608A..56G},
its enhancement in the outflow is not seen in NGC 3256.
This case is similar to NGC 1097, where no particular enhancement was seen around the AGN for CCH \citep{2015A&A...573A.116M}.
Their abundances in NGC 3256 is highest in the N nucleus if the column density ratios over $^{13}$CO, but there is less variation among positions 
within NGC 3256 for ratios over CS.

\subsection{Comparison with other galactic nuclei}
We next compare the chemistry in the nuclear positions in NGC 3256 and that of NGC 253 and Arp 220,
galaxies with different star formation rates and efficiencies. The properties of these galaxies are 
summarized in Table \ref{tab:gal_prop}.

NGC 253 is a well-studied, nearby starburst galaxy. The molecular gas distribution of the central molecular zone (CMZ)
 is similar to that of the Milky Way. However, the star formation rate inside is much higher than the Milky Way CMZ.
 Yet, NGC 3256 has even higher star formation rate within its CMZs.
For NGC 253, we use the results by \citet{2015A&A...579A.101A} taken by IRAM 30-meter telescope.
The angular resolution of this IRAM study is $17'' - 24''$, which converts into $300-400\,$pc.
This spatial scale is equivalent to the spatial resolution in our analysis of NGC 3256, which is $\sim 300\,$pc.

Like NGC 3256, Arp 220 is also a merger, but has even higher luminosity than NGC 3256. 
The nuclear separation of the two nuclei in Arp 220 is $\sim 400\,$pc on the sky, which is smaller than in the NGC3256.
The column densities in Arp 220 are taken from results by Mart\'in et al. (in preparation) where
they used ALMA for the spectral scan of Bands 6 and 7 \citep[see ][ for some description of these data]{2016A&A...590A..25M}.
The exact values of column densities and errors will be found in Mart\'in et al. (in preparation).\footnote{We note that the analysis
 in Mart\'in et al. (in preparation) did not use the Gaussian fit to the observed spectra for the calculation of column densities
because they are severely affected by absorption at the line center. They use the side of the lines to calculate 
column densities. Therefore, we claim that the error of column densities from absorption is minimized,
but it cannot be claimed that those values are free from absorption effects. } 
Their spatial resolution is also around 300 pc.

\begin{figure}
\includegraphics[width=.5\textwidth, trim =  0 0 0 0]{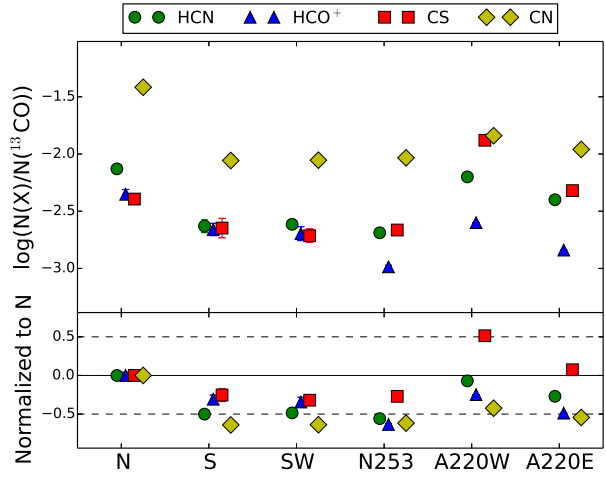} 
\caption{({\it Top}) Column density ratios of HCN, HCO$^+$, CS, and CN over $^{13}$CO and over CS at nuclear positions N, S, and SW in NGC 3256, in NGC 253, and W and E in Arp 220. 
({\it Bottom})The same figure as the top figure, but all the values are normalized to the position ``N". All the values are shown in a log scale.\label{fig:dense_nuc}}
\end{figure}

\begin{figure*}
\centerline{
\includegraphics[width=.5\textwidth, trim =  0 0 0 0]{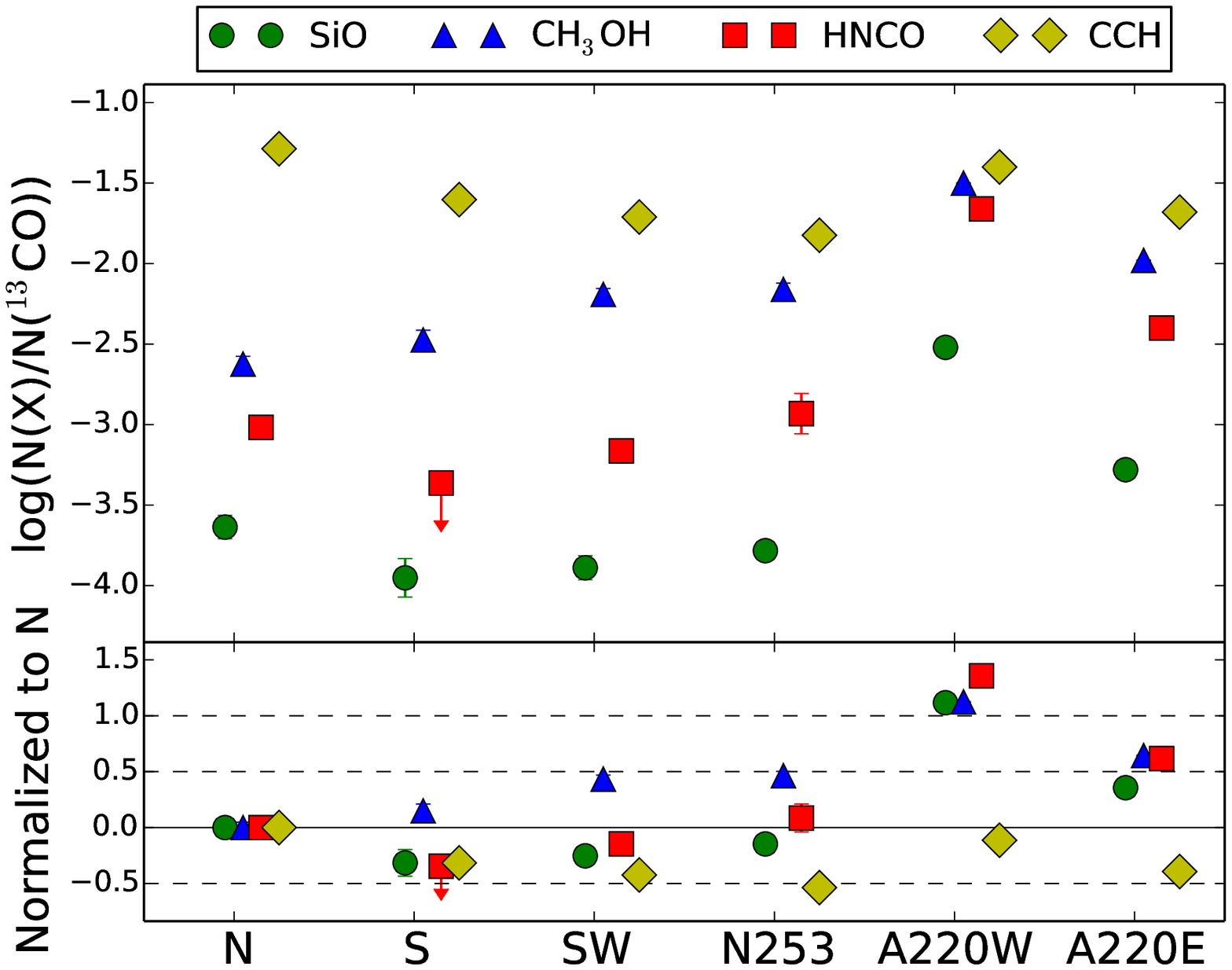} 
\includegraphics[width=.5\textwidth, trim =  0 0 0 0]{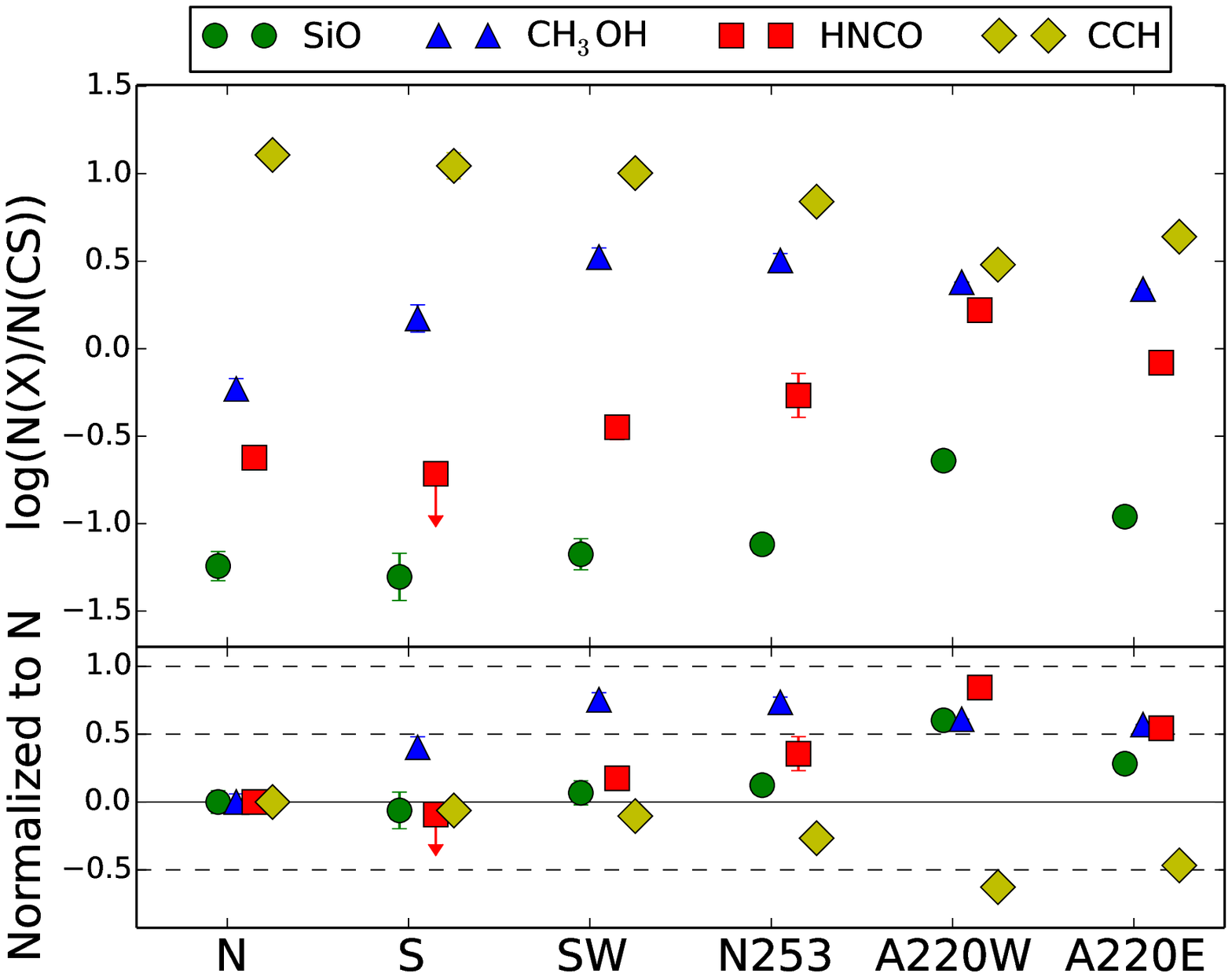} 
}
\centerline{
\includegraphics[width=.5\textwidth, trim =  0 0 0 0]{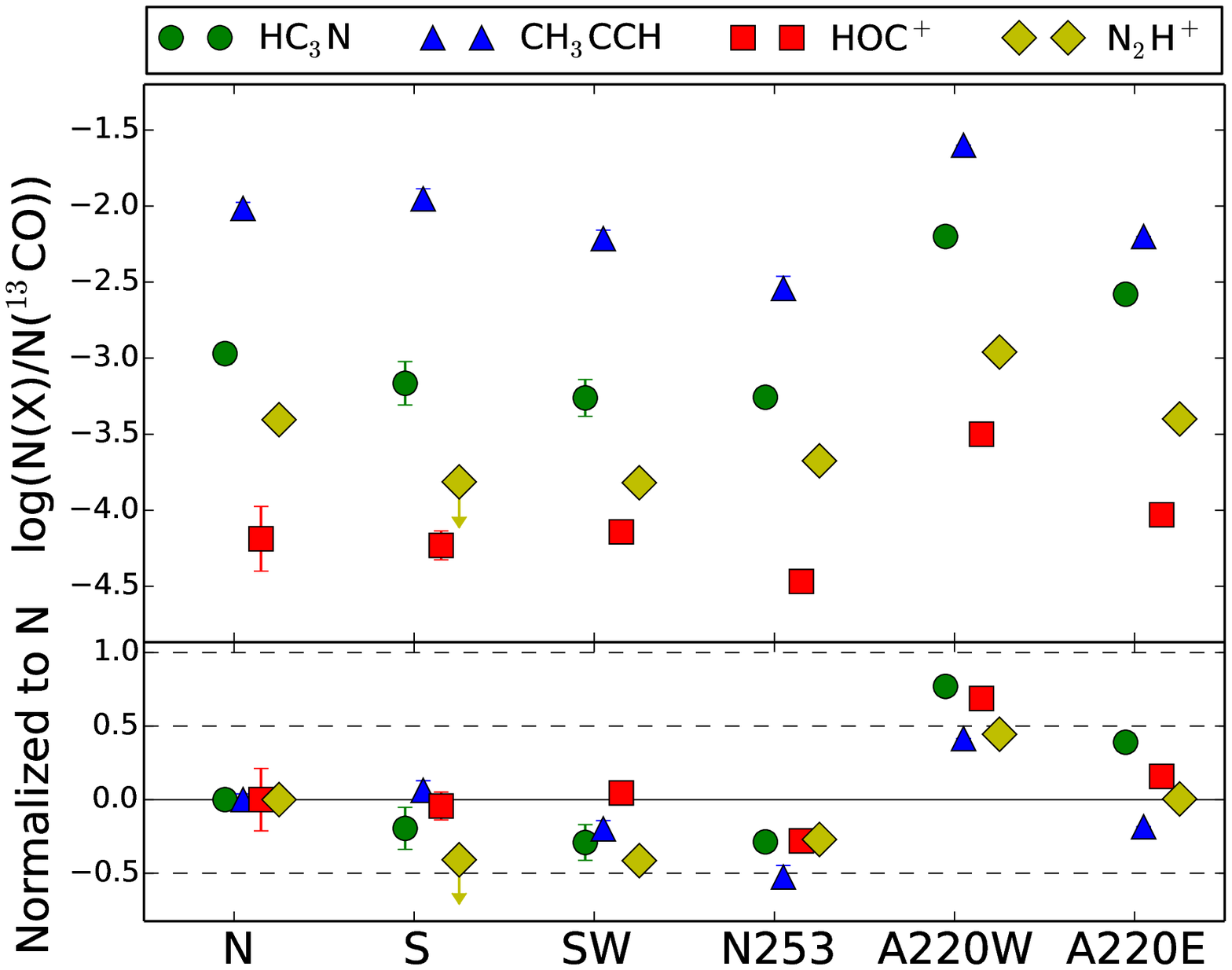} 
\includegraphics[width=.5\textwidth, trim =  0 0 0 0]{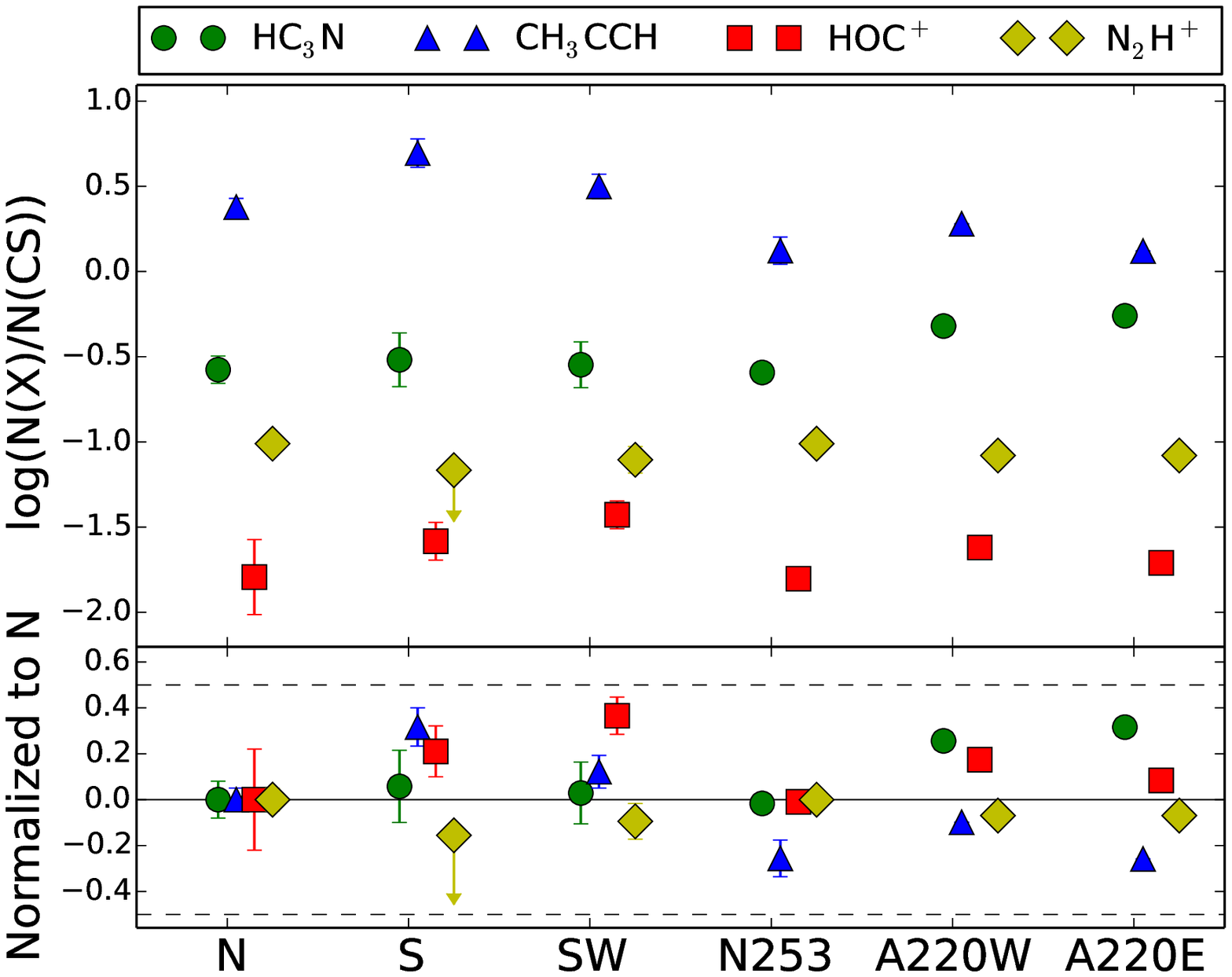} 
}
\caption{({\it Top}) Column density ratios of selected species over $^{13}$CO (left panels) and over CS (right panels) at nuclear positions N, S, and SW in NGC 3256, in NGC 253, and W and E in Arp 220.({\it Bottom})The same figure as the top figure, but all the values are normalized to the position ``N". All the values are shown in a log scale.\label{fig:ratios_nuc}}
\end{figure*}

The column density ratios of the nuclei in NGC 253, NGC 3256, and Arp 220 are plotted in Figures \ref{fig:dense_nuc} and \ref{fig:ratios_nuc},
and these ratios are discussed below.

{\bf HCN, HCO$^+$, CS, and CN:} The column density ratios of HCN, HCO$^+$, CS, and CN seem to follow similar trends;
ratios are similar in NGC 3256(S \& SW) and NGC 253, but values in NGC 3256(N), Arp 220 (E \& W) are a factor of a few higher.
Some exceptions are suppressed CN and HCO$^+$ by a factor of a few in Arp 220 compared with other species, 
and slightly enhanced CS in Arp 220 (W).

{\bf SiO:} Among the galactic nuclei, Arp 220 has the highest column density ratios of SiO both over $^{13}$CO and CS.
The spectral shapes observed in Arp 220 strongly suggests that it has molecular outflows from the both nuclei \citep{2009ApJ...700L.104S} with the one from the western 
nucleus being more prominent \citep{2017ApJ...849...14S}.
It has been already discussed in Section \ref{sec:comp_n3256} that the outflow from the N nucleus in NGC 3256 does not contribute to high SiO abundances.
Unlike the outflow from the S nucleus in NGC 3256, each outflow in Arp 220 should be within the beam that covers each nucleus. 
These outflows might be the reason for the higher SiO abundances in Arp 220.
An alternative explanation is that there are more cloud-cloud collisions in the dense and compact nuclei of Arp 220.

{\bf CH$_3$OH and HNCO:} Column density ratios of CH$_3$OH and HNCO are lowest in the NGC 3256 nuclei than NGC 253 and Arp 220.
As mentioned above, there are nuclear outflows in Arp 220, which can enhance the methanol and HNCO abundance. 
At the same time, NGC 253 also has evidence of an outflow from the starburst \citep{2006ApJ...636..685S,2013Natur.499..450B}.
Alternatively, an enhancement in NGC 253 may be due to the cloud-cloud collision at the intersection of galactic orbits.
Similar to SiO, the N nucleus of NGC 3256 does not seem to enhance CH$_3$OH {\bf and HNCO} by large amount.

{\bf N$_2$H$^+$:} Although the column density ratios of $N({\rm N_2H^+})/N(\rm{^{13}CO})$ have variation among the galactic nuclei, 
ratios $N({\rm N_2H^+})/N({\rm CS})$ have very little variation. This lack of variation may indicate that the column density 
ratios of $N({\rm N_2H^+})/N(\rm{^{13}CO})$ may be the direct measure of dense cloud fraction,
and that, within the dense gas, N$_2$H$^+$ abundance does not vary much with respect to other molecules tracing the dense gas.

{\bf HC$_{3}$N:}  Here we can examine the relation of HC$_3$N fractional abundances and the star formation efficiency ($= \tau_{\rm dep}^{-1}$, see Table \ref{tab:gal_prop}). 
Although galactic nuclei with higher star formation efficiency tend to
have higher HC$_3$N/$^{13}$CO ratios, variation among sources are not as large as the star formation efficiency itself (see also discussion in Section \ref{sec:disc_sfr}).

{\bf CH$_3$CCH:} Column density ratios CH$_3$CCH/$^{13}$CO appear to be higher in sources with higher
star formation efficiencies (see Section \ref{sec:disc_sfr}). However, as mentioned earlier, the behavior of this molecule is not well-known,
and it is unclear what mechanism is causing the enhancement with higher star formation activity.

\begin{figure} 
\includegraphics[width=.5\textwidth, trim =  0 0 0 0]{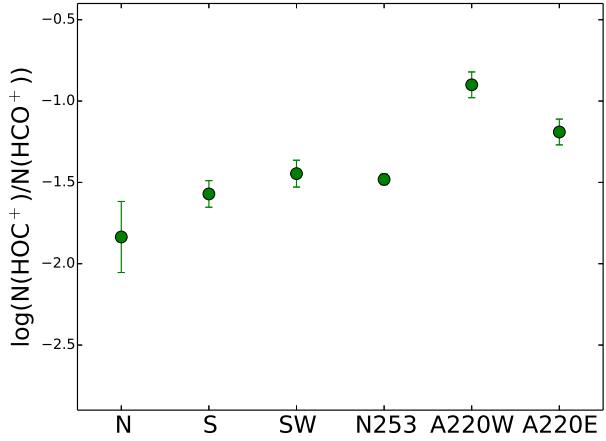} 
\caption{Column density ratios of HOC$^+$ over HCO$^+$ at nuclear 
positions N, S, and SW in NGC 3256, in NGC 253, and W and E in Arp 220.
 \label{fig:hoc+_nuc}}
\end{figure}

{\bf HOC$^+$:} Ratios $N({\rm HOC^+})/N({\rm ^{13}CO})$ is the highest in Arp220W, 
but there is little variation among the galactic nuclei for $N({\rm HOC^+})/N({\rm CS})$.
However, if $N({\rm HOC^+})/N({\rm HCO^+})$ is plotted, it is also the highest in Arp220W (Figure \ref{fig:hoc+_nuc}).
Since $N({\rm HOC^+})/N({\rm HCO^+})$ is more reliable because it is not affected by the difference of elemental abundance,
HOC$^+$ abundance in Arp 220W is likely higher than in other sources.
Since Arp220W has a prominent outflow, the connection of HOC$^+$ with outflows may be seen here again,
but it needs to be confirmed with high-resolution imaging of Arp 220 in HOC$^+$.
The effect of starburst in Arp 220 may also be the reason of increased $N({\rm HOC^+})/N({\rm HCO^+})$.

{\bf CCH:} Since CCH is known to be a PDR tracer, one might expect higher fractional abundances of CCH in galaxies with 
higher star formation rates. However, this is not necessarily the case. A possible reason is the differences in the mean density. 
When the mean density in the ISM is very high, only a fraction of volume can be influenced by UV-photons because those photons become
attenuated before reaching far from the vicinity of OB stars, and the enhancement of PDR tracers cannot be seen in a large scale.

\subsection{Comparison with a Galactic spiral arm region}
Here we compare the chemical compositions of the tidal arm positions in our study with the one in W51
for relevant comparison among spiral arm positions.
W51 is a molecular cloud complex in the Sagittarius arm in the Milky Way at the distance of 5.4\,kpc \citep{2010ApJ...720.1055S}.
Within W51, there is an active star-forming region, which include a hot core W51 e1/e2.
\citet{2017ApJ...845..116W} mapped various molecular species in $\sim 40$\,pc$\times 50\,$pc region in W51,
and studied the average chemistry within this region so that their results can be compared 
with extragalactic interferometric observations taken with beam sizes of at least tens of parsecs.
We do not discuss HOC$^+$ here because this species was not detected in the averaged spectra of W51.
Although their spatial scale is a factor of several smaller than our resolution, we assume that there are molecular clouds with
sizes similar to that of W51 within our beam, and compare the chemistry in the NGC 3256 arm positions and W51.
Since \citet{2017ApJ...845..116W} only had data in the 3-mm band, they derived column densities using excitation temperatures of 
$T_{\rm ex}=10,15,20\,$K. Because the excitation temperatures of our results are $T_{\rm ex}=10\,$K or less, we use their column densities
using $T_{\rm ex}=10\,$K. Although $T_{\rm ex}$ of some molecules are $\sim5\,$K or so, we checked that the derived column densities
do not vary significantly if we derive them by assuming $T_{\rm ex}=10\,$K for those molecules.

\begin{figure}
\includegraphics[width=.5\textwidth, trim =  0 0 0 0]{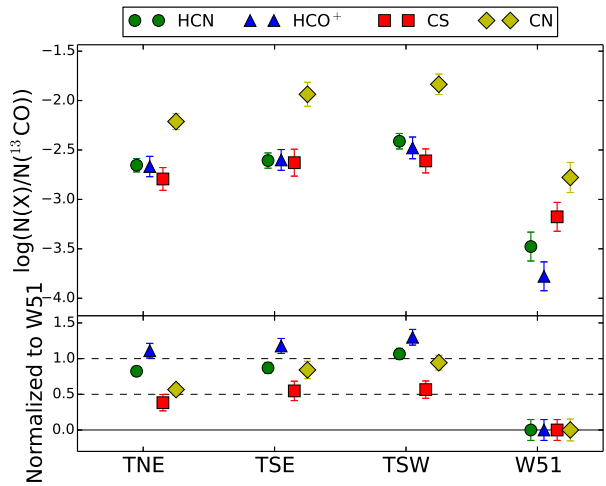} 
\caption{({\it Top}) Column density ratios of HCN, HCO$^+$, CS, and CN over $^{13}$CO at tidal arm positions in NGC 3256 (TNE, TSE, and TSW) and W51. 
({\it Bottom})The same figure as the top figure, but all the values are normalized to W51. All the values are shown in a log scale.\label{fig:dense_arm}}
\end{figure}

{\bf HCN, HCO$^+$, CS, and CN:}
Molecules with higher critical densities have higher column densities ratios over $^{13}$CO
at the tidal arm positions in NGC 3256 than in W 51 (Figure \ref{fig:dense_arm}).
This trend may indicate the higher fraction of dense cloud in the beam in NGC 3256 tidal arms as 
discussed at the beginning of Section \ref{sec:column}, where we suggest the variation of 
these ratios come from the variation of dense gas fraction within NGC 3256.

\begin{figure*}
\centerline{
\includegraphics[width=.5\textwidth, trim =  0 0 0 0]{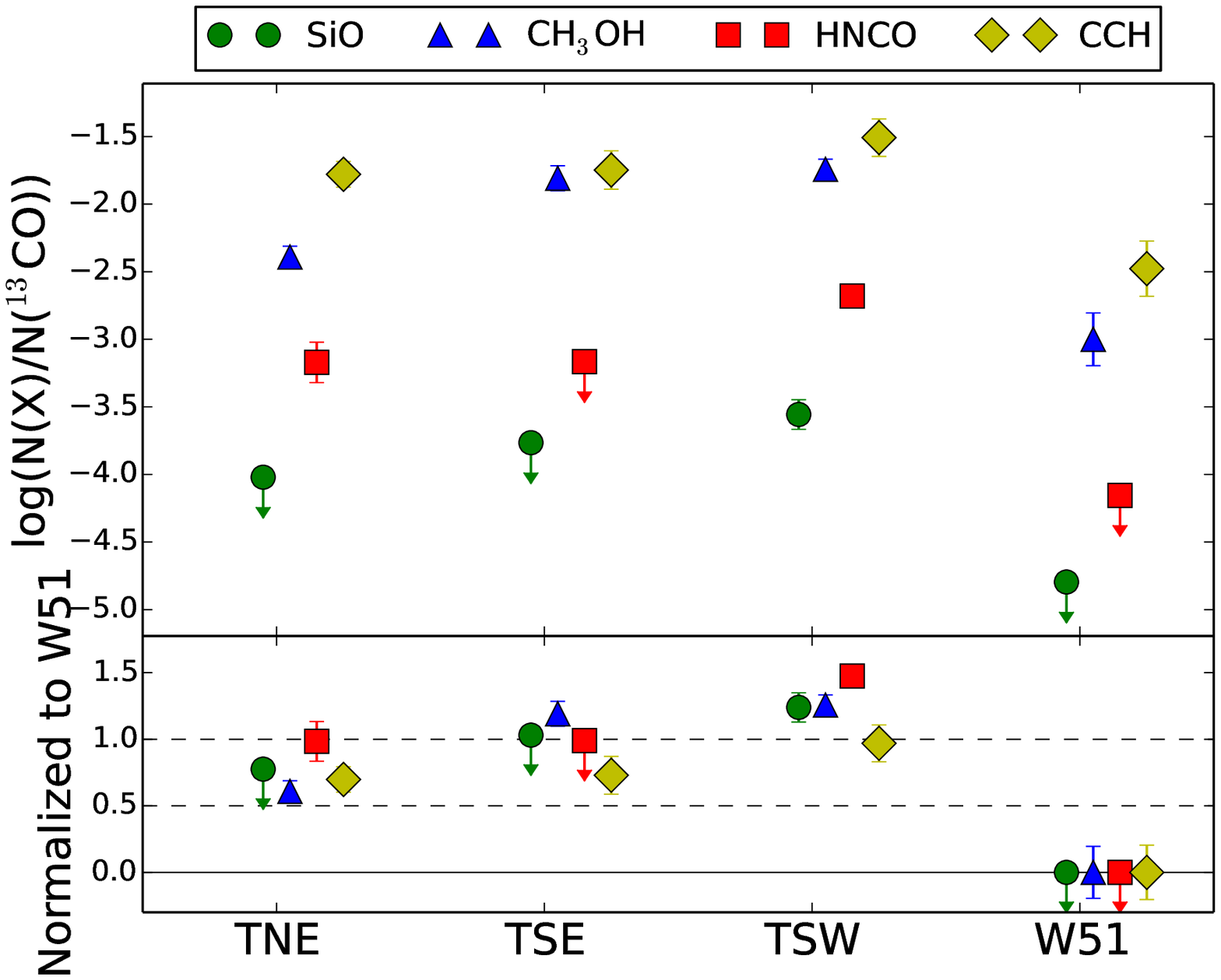} 
\includegraphics[width=.5\textwidth, trim =  0 0 0 0]{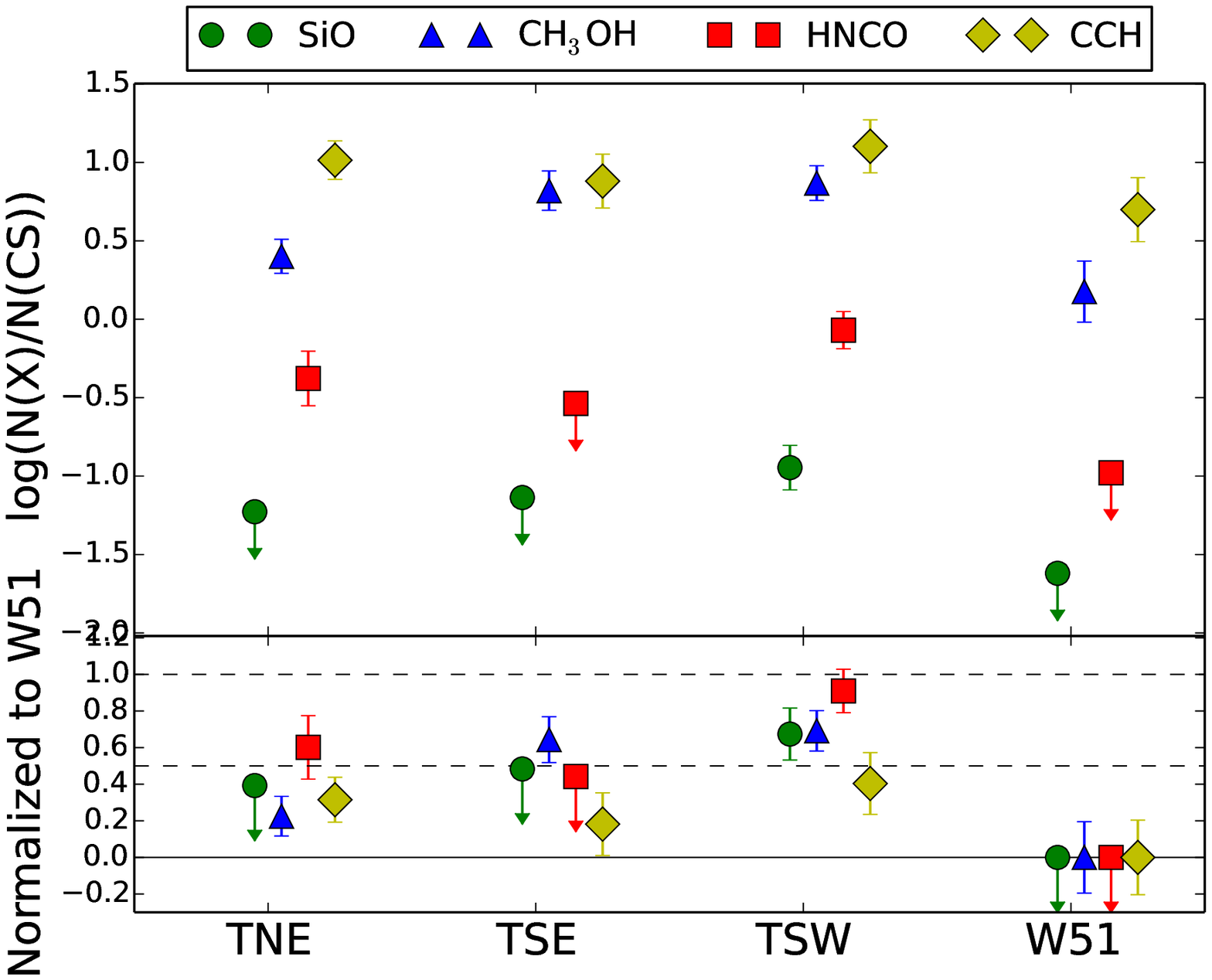} 
}
\centerline{
\includegraphics[width=.5\textwidth, trim =  0 0 0 0]{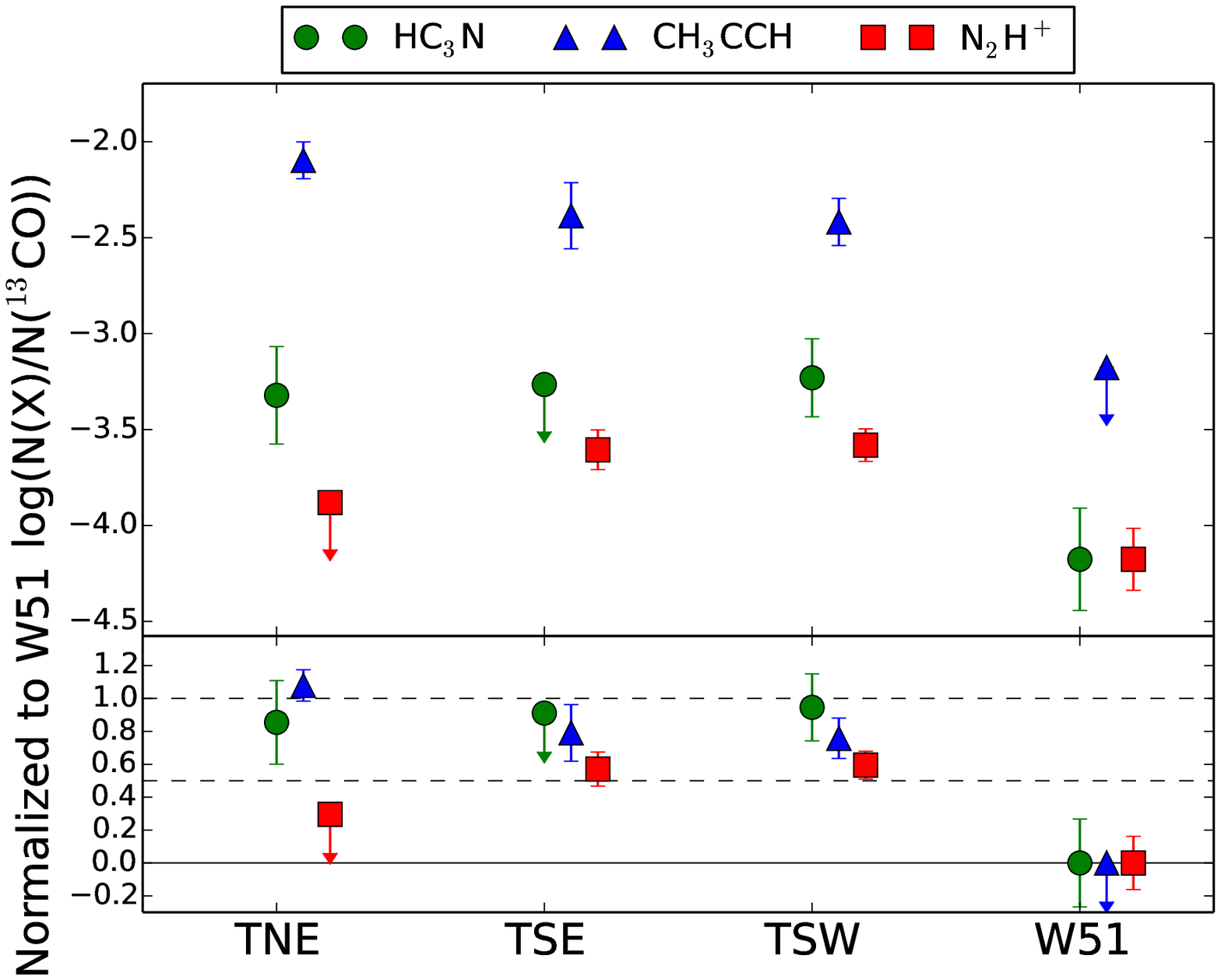} 
\includegraphics[width=.5\textwidth, trim =  0 0 0 0]{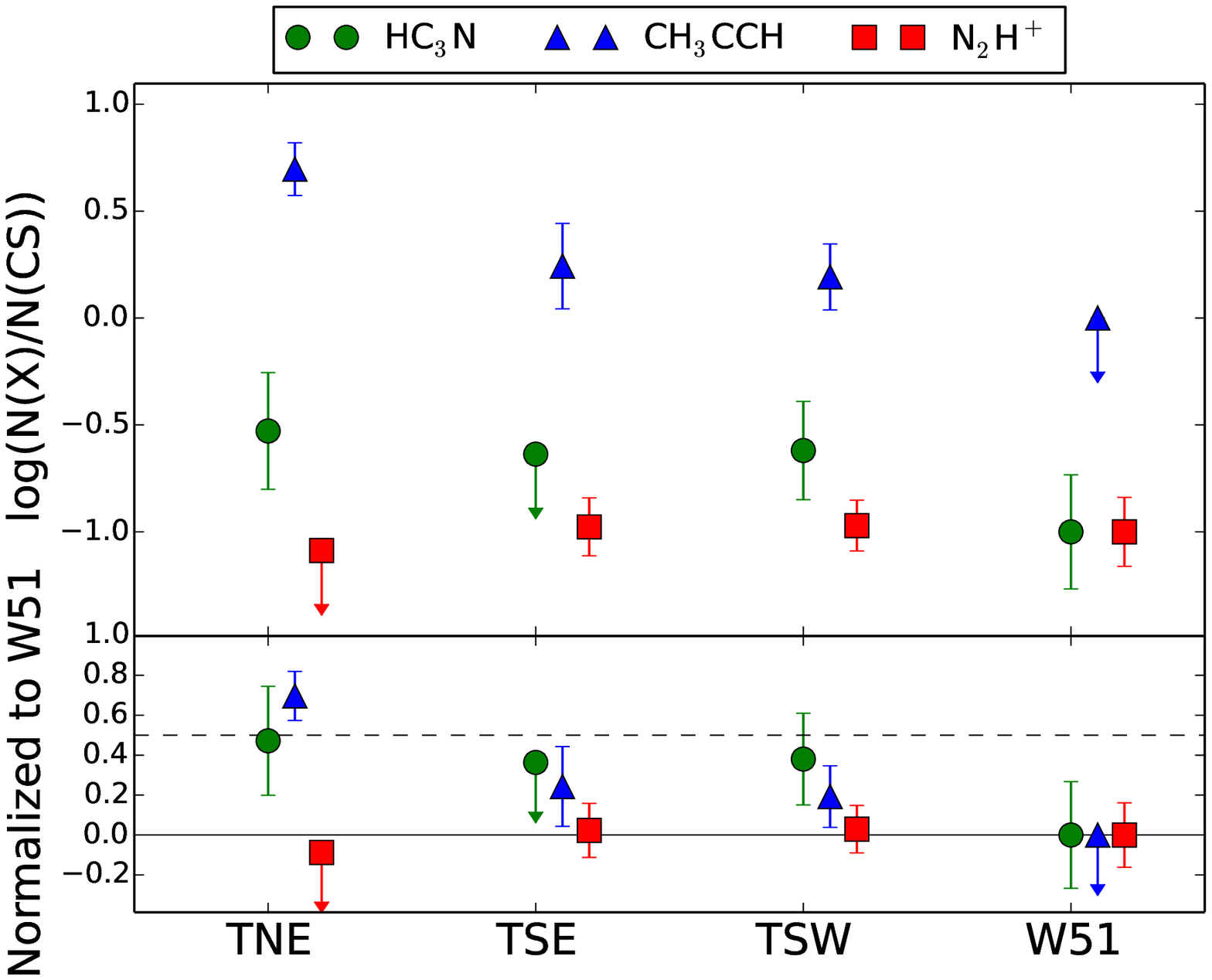} 
}
\caption{({\it Top}) Column density ratios of selected species over $^{13}$CO (left panels) and over CS (right panels) at tidal arm positions in NGC 3256 (TNE, TSE, and TSW) and W51. ({\it Bottom})The same figure as the top figure, but all the values are normalized to W51. All the values are shown in a log scale. \label{fig:ratios_arm}}
\end{figure*}

Column density ratios of other species are plotted in Figure \ref{fig:ratios_arm}.

{\bf SiO:} The column density ratios of SiO both over $^{13}$CO and CS at the position TSW are much higher than the upper limit
in W51. This high abundance of SiO indicates more strong shocks in TSW of NGC 3256, possibly caused by the merger interaction.
The upper limits at positions TNE and TSE are higher than that of W51, 
hence it is unknown whether these positions have more influence from shocks.

{\bf CH$_3$OH and HNCO:} Similarly to SiO, higher column density ratios of methanol and HNCO in NGC 3256 than in W51 
are seen for ratios both over $^{13}$CO and CS.
Unlike SiO and HNCO, the enhancement is also seen in position TSE as well as in TSW for methanol.

{\bf N$_2$H$^+$:} The trend of column density ratios of N$_2$H$^+$ over $^{13}$CO and CS is similar to that of HC$_3$N.
The ratios over $^{13}$CO are higher in NGC 3256 positions than that in W 51, but there is no obvious variation among positions for ratios over CS.
Since N$_2$H$^+$ is also abundant in more compact regions just like HC$_3$N, the same explanation for HC$_3$N can possibly used for this trend.

{\bf HC$_{3}$N:} Ratios $N({\rm HC_3N})/N(^{13}{\rm CO})$ are higher in the tidal arm positions in NGC 3256 than in W 51.
However, the values of $N({\rm HC_3N})/N({\rm CS})$ have very little variation among sources.
In Galactic GMCs, HC$_3$N comes from more compact regions than CS. 
If this is the case also in NGC 3256, our results indicate that within a CS-emitting cloud, 
there is a similar fraction of HC$_3$N-emitting cores both in NGC 3256 and in W 51.

{\bf CH$_3$CCH:} In the tidal arm positions in NGC 3256, $N({\rm CH_3CCH})/N(^{13}{\rm CO})$ and $N({\rm CH_3CCH})/N({\rm CS})$ are higher than in W 51.
The ratio is the highest in TNE, the only off-nucleus position where H40$\alpha$ was detected.
A hint of positive correlation between the star formation rate and CH$_3$CCH is again seen here, but the reason behind needs more investigation.

{\bf CCH:} Ratios $N({\rm CCH})/N(^{13}{\rm CO})$ have higher values at the tidal arm position of NGC 3256 than in W 51, and 
 $N({\rm CCH})/N({\rm CS})$ does not vary much among those positions. This trend is similar to N$_2$H$^+$ and HC$_3$N.
 However, CCH having this similarity with N$_2$H$^+$ and HC$_3$N is puzzling
 because CCH is usually thought to come from rather extended regions.

\section{Discussion}\label{sec:disc}

\subsection{$^{13}$CO and C$^{18}$O anomaly in LIRGs}
It has been pointed out that the $^{12}$CO/$^{13}$CO and $^{12}$C$^{16}$O/C$^{18}$O intensity ratios in luminous infrared galaxies 
are higher than in normal spiral galaxies \citep{1991A&A...249..323A,1992A&A...264...49C}.
\citet{1992A&A...264...49C} concluded that $^{13}$CO and C$^{18}$O lines are about four times weaker in NGC 3256 than 
what is normally expected from $^{12}$CO in their single-dish observations.
This variation in intensity ratios can be caused by excitation effects such as high temperature or turbulence 
\citep{1991A&A...249..323A, 2010A&A...522A..59A, 2014MNRAS.445.2378D}. 
Turbulence can increase $^{12}$CO/$^{13}$CO by causing $^{12}$CO to be less optically thick,
or turbulence occurs at the position where the low-metallicity gas inflows \citep{2014A&A...565A...3H,2016A&A...594A..70K}.
If this is the case, our chemistry analysis using column density ratios $N(X)/N({\rm ^{13}CO})$ is not affected.

However, it can also be caused by abundance deficit of $^{13}$C due to very 
young starburst or top-heavy initial mass function as proposed by \citet{2017ApJ...840L..11S}.
The abundance ratio of $^{12}$CO/$^{13}$CO can also increase by the selective photo-desorption of $^{13}$CO
because $^{13}$CO is less likely to be self-shielded. 
If it is caused by the elemental abundances of $^{12}$C/$^{13}$C, our derived column density ratios $N(X)/N({\rm ^{13}CO})$
may not reflect variation of only $N(X)$, but also of $N({\rm ^{13}CO})$.
If $^{12}$CO/$^{13}$CO column density ratios are higher in the tidal arm positions of NGC 3256 than in W 51,
the higher ratios of $N(X)/N({\rm ^{13}CO})$ for most species in NGC 3256 shown in Figures \ref{fig:dense_arm} and \ref{fig:ratios_arm}
are due to variation of $N({\rm ^{13}CO})$, and not $N(X)$. 
Further observations are needed to obtain the spatial dependence of $^{12}$C/$^{13}$C,
and we leave it as future work.

\subsection{Excitation or abundance?}
From the variation of column density ratios of molecules with high critical densities over $^{13}$CO,
we have argued that this variation should be a reflection of multiple components of different excitation conditions within the beam 
showing up as variation of apparent abundances.
However, we cannot exclude the possibility that intrinsic variation of abundances is playing a role. 
The molecules we show in Figure \ref{fig:dense_n3256}, HCN, HCO$^+$, CS, and CN
do have dependence on the temperature, the density, the UV or cosmic-ray ionization rate, 
but it is difficult to think that those molecules all have similar variation among positions.
Therefore, it is more likely that the density distribution function is the largest contributing factor of this variation.


\subsection{Effects of starburst on the chemical composition}\label{sec:disc_sfr}
\begin{figure} 
\includegraphics[width=.5\textwidth, trim =  0 0 0 0]{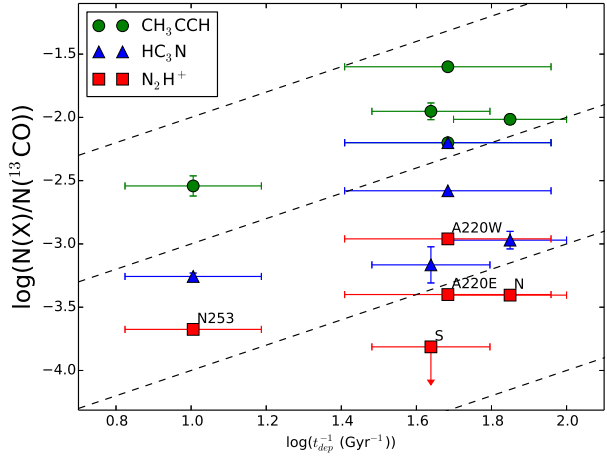} 
\caption{Column density ratios of CH$_3$CCH, HC$_3$N, and N$_2$H$^+$ over $^{13}$CO are shown as a function of an inverse of the depletion times,
which is proportional to the star formation efficiencies. Note that separate values of star formation efficiencies in the two nuclei of Arp 220 could not be obtained , and 
the same value is used for the Eastern and Western nuclei. The case of $N(X)/N(^{13}CO) \propto \tau_{\rm dep}^{-1}$ are shown in dotted lines for a reference. \label{fig:set2_sfe}}
\end{figure}

\begin{figure*} 
\includegraphics[angle=0,width=1.\textwidth,trim= 0 0 0 0]{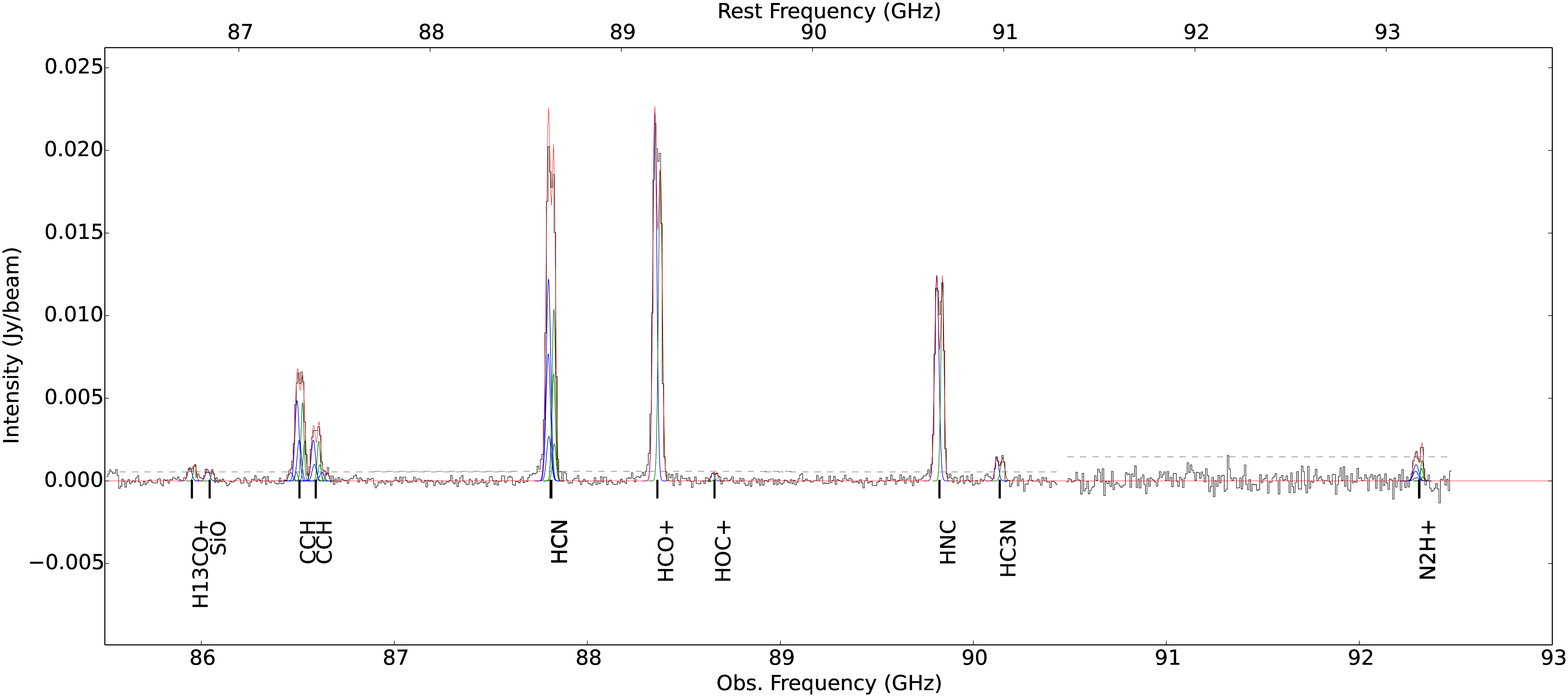} 
\includegraphics[angle=0,width=1.\textwidth,trim= 0 0 0 0]{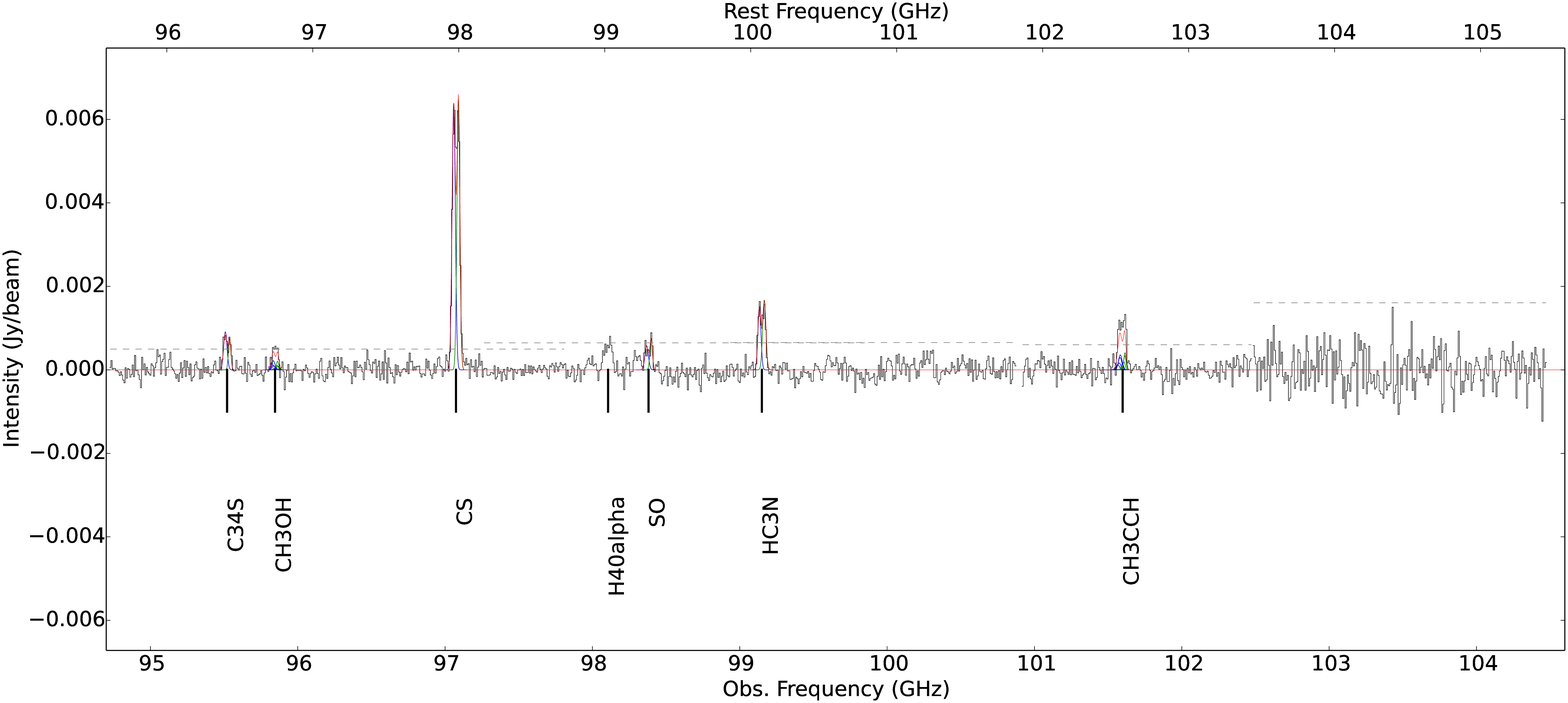} 
\includegraphics[angle=0,width=1.\textwidth,trim= 0 0 0 0]{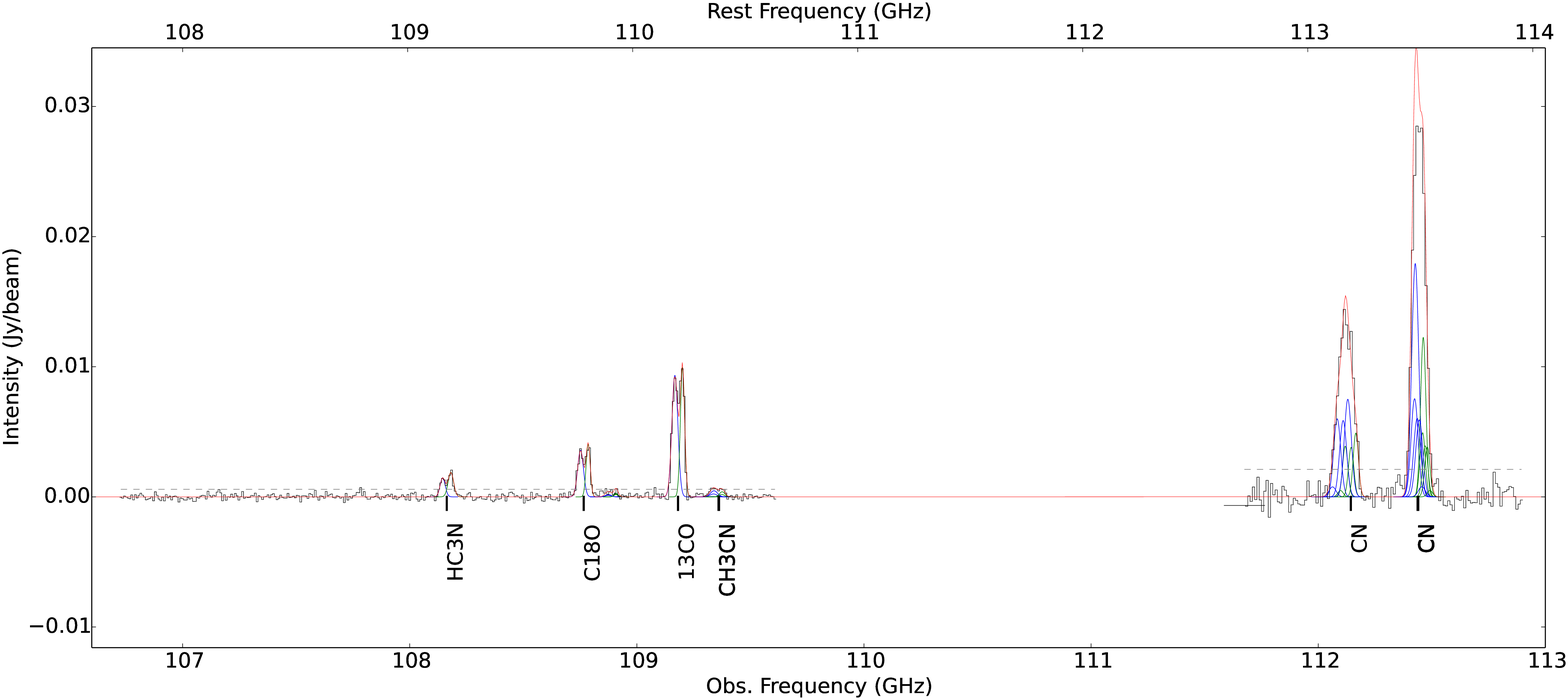} 
\caption{\label{fig:N_b3} Observed spectra from 1.7$''$-resolution cubes at position N in Band 3. 
Fitted spectra for individual transitions are shown in blue for component 1 and in green for component 2,
while the summed intensities of individual transitions are shown in red. Grey dotted lines show the levels of $3\sigma$.
The complete figure set (32 images) is available in the online journal of the published version. } 
\end{figure*}

\begin{figure*} 
\includegraphics[angle=0,width=1.\textwidth,trim= 0 0 0 0]{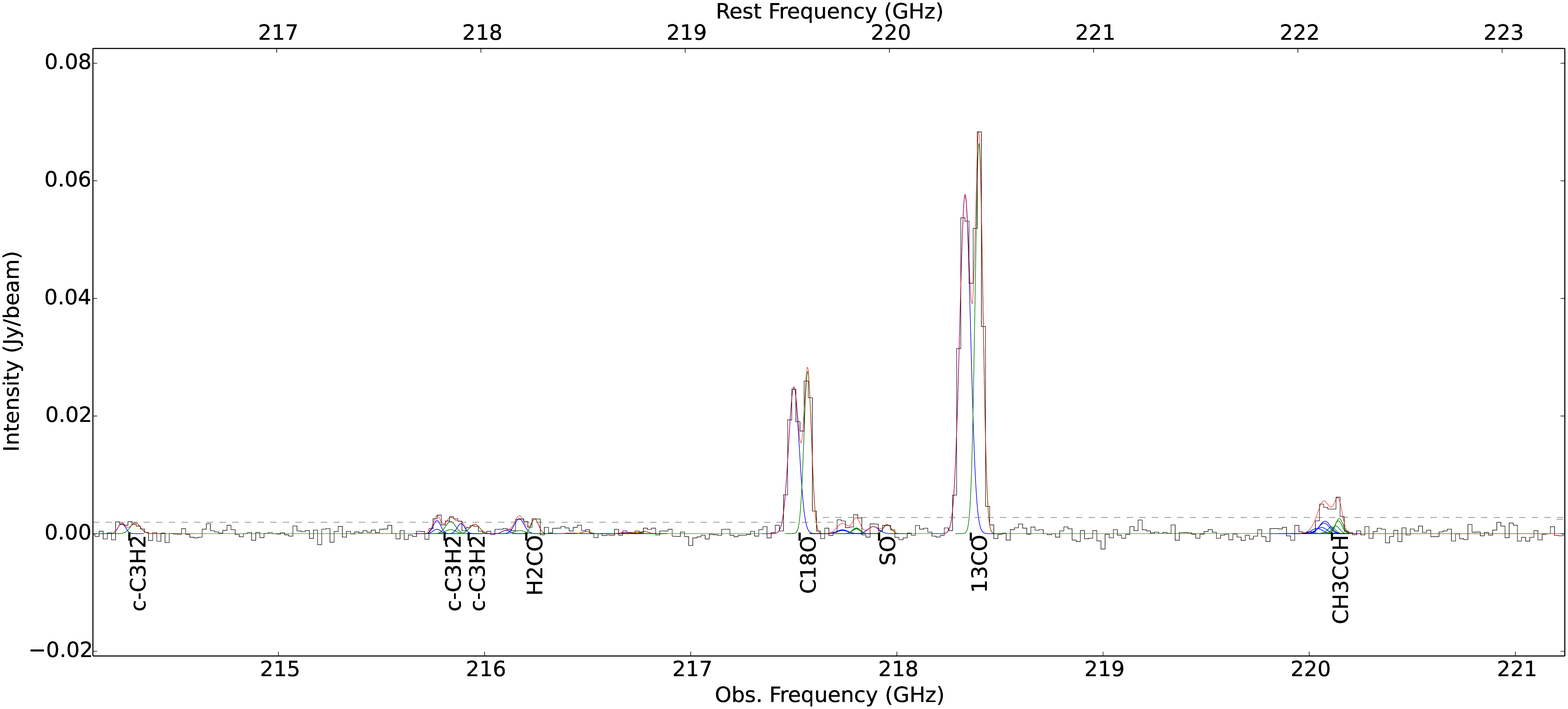} 
\includegraphics[angle=0,width=1.\textwidth,trim= 0 0 0 0]{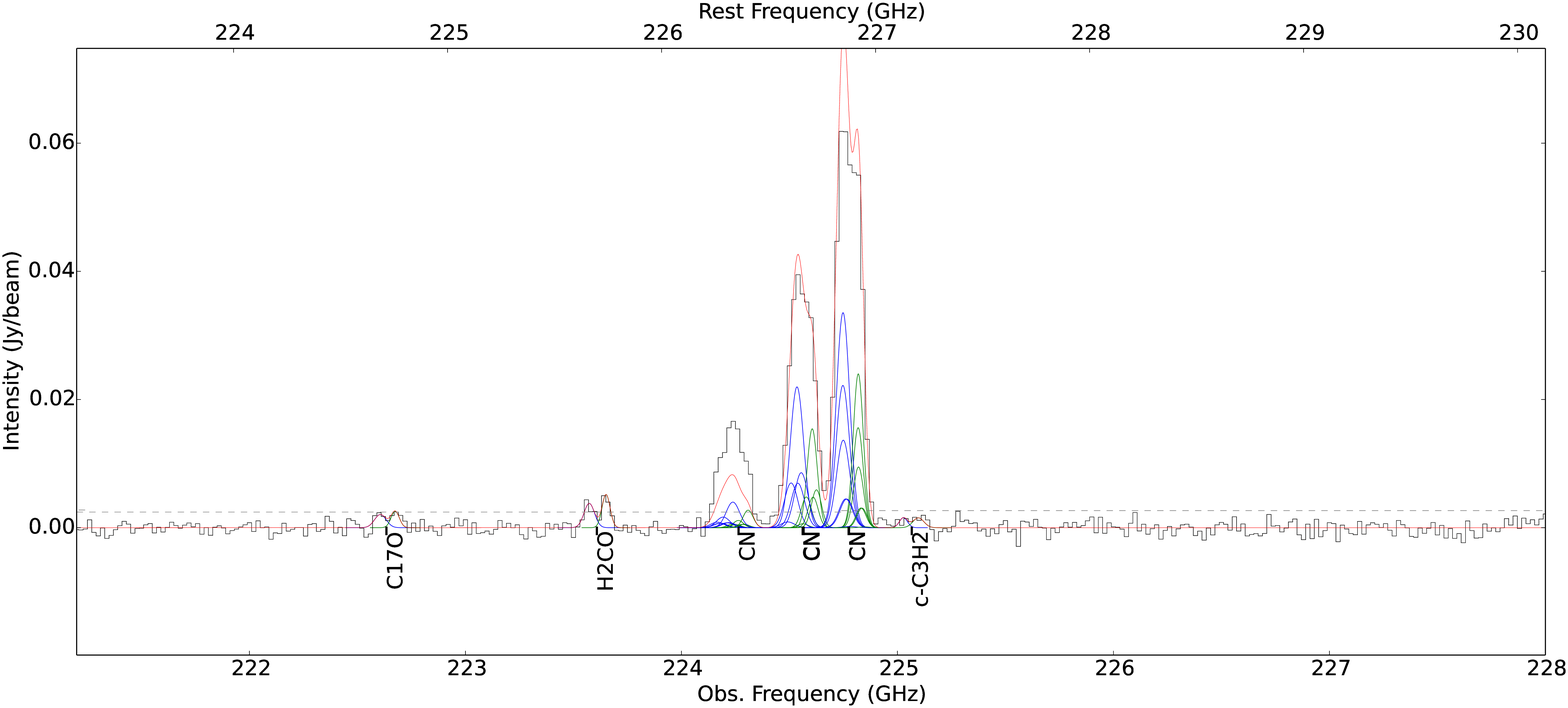} 
\includegraphics[angle=0,width=1.\textwidth,trim= 0 0 0 0]{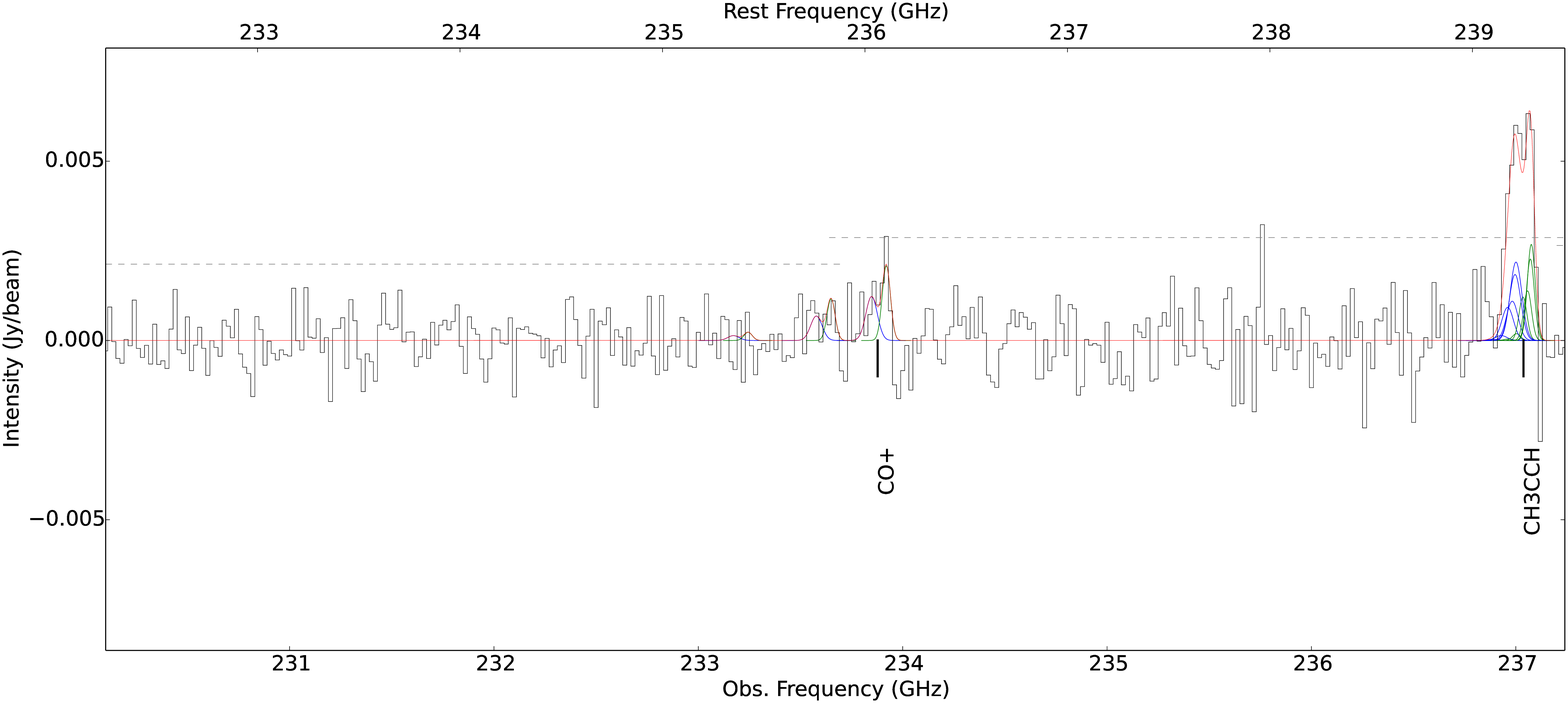} 
\caption{\label{fig:N_b6_1} Same as Fig \ref{fig:N_b3}, but for the Band 6 with sky frequencies 214.1-237.24 (GHz). } 
\end{figure*}

\begin{figure*} 
\includegraphics[angle=0,width=1.\textwidth,trim= 0 0 0 0]{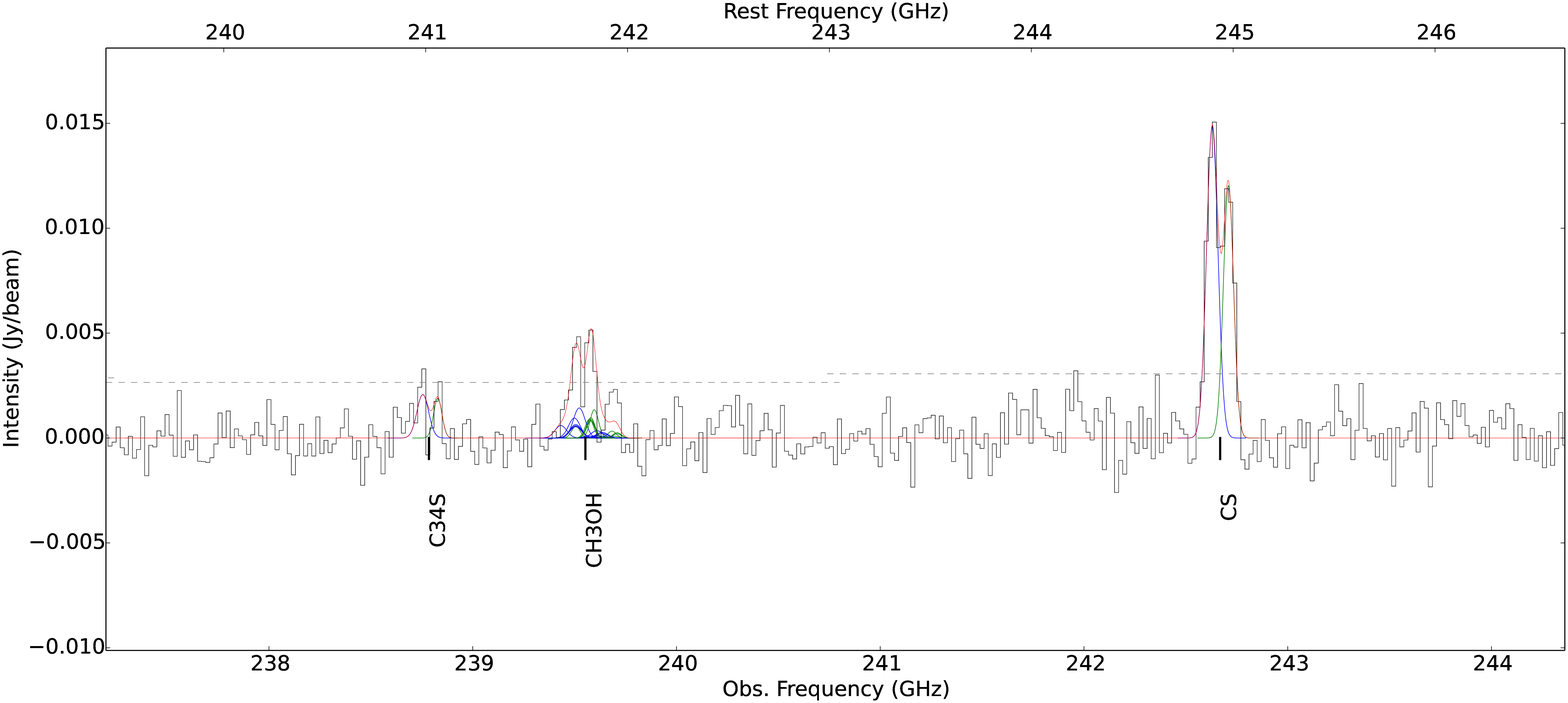} 
\includegraphics[angle=0,width=1.\textwidth,trim= 0 0 0 0]{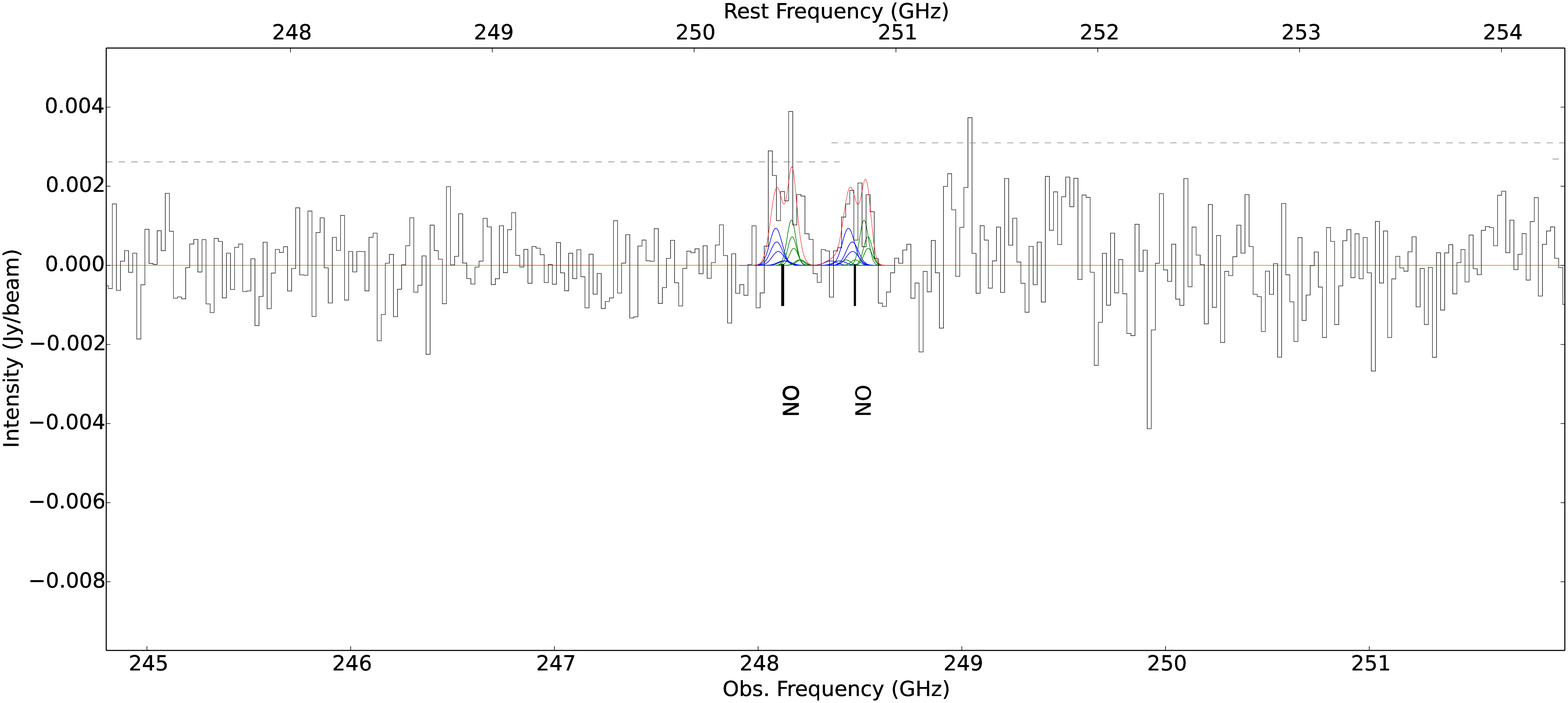} 
\includegraphics[angle=0,width=1.\textwidth,trim= 0 0 0 0]{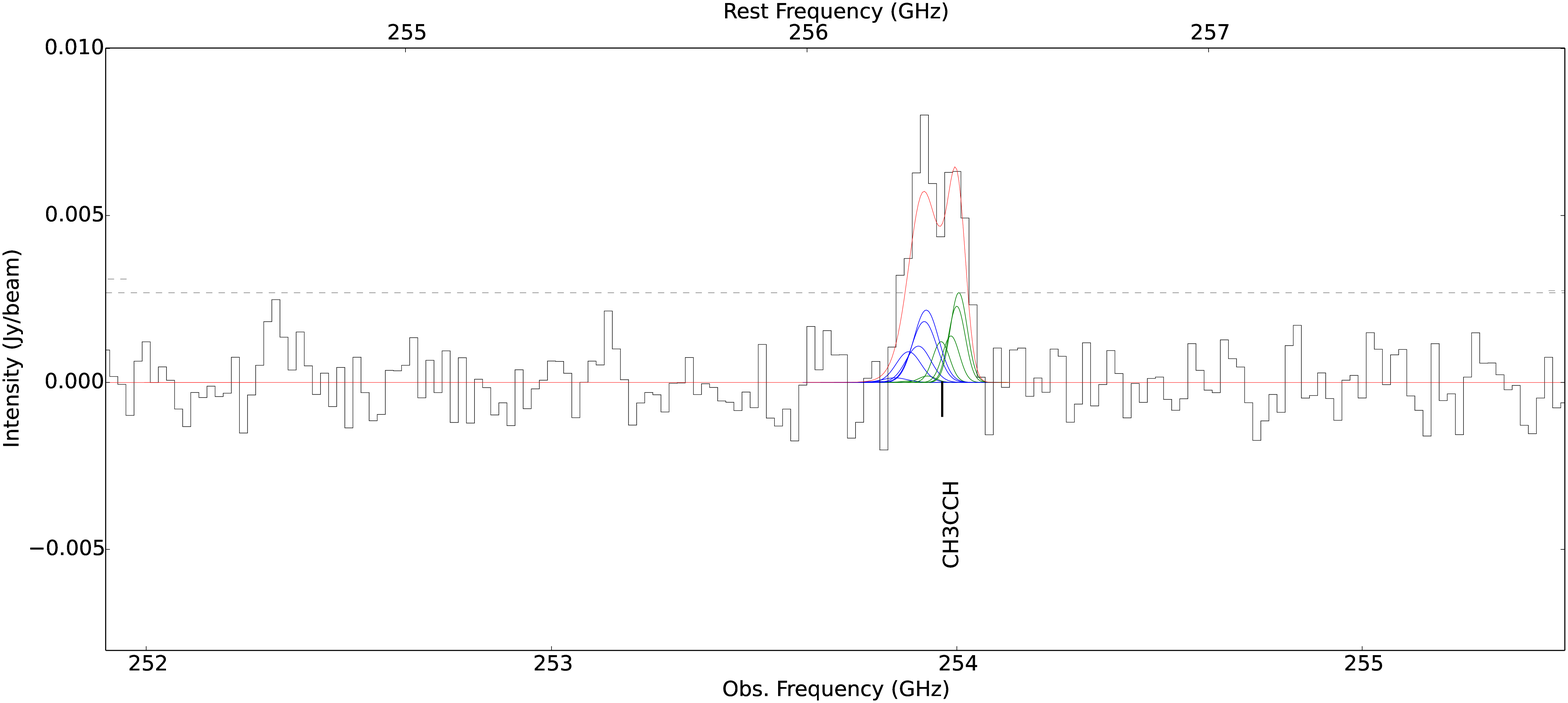} 
\caption{\label{fig:N_b6_2} Same as Fig \ref{fig:N_b3}, but for the Band 6 with sky frequencies 237.2-255.5 (GHz). } 
\end{figure*}

\begin{figure*} 
\includegraphics[angle=0,width=1.\textwidth,trim= 0 0 0 0]{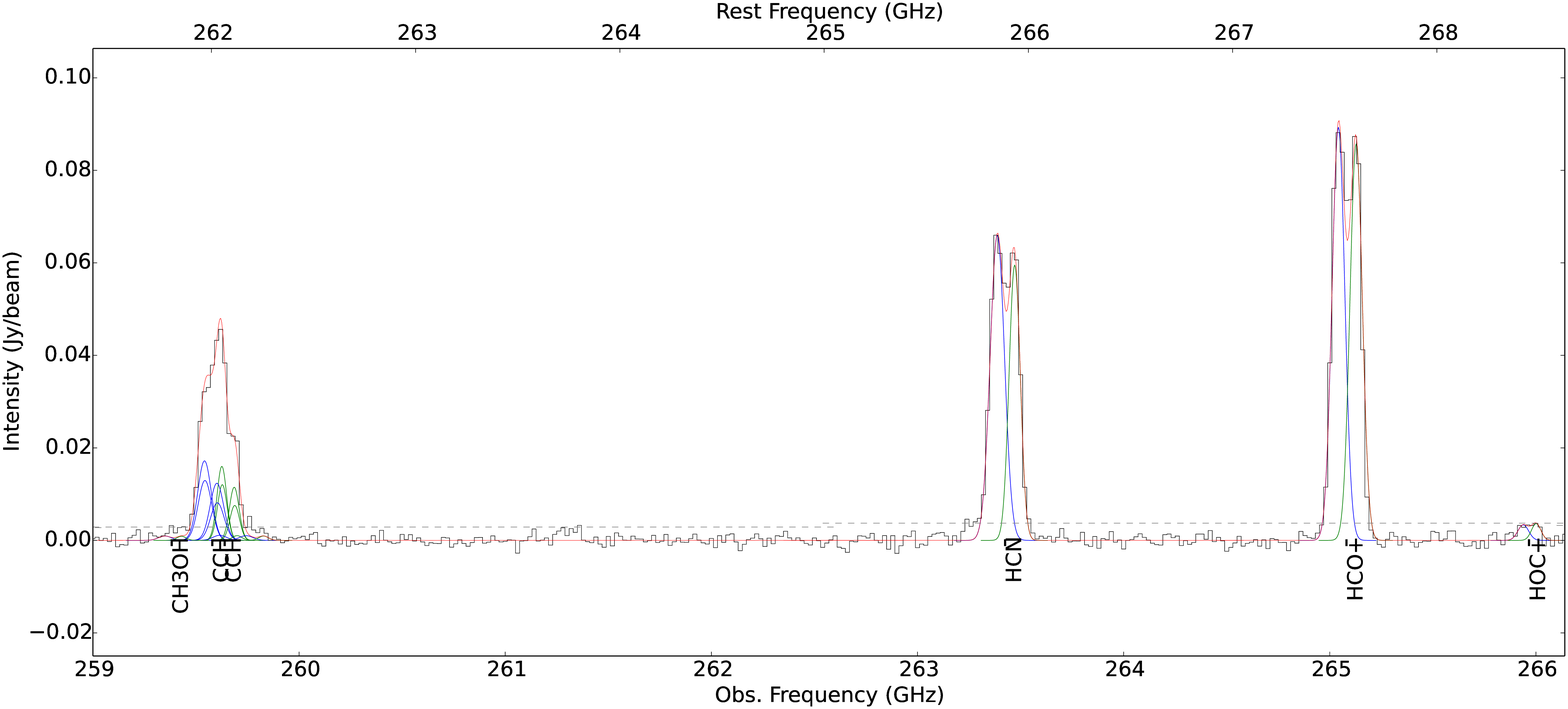} 
\includegraphics[angle=0,width=1.\textwidth,trim= 0 0 0 0]{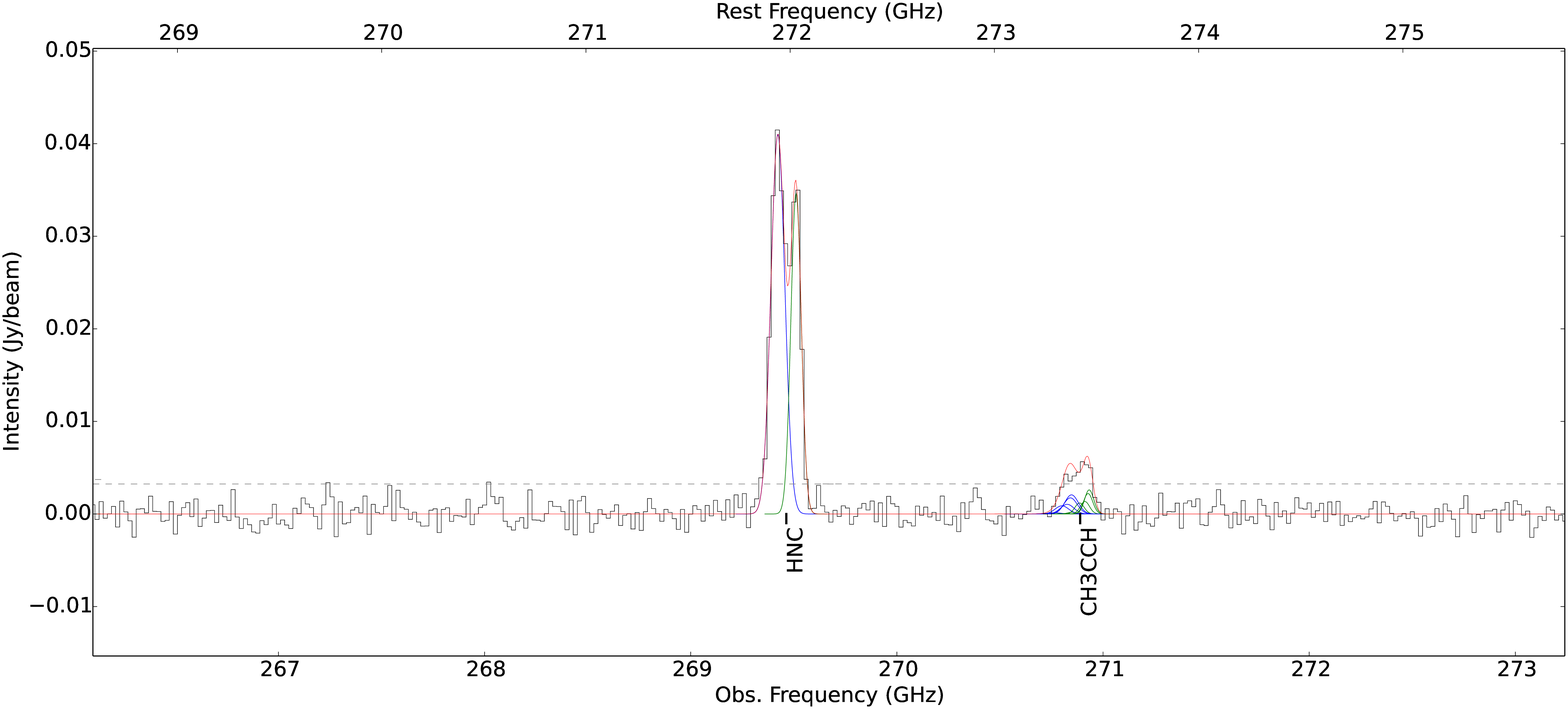} 
\caption{\label{fig:N_b6_3} Same as Fig \ref{fig:N_b3}, but for the Band 6 with sky frequencies 259.0-273.24 (GHz).} 
\end{figure*}

If some molecules emit only from compact star-forming regions, then the column densities of these molecules are proportional to the surface
number density of the star-forming regions and hence should be proportional to the star formation rate in the observed area.
For such molecules $N(X)/N(^{13}{\rm CO})$ should be proportional to the star formation efficiencies assuming that $N(^{13}{\rm CO}) \propto N({\rm H_2})$.
Some column density ratios in Figure \ref{fig:ratios_nuc} are rearranged by the star formation efficiencies in Figure \ref{fig:set2_sfe}.
As obviously seen in Figure \ref{fig:set2_sfe}, error bars on the star formation efficiencies are large due to uncertainty in molecular mass. 
Therefore, the precise relation cannot be determined, but there are some trends that can be discussed as below.
Although HC$_3$N and N$_2$H$^+$ are molecules found preferentially in denser regions, 
those molecules do not have strong correlation with the star formation efficiency when compared among different galactic nuclei.
NGC 3256 has at least a factor of a few, but more likely about an order of magnitude, star formation efficiency than NGC 253,
while Arp 220 and NGC 3256 have similar star formation efficiencies.
The differences of $N({\rm HC_3N})/N(^{13}{\rm CO})$ and $N({\rm N_2H^+})/N(^{13}{\rm CO})$ among
 the three galactic nuclei are not as large as the variation of star formation efficiency.

There are a few possible reasons for this lack of correlation between the abundances of those molecules and the star formation efficiencies.
First, those molecules may not be tracing the star-forming regions. 
For example, HC$_3$N may also be enhanced in shocks \citep[e.g.,][]{2015A&A...584A.102H}.
Another explanation is that HC$_3$N is prone to dissociation by UV-photons, and may decrease in abundances in PDRs.
N$_2$H$^+$ abundances may also be influenced by the cosmic-ray ionization rate.
Second, molecular clouds may already be reflecting the next time step of star formation because those clouds are 
places of future star formation.
If the feedback form the existing stars is suppressing the on-going star formation, the chemistry is unlikely to reflect 
the star formation rate.
However, it is hard to assume that NGC 3256 is already quenching star formation before the final stage of the merging event.

The column density ratios $N({\rm CH_3CCH})/N(^{13}{\rm CO})$ do seem to vary as the star formation efficiency,
but the behavior of this molecule still needs to be understood.

\subsection{Effects of a merging event on the properties of the ISM}
From our observations of SiO and CH$_3$OH, it is likely that frequent shocks are occurring in tidal arms of NGC 3256.
Such shocked ISM is also proposed to be present galaxy-wide in NGC 3256 from observations of optical emission lines by \citet{2011ApJ...734...87R},
and they suggested shocks as important mechanism to dissipate kinetic energy and angular momentum,
which can promote the transport of gas into the central regions.
Our observations also suggest higher fraction of dense clouds in the tidal arms of NGC 3256 than in W 51,
likely due to the compression from the shock. If such compression occurs, one could ask a question
whether this compression can induce star formation by cloud-cloud collision.
Judging from the IR observations and the radio recombination line in our observations, 
there is no such evidence of enhancement. In fact, positions with high abundance of shock molecular tracers such as TSE or TSW
are the positions that lack the evidence of star formation.

\section{Summary}\label{sec:summary}
We have conducted a molecular line survey using ALMA Band 3 and Band 6 in an infrared-luminous merger NGC 3256.
This paper first presents continuum images, velocity-integrated intensity maps, and intensity ratio maps.
From the observed intensities, 
column densities of detected molecules are derived in 8 positions of interest including the two nuclei, tidal arms positions, and outflow positions.
We have compared the derived molecular compositions within NGC 3256, among some galactic nuclei in NGC 3256, Arp 220, and NGC 253,
and between the tidal arms of NGC 3256 and a spiral arm in our Galaxy at W 51.

Here are our main findings:
\begin{itemize}
\item The intensity ratios of HCN($1-0$)/HCO$^+$($1-0$) and HCN($1-0$)/CS($2-1$) are lower in the S nucleus than 
in the N nucleus. The S nucleus has some signs of an AGN while the N nucleus does not have any evidence of an AGN. 
Previous statistical studies of HCN/HCO$^+$ and HCN/CS in other galaxies showed enhanced ratios in AGN-containing galaxies 
\citep{2001ASPC..249..672K,2014AJ....148....9I,2016ApJ...818...42I}, and our results in NGC 3256 do not follow this trend.
\item Higher influence of shock is found in the outflow position from the southern galaxy of NGC 3256 shown in SiO observations.
\item Comparing the chemistry of NGC 3256, NGC 253, and Arp 220, Arp 220 shows an enhancement in $N({\rm HC_3N})/N({\rm ^{13}CO})$
and $N({\rm SiO})/N({\rm ^{13}CO})$. The enhancement of HC$_3$N is likely to be caused by the hot and dense ISM of Arp 220,
while SiO abundances may be increased due to the shock from the outflows.
\item We examined the relationship with the column density ratios over $^{13}$CO of observed species among the galactic nuclei of NGC 3256, NGC 253, and Arp 220.
 The only ratio that seems to positively correlate with the star formation efficiency is $N({\rm CH_3CCH})/N({\rm ^{13}CO})$, 
 which needs more understanding of major formation and destruction routes.
\item The tidal arm positions in NGC 3256 also have strong influence of shock compared with Galactic spiral arm position.
 They also show hints of compression due to higher apparent 
column density ratios of molecules with higher critical densities over $^{13}$CO, 
but these ratios may also be due to lower $^{13}$C elemental abundances.
\end{itemize}

Our line survey in two frequency bands has highlighted the chemical and physical variation of the ISM within NGC 3256 and among galaxies.
Further analysis of the physical conditions such as the temperature and the density using the large velocity gradient analysis would 
help our understanding of the chemistry. The physics and the chemistry of outflow features are worth future follow-up studies. 

\acknowledgments

\sloppypar{We are extremely grateful to the ALMA staff for their service in observations, quality assessment, and local assistance
 at Taiwanese ALMA regional center (ARC). 
 This paper makes use of the following ALMA data: ADS/JAO.ALMA\#2015.1.00412.S, ADS/JAO.ALMA\#2016.1.00965.S, ADS/JAO.ALMA\#2015.1.00993.S, 
 and ADS/JAO.ALMA\#2011.0.00525.S. ALMA is a partnership of ESO (representing its member states), NSF (USA) and NINS (Japan), together with NRC (Canada), 
 MOST and ASIAA (Taiwan), and KASI (Republic of Korea), in cooperation with the Republic of Chile. The Joint ALMA Observatory is operated by ESO, AUI/NRAO and NAOJ.}
 KSa and NH were supported by the grant MOST 106-2119-M-001-025 from the Ministry of Science and Technology, Taiwan. 

{\it Facilities: ALMA}

{\it Software: CASA \citep[v4.5.3, v4.6, and v4.7][]{2007ASPC..376..127M}, MADCUBA (http://cab.inta-csic.es/madcuba/Portada.html)}



\clearpage

\clearpage

\begin{deluxetable}{cccccccccc}
\tablecolumns{8}
\tablewidth{0pc}
 \tabletypesize{\scriptsize}
\tablecaption{Observational Parameters\label{tab:obs_param}.}
\tablehead{\colhead{ID} &\colhead{Array} &\colhead{Cycle} &\colhead{config}  &\colhead{Observation date}  &\colhead{$N_{ant}$}  &\colhead{Baseline} &\colhead{T$_{on}$}
 &\colhead{LSB range} &\colhead{USB range}\\
 \colhead{} &\colhead{} &\colhead{} &\colhead{}  &\colhead{}  &\colhead{}  &\colhead{Range (m)} &\colhead{(min)}
 &\colhead{(GHz)} &\colhead{(GHz)}\\}
\startdata
  \multicolumn{7}{c}{PI data (2015.1.00412.S, 2016.1.00965.S)}\\
  \hline
Band3\_a  &12 m&3 &C40-3  &2016-5-28  &39  &15   - 704.1&47.3 &85.49-89.07&97.24-100.89\\
                  &12 m &4&C40-6&2016-10-29&40&18.4 - 1107.2&46.7\\
Band3\_b  &12m&3 &C40-3  &2016-5-28  &36  &15.7  -649.4&46.2 &88.91-92.49&100.91-104.49\\
 & 12 m&4&C40-6&2016-10-22&39&18.2-1291.2&13.1\\
Band3\_c  &12 m&3 &C36-2/3  &2016-4-30  &40  &14.6  -627.6&40.8 &94.71-98.29&106.71-110.29\\
Band6\_a  &12 m&3 &C36-2/3  &2016-4-1  &43  &14.0  -431.0&11.4 &214.06-217.71&230.06-233.71\\
 &7m &4&\nodata &2016-10-08&10&7.4-44.2&18.1 \\
Band6\_b  &12 m&3 &C36-2/3  &2016-4-2  &42  &14.1  -429.7&9.9 &217.61-221.26&233.61-237.26\\
Band6\_c  &12 m&3 &C36-2/3  &2016-4-2  &42  &14.3  -425&9.9 &221.16-224.81&237.16-240.81\\
Band6\_d  &12 m&3 &C36-2/3  &2016-4-2  &42  &14.0  -418.9&9.9 &224.71-228.36&240.71-244.36\\
Band6\_e  &12 m&3 &C36-2/3  &2016-4-2  &42  &13.4  -404.0&10.9 &244.79- 248.43&258.98-262.62\\
  &12 m&4 &C40-4  &2016-11-28  &45  &16.1-701.2 &10.9\\
   &7m&4 &\nodata&2016-10-14&9&9.0-43.6&18.1\\
Band6\_f  &12 m&3 &C36-2/3  &2016-4-2  &42  &13.7  -412&10.4 &248.32-251.96&262.51-266.15\\
 &7m&4 &\nodata&2016-10-11&10&9.0 - 44.5&17.6\\
Band6\_g  &12 m&3 &C36-2/3  &2016-4-9  &42  &12.9  -395.5&10.4 &251.87-255.51&266.06-269.70\\
 &7m&4 &\nodata&2016-10-05&8&7.5-39.6&17.1\\
Band6\_h  &12 m&3 &C36-2/3  &2016-4-10  &40  &14.1  -420.1&10.4 &255.44-259.08&269.63-273.27\\
 &12 m&4 &C40-4&2016-11-29&43&13.5 - 632.0&10.4\\
  &7m&4 &\nodata&2016-10-21&9&7.6-44.1&17.6\\
  \hline
  \multicolumn{7}{c}{Archive Data (2015.1.00993.S)}&\\
  \hline
  Band3\_a  &12 m&3 &C36-1/2 &2016-3-4  &42  &13.6-429.4&33.3 &84.26-87.63 &96.30-99.86\\
     &12 m&3 &C36-1/2 &2016-3-7  &40  &13.7-429.4&33.3\\
     Band3\_b &12m &3 &C36-1/2&2016-3-7&40 &12.6 - 428.0&32.8 &86.89 - 90.45 &98.93-102.49\\
     &12 m&3&C36-2/3&2016-3-8&44&13.1-438.9&32.8\\
     Band3\_c &12 m &3&C36-2/3&2016-3-9&41&13.9 - 428.2&33.8 &94.56 - 97.83 &106.06 - 109.62\\
     &12 m&3&C36-2/3&2016-3-11&39&13.6 - 426.3&33.8\\
\enddata

\end{deluxetable}

\begin{deluxetable}{lccc}
\tablecolumns{4}
\tablewidth{0pc}
\tablecaption{Image parameters used to create the moment 0 maps in Figures \ref{fig:mom0}-\ref{fig:mom0-2} and their cube properties. 
For the weighting, ${\rm rob}a$ indicates that it was imaged with a robust parameter $a$, and ${\rm rob}a{\rm tp}b$ indicates that it was imaged with 
a robust parameter $a$ while using uv-taper of outer baselines less than $b ''$. \label{tab:imparline}}
\tablehead{\colhead{Line} &\colhead{weighting} &\colhead{beam size ($''$)} &\colhead{rms (mJy/beam km/s)}}
\startdata
$^{13}$CO($1-0$) &rob1 &1.50 $\times$ 1.44 &31.7\\
$^{13}$CO($2-1$) &rob2tp0.5 &1.06 $\times$ 0.98 &82.9\\
C$^{18}$O($1-0$) &rob1 &1.51 $\times$ 1.44 &19.8\\
C$^{18}$O($2-1$) &rob2tp0.5 &1.07 $\times$ 1.01 &56.1\\
HCN($1-0$) &rob0 &0.73 $\times$ 0.61 &28.5\\
HCN($3-2$) &rob2tp0.5 &0.94 $\times$ 0.91 &81.6\\
HCO$^+$($1-0$) &rob0 &0.73 $\times$ 0.61 &30.4\\
HCO$^+$($3-2$) &rob2tp0.5 &0.94 $\times$ 0.90 &113.6\\
HNC($1-0$) &rob2&1.69 $\times$ 1.48 &21.0\\
HNC($3-2$) &rob2tp0.5&0.98 $\times$ 0.87&71.6 \\
CS($2-1$) &rob1&1.65 $\times$ 1.60 &20.7\\
CS($5-4$) &rob2tp0.5&1.00 $\times$ 0.94&64.6\\
CH$_3$OH($2_k-1_k$) &rob1&1.68 $\times$ 1.62 &17.9\\
CH$_3$OH($5_k-4_k$) &rob2tp0.5&0.99 $\times$ 0.94 &67.8\\
CCH($1-0$) &rob2&1.37 $\times$ 1.16&31.2\\
CCH($3-2$) &rob2tp0.5&0.81 $\times$ 0.73&64.1\\
CN($2-1$)&rob2tp0.5&1.03 $\times$ 1.01&129.1\\
SiO($2-1$)&rob2&1.56 $\times$ 1.32 &16.4\\
H40$\alpha$ &rob2&1.38 $\times$ 1.22 &12.4\\
N$_2$H$^+$($1-0$) &rob2& 1.15 $\times$ 1.06 &35.4 \\
CH$_3$CCH($6_k-5_k$)  &rob2&1.50 $\times$ 1.32&19.7\\
CH$_3$CCH($13_k-12_k$) &rob2tp0.5&1.07 $\times$ 1.01&63.6\\
HC$_3$N($12-11$) &rob1&1.50 $\times$ 1.45&18.6\\
HC$_3$N($10-9$) &rob2&1.68 $\times$ 1.47&19.6\\
\enddata
\end{deluxetable}

\begin{deluxetable}{cccc}
\tablecolumns{4}
\tablewidth{0pc}
\tablecaption{Values of rms in each scheduling block after the beam was convolved to $1.7'' \times 1.7''$ \label{tab:imparam}}
\tablehead{\colhead{ID} &\colhead{$f_{range}$(GHz)} &\colhead{band}  &\colhead{rms (mJy/beam)}}
\startdata
Band3\_a1 &85.49 - 87.63  &LSB &0.181\\
Band3\_a1 &97.24 - 99.86   &USB &0.215\\
Band3\_a2 &87.63 - 89.07  &LSB &0.193\\
Band3\_a2 &99.86- 100.89 &USB &0.217\\
Band3\_b1 &88.91- 90.45 &LSB &0.181 \\
Band3\_b1 &100.91 - 102.49 &USB &0.200 \\
Band3\_b2 &90.45 - 92.49 &LSB &0.484\\
Band3\_b2 &102.49 - 104.49 &USB &0.533\\
Band3\_c  &94.71 - 97.83 &LSB &0.165\\
Band3\_c  &106.71- 109.62 &USB &0.192\\
Band6\_a  &214.06-217.71 &LSB &0.633\\
Band6\_a  &230.06-233.71 &USB &0.692\\
Band6\_b  &217.61-221.26 &LSB &0.898\\
Band6\_b  &233.61-237.26 &USB &0.932\\
Band6\_c  &221.16-224.81 &LSB &0.793\\
Band6\_c  &237.16-240.81 &USB &0.860\\
Band6\_d  &224.71-228.36&LSB & 0.867\\
Band6\_d  &240.71-244.36  &USB & 0.993\\
Band6\_e  &244.79- 248.43&LSB &0.847\\
Band6\_e  &258.98-262.62  &USB &0.931\\
Band6\_f  &248.32-251.96 &LSB &1.002\\
Band6\_f  &262.51-266.15 &USB &1.202\\
Band6\_g  &251.87-255.51 &LSB &0.869\\
Band6\_g  &266.06-269.70 &USB &1.043\\
Band6\_h  &255.44-259.08 &LSB &0.888\\
Band6\_h  &269.63-273.27  &USB &1.047\\
\enddata
\end{deluxetable}

\begin{deluxetable}{ccccc}
\tablecolumns{5}
\tablewidth{0pc}
\tablecaption{Properties of the central molecular zones of NGC 253, NGC 3256, and Arp 220. 
Distances from the earth $D$, star formation rates SFR, molecular masses $M_{\rm mol}$, and depletion times $\tau_{dep}$ 
($\equiv M_{\rm mol}/SFR$; $\tau_{dep}^{-1}$ is an indication of star formation efficiency).
References are a) \citet{2013AJ....146...86T} b) \citet{2003AJ....126.1607S}, c) \citet{2015ApJ...799...10B} d) \citet{2015ApJ...801...25L} 
e) \citet{2008MNRAS.384..316L} f) \citet{2000PASJ...52..785S} g) \citet{2014ApJ...797...90S} \label{tab:gal_prop}}
\tablehead{\colhead{Galaxy} &\colhead{$D$\,(Mpc)}  &\colhead{SFR($M_{\odot}$ yr$^{-1}$)} &\colhead{$M_{\rm mol}$($M_{\odot}$)} &\colhead{$\tau_{\rm dep}$ (yr)}}
\startdata
NGC 253 &3.5$^a$ &2$^d$&$(1.3-3)\times 10^{8d,f}$&$(0.65-1.5)\times 10^8$\\
NGC 3256 &35$^b$&15(N)$^e$& $(1.5-3) \times 10^{8,g}$&$(1-2)\times 10^7$\\
 &&6(S)$^e$&$(1-2)\times 10^{8,g}$&$(1.6-3.3) \times 10^7$\\
Arp220 &80$^b$&180$^c$&$(2-7)\times 10^{9c}$&$(1.1- 3.9)\times 10^7$\\
\enddata
\end{deluxetable}

\begin{deluxetable}{ccc} 
\tablecolumns{3} 
\tablewidth{0pc} 
\tablecaption{Column densities at N\label{tab:colN}} 
\tablehead{\colhead{Molecule} & \colhead{N (cm$^{-2}$)} & \colhead{T$_{\rm ex}$ (K)} } 
\startdata 
$^{13}$CO & $(3.0 \pm 0.1) \times 10^{16.0}$  & $10.5 \pm 0.6$  \\ 
 & $(2.3 \pm 0.1) \times 10^{16.0}$  & $11.5 \pm 0.7$  \\ 
C$^{18}$O & $(1.2 \pm 0.1) \times 10^{16.0}$  & $11.8 \pm 1.4$  \\ 
 & $(9.1 \pm 0.6) \times 10^{15.0}$  & $11.5 \pm 1.3$  \\ 
C$^{17}$O & $(9.8 \pm 1.8) \times 10^{14.0}$  & $10.5 \pm ...$  \\ 
 & $(1.1 \pm 0.1) \times 10^{15.0}$  & $11.3 \pm ...$  \\ 
HCN & $(2.4 \pm 0.1) \times 10^{14.0}$  & $6.0 \pm 0.0$  \\ 
 & $(1.5 \pm 0.1) \times 10^{14.0}$  & $6.1 \pm 0.0$  \\ 
HCO$^+$ & $(1.3 \pm 0.1) \times 10^{14.0}$  & $6.5 \pm 0.1$  \\ 
 & $(1.0 \pm 0.1) \times 10^{14.0}$  & $6.9 \pm 0.1$  \\ 
HNC & $(1.1 \pm 0.1) \times 10^{14.0}$  & $6.3 \pm 0.1$  \\ 
 & $(8.3 \pm 0.5) \times 10^{13.0}$  & $6.0 \pm 0.1$  \\ 
CS & $(1.1 \pm 0.1) \times 10^{14.0}$  & $9.6 \pm 0.2$  \\ 
 & $(1.0 \pm 0.1) \times 10^{14.0}$  & $8.8 \pm 0.2$  \\ 
CN & $(1.4 \pm 0.1) \times 10^{15.0}$  & $5.0 \pm 0.1$  \\ 
 & $(6.4 \pm 0.6) \times 10^{14.0}$  & $5.2 \pm 0.1$  \\ 
CCH & $(1.6 \pm 0.1) \times 10^{15.0}$  & $6.8 \pm 0.1$  \\ 
 & $(1.1 \pm 0.1) \times 10^{15.0}$  & $6.7 \pm 0.1$  \\ 
CH$_3$OH & $(7.6 \pm 0.5) \times 10^{13.0}$  & $15.0 \pm ...$  \\ 
 & $(5.1 \pm 0.5) \times 10^{13.0}$  & $15.0 \pm ...$  \\ 
H$_2$CO & $(1.2 \pm 0.1) \times 10^{13.0}$  & $18.5 \pm 4.3$  \\ 
 & $(1.4 \pm 0.3) \times 10^{13.0}$  & $35.1 \pm 13.2$  \\ 
NNH$^+$ & $(1.3 \pm 0.1) \times 10^{13.0}$  & $10.0 \pm ...$  \\ 
 & $(8.4 \pm 0.5) \times 10^{12.0}$  & $10.0 \pm ...$  \\ 
HCCCN & $(3.2 \pm 0.5) \times 10^{13.0}$  & $13.4 \pm 1.5$  \\ 
 & $(2.5 \pm 0.3) \times 10^{13.0}$  & $22.3 \pm 3.8$  \\ 
CH3CCH & $(2.8 \pm 0.1) \times 10^{14.0}$  & $43.0 \pm 1.7$  \\ 
 & $(2.3 \pm 0.1) \times 10^{14.0}$  & $45.0 \pm 2.4$  \\ 
CH3CN & $(1.0 \pm 0.0) \times 10^{13.0}$  & $10.0 \pm ...$  \\ 
 & $(6.6 \pm 0.3) \times 10^{12.0}$  & $10.0 \pm ...$  \\ 
SiO & $(7.6 \pm 0.9) \times 10^{12.0}$  & $10.0 \pm ...$  \\ 
 & $(4.7 \pm 0.8) \times 10^{12.0}$  & $10.0 \pm ...$  \\ 
H$^{13}$CO$^+$ & $(4.6 \pm 1.8) \times 10^{12.0}$  & $6.0 \pm 0.7$  \\ 
 & $(4.3 \pm 1.8) \times 10^{12.0}$  & $5.1 \pm 0.5$  \\ 
HOC$^+$ & $(1.8 \pm 1.1) \times 10^{12.0}$  & $8.3 \pm 2.1$  \\ 
 & $(1.6 \pm 1.0) \times 10^{12.0}$  & $8.5 \pm 2.4$  \\ 
c-C$_3$H$_2$ & $(8.5 \pm 1.6) \times 10^{12.0}$  & $17.1 \pm 3.1$  \\ 
 & $(1.2 \pm 0.2) \times 10^{13.0}$  & $15.3 \pm 1.6$  \\ 
SO & $(2.7 \pm 1.7) \times 10^{13.0}$  & $10.5 \pm 3.7$  \\ 
 & $(2.5 \pm 1.4) \times 10^{13.0}$  & $10.4 \pm 3.3$  \\ 
C$^{34}$S & $(1.7 \pm 0.7) \times 10^{13.0}$  & $9.4 \pm 1.4$  \\ 
 & $(1.0 \pm 0.5) \times 10^{13.0}$  & $9.8 \pm 2.1$  \\ 
NO & $(1.0 \pm 0.1) \times 10^{15.0}$  & $10.0 \pm ...$  \\ 
 & $(8.8 \pm 0.6) \times 10^{14.0}$  & $10.0 \pm ...$  \\ 
CO$^+$ & $(1.1 \pm 0.3) \times 10^{13.0}$  & $10.0 \pm ...$  \\ 
 & $(1.3 \pm 0.2) \times 10^{13.0}$  & $10.0 \pm ...$  \\ 
HNCO & $(2.5 \pm 0.1) \times 10^{13.0}$  & $30.0 \pm ...$  \\ 
 & $(2.6 \pm 0.1) \times 10^{13.0}$  & $30.0 \pm ...$  \\ 
\enddata 
\end{deluxetable} 
\begin{deluxetable}{ccc} 
\tablecolumns{3} 
\tablewidth{0pc} 
\tablecaption{Column densities at S \label{tab:colS}} 
\tablehead{\colhead{Molecule} & \colhead{N (cm$^{-2}$)} & \colhead{T$_{\rm ex}$ (K)} } 
\startdata 
$^{13}$CO & $(2.5 \pm 0.2) \times 10^{16.0}$  & $10.8 \pm 1.1$  \\ 
 & $(3.4 \pm 0.3) \times 10^{16.0}$  & $8.5 \pm 0.6$  \\ 
C$^{18}$O & $(3.8 \pm 0.7) \times 10^{15.0}$  & $10.6 \pm 2.8$  \\ 
 & $(6.1 \pm 1.3) \times 10^{15.0}$  & $6.7 \pm 0.9$  \\ 
C$^{17}$O & $(8.1 \pm 1.2) \times 10^{14.0}$  & $10.8 \pm ...$  \\ 
 & $(1.0 \pm 0.2) \times 10^{15.0}$  & $8.5 \pm ...$  \\ 
HCN & $(8.8 \pm 0.4) \times 10^{13.0}$  & $5.6 \pm 0.1$  \\ 
 & $(4.9 \pm 0.4) \times 10^{13.0}$  & $5.8 \pm 0.1$  \\ 
HCO$^+$ & $(8.5 \pm 0.3) \times 10^{13.0}$  & $6.7 \pm 0.1$  \\ 
 & $(4.3 \pm 0.3) \times 10^{13.0}$  & $7.3 \pm 0.2$  \\ 
HNC & $(4.3 \pm 0.4) \times 10^{13.0}$  & $6.1 \pm 0.2$  \\ 
 & $(1.3 \pm 0.3) \times 10^{13.0}$  & $6.2 \pm 0.4$  \\ 
CS & $(7.8 \pm 0.9) \times 10^{13.0}$  & $8.2 \pm 0.2$  \\ 
 & $(5.3 \pm 0.8) \times 10^{13.0}$  & $8.6 \pm 0.3$  \\ 
CN & $(3.8 \pm 0.1) \times 10^{14.0}$  & $5.1 \pm 0.1$  \\ 
 & $(1.3 \pm 0.1) \times 10^{14.0}$  & $5.3 \pm 0.1$  \\ 
CCH & $(9.5 \pm 0.4) \times 10^{14.0}$  & $6.9 \pm 0.1$  \\ 
 & $(5.1 \pm 0.4) \times 10^{14.0}$  & $7.0 \pm 0.2$  \\ 
CH$_3$OH & $(4.8 \pm 0.3) \times 10^{13.0}$  & $18.8 \pm 1.4$  \\ 
 & $(1.5 \pm 0.1) \times 10^{14.0}$  & $9.2 \pm 0.4$  \\ 
H$_2$CO & $(1.5 \pm 0.9) \times 10^{13.0}$  & $40.7 \pm 25.6$  \\ 
 & $(6.0 \pm 1.4) \times 10^{12.0}$  & $20.1 \pm 8.9$  \\ 
NNH$^+$ & $<4.0\times 10^{12.0}$  & $...$  \\ 
 & $<5.0 \times 10^{12.0}$  & $...$  \\ 
HCCCN & $(1.7 \pm 0.5) \times 10^{13.0}$  & $13.0 \pm 2.6$  \\ 
 & $(2.2 \pm 0.7) \times 10^{13.0}$  & $10.2 \pm 1.6$  \\ 
CH3CCH & $(2.7 \pm 0.2) \times 10^{14.0}$  & $36.3 \pm 1.5$  \\ 
 & $(3.8 \pm 0.3) \times 10^{14.0}$  & $31.8 \pm 1.4$  \\ 
CH3CN & $<3.6\times 10^{12.0}$  & $...$  \\ 
 & $<4.8 \times 10^{12.0}$  & $...$  \\ 
SiO & $(1.9 \pm 0.7) \times 10^{12.0}$  & $10.0 \pm ...$  \\ 
 & $(4.6 \pm 0.8) \times 10^{12.0}$  & $10.0 \pm ...$  \\ 
H$^{13}$CO$^+$ & $<2.1\times 10^{12.0}$  & $...$  \\ 
 & $<2.1 \times 10^{12.0}$  & $...$  \\ 
HOC$^+$ & $(1.6 \pm 0.3) \times 10^{12.0}$  & $10.0 \pm ...$  \\ 
 & $(1.8 \pm 0.3) \times 10^{12.0}$  & $10.0 \pm ...$  \\ 
c-C$_3$H$_2$ & $<8.3\times 10^{12.0}$  & $...$  \\ 
 & $<1.1 \times 10^{13.0}$  & $...$  \\ 
SO & $(3.3 \pm 0.6) \times 10^{13.0}$  & $10.0 \pm ...$  \\ 
 & $(4.1 \pm 0.5) \times 10^{13.0}$  & $10.0 \pm ...$  \\ 
C$^{34}$S & $(2.8 \pm 1.3) \times 10^{12.0}$  & $8.2 \pm ...$  \\ 
 & $(3.7 \pm 1.5) \times 10^{12.0}$  & $8.6 \pm ...$  \\ 
NO & $(5.5 \pm 0.6) \times 10^{14.0}$  & $10.0 \pm ...$  \\ 
 & $(8.6 \pm 0.8) \times 10^{14.0}$  & $10.0 \pm ...$  \\ 
CO$^+$ & $<1.1\times 10^{13.0}$  & $...$  \\ 
 & $<1.4 \times 10^{13.0}$  & $...$  \\ 
HNCO & $<1.1\times 10^{13.0}$  & $...$  \\ 
 & $<1.4 \times 10^{13.0}$  & $...$  \\ 
\enddata 
\end{deluxetable} 
\begin{deluxetable}{ccc} 
\tablecolumns{3} 
\tablewidth{0pc} 
\tablecaption{Column densities at C \label{tab:colC}} 
\tablehead{\colhead{Molecule} & \colhead{N (cm$^{-2}$)} & \colhead{T$_{\rm ex}$ (K)} } 
\startdata 
$^{13}$CO & $(3.7 \pm 0.3) \times 10^{16.0}$  & $8.1 \pm 0.6$  \\ 
C$^{18}$O & $(1.1 \pm 0.1) \times 10^{16.0}$  & $8.4 \pm 0.7$  \\ 
C$^{17}$O & $(10.0 \pm 3.5) \times 10^{14.0}$  & $8.1 \pm ...$  \\ 
HCN & $(1.8 \pm 0.1) \times 10^{14.0}$  & $5.3 \pm 0.1$  \\ 
HCO$^+$ & $(1.4 \pm 0.1) \times 10^{14.0}$  & $6.1 \pm 0.1$  \\ 
HNC & $(6.6 \pm 0.7) \times 10^{13.0}$  & $4.9 \pm 0.1$  \\ 
CS & $(1.3 \pm 0.1) \times 10^{14.0}$  & $7.5 \pm 0.2$  \\ 
CN & $(7.8 \pm 0.3) \times 10^{14.0}$  & $4.2 \pm 0.0$  \\ 
CCH & $(1.4 \pm 0.1) \times 10^{15.0}$  & $5.3 \pm 0.1$  \\ 
CH$_3$OH & $(4.3 \pm 0.2) \times 10^{14.0}$  & $9.6 \pm 0.3$  \\ 
H$_2$CO & $(9.3 \pm 1.5) \times 10^{12.0}$  & $18.3 \pm 6.3$  \\ 
NNH$^+$ & $(6.5 \pm 0.2) \times 10^{12.0}$  & $10.0 \pm ...$  \\ 
HCCCN & $(4.4 \pm 1.4) \times 10^{13.0}$  & $11.5 \pm 2.1$  \\ 
CH3CCH & $(2.1 \pm 0.3) \times 10^{14.0}$  & $28.8 \pm 1.8$  \\ 
CH3CN & $(7.5 \pm 0.5) \times 10^{12.0}$  & $10.0 \pm ...$  \\ 
SiO & $(8.5 \pm 1.2) \times 10^{12.0}$  & $10.0 \pm ...$  \\ 
H$^{13}$CO$^+$ & $(2.6 \pm 0.6) \times 10^{12.0}$  & $6.1 \pm ...$  \\ 
HOC$^+$ & $(5.6 \pm 0.4) \times 10^{12.0}$  & $10.0 \pm ...$  \\ 
c-C$_3$H$_2$ & $<7.9\times 10^{12.0}$  & $...$  \\ 
SO & $(6.6 \pm 0.7) \times 10^{13.0}$  & $10.0 \pm ...$  \\ 
C$^{34}$S & $(9.5 \pm 2.1) \times 10^{12.0}$  & $7.5 \pm ...$  \\ 
NO & $(1.1 \pm 0.1) \times 10^{15.0}$  & $10.0 \pm ...$  \\ 
CO$^+$ & $<1.3\times 10^{13.0}$  & $...$  \\ 
HNCO & $(6.2 \pm 0.5) \times 10^{13.0}$  & $17.9 \pm 5.7$  \\ 
\enddata 
\end{deluxetable} 
\begin{deluxetable}{ccc} 
\tablecolumns{3} 
\tablewidth{0pc} 
\tablecaption{Column densities at TNE \label{tab:colTNE}} 
\tablehead{\colhead{Molecule} & \colhead{N (cm$^{-2}$)} & \colhead{T$_{\rm ex}$ (K)} } 
\startdata 
$^{13}$CO & $(3.5 \pm 0.4) \times 10^{16.0}$  & $8.3 \pm 0.8$  \\ 
C$^{18}$O & $(5.4 \pm 1.0) \times 10^{15.0}$  & $8.8 \pm 1.6$  \\ 
C$^{17}$O & $<7.9\times 10^{14.0}$  & $...$  \\ 
HCN & $(7.7 \pm 0.4) \times 10^{13.0}$  & $5.0 \pm 0.1$  \\ 
HCO$^+$ & $(7.5 \pm 1.2) \times 10^{13.0}$  & $6.3 \pm 0.3$  \\ 
HNC & $(3.2 \pm 0.7) \times 10^{13.0}$  & $5.2 \pm 0.3$  \\ 
CS & $(5.6 \pm 1.1) \times 10^{13.0}$  & $6.8 \pm 0.4$  \\ 
CN & $(2.1 \pm 0.2) \times 10^{14.0}$  & $4.5 \pm 0.1$  \\ 
CCH & $(5.8 \pm 0.8) \times 10^{14.0}$  & $5.8 \pm 0.2$  \\ 
CH$_3$OH & $(1.4 \pm 0.1) \times 10^{14.0}$  & $9.8 \pm 0.6$  \\
H$_2$CO & $(8.9 \pm 1.5) \times 10^{12.0}$  & $18.3 \pm 6.8$  \\ 
NNH$^+$ & $<4.6\times 10^{12.0}$  & $...$  \\ 
HCCCN & $(1.7 \pm 1.1) \times 10^{13.0}$  & $10.7 \pm 3.9$  \\ 
CH3CCH & $(2.8 \pm 0.4) \times 10^{14.0}$  & $29.3 \pm 1.9$  \\ 
CH3CN & $<4.6\times 10^{12.0}$  & $...$  \\ 
SiO & $<3.3\times 10^{12.0}$  & $...$  \\ 
H$^{13}$CO$^+$ & $(2.5 \pm 0.6) \times 10^{12.0}$  & $6.3 \pm ...$  \\ 
HOC$^+$ & $(1.9 \pm 0.2) \times 10^{12.0}$  & $10.0 \pm ...$  \\ 
c-C$_3$H$_2$ & $(7.1 \pm 3.4) \times 10^{12.0}$  & $10.0 \pm ...$  \\ 
SO & $(4.4 \pm 0.5) \times 10^{13.0}$  & $20.0 \pm ...$  \\ 
C$^{34}$S & $<6.3\times 10^{12.0}$  & $...$  \\ 
NO & $(8.0 \pm 1.1) \times 10^{14.0}$  & $10.0 \pm ...$  \\ 
CO$^+$ & $<8.7\times 10^{12.0}$  & $...$  \\ 
HNCO & $(2.3 \pm 0.7) \times 10^{13.0}$  & $7.3 \pm 2.3$  \\ 
\enddata 
\end{deluxetable} 
\begin{deluxetable}{ccc} 
\tablecolumns{3} 
\tablewidth{0pc} 
\tablecaption{Column densities at TSE\label{tab:colTSE}} 
\tablehead{\colhead{Molecule} & \colhead{N (cm$^{-2}$)} & \colhead{T$_{\rm ex}$ (K)} } 
\startdata 
$^{13}$CO & $(2.9 \pm 0.4) \times 10^{16.0}$  & $6.9 \pm 0.6$  \\ 
C$^{18}$O & $(5.3 \pm 1.5) \times 10^{15.0}$  & $5.4 \pm 0.7$  \\ 
C$^{17}$O & $<1.0\times 10^{15.0}$  & $...$  \\ 
HCN & $(7.2 \pm 0.4) \times 10^{13.0}$  & $5.1 \pm 0.1$  \\ 
HCO$^+$ & $(7.3 \pm 1.0) \times 10^{13.0}$  & $5.7 \pm 0.2$  \\ 
HNC & $(2.5 \pm 1.1) \times 10^{13.0}$  & $4.8 \pm 0.5$  \\ 
CS & $(6.9 \pm 1.6) \times 10^{13.0}$  & $7.0 \pm 0.6$  \\ 
CN & $(3.4 \pm 0.6) \times 10^{14.0}$  & $3.4 \pm 0.1$  \\ 
CCH & $(5.2 \pm 1.3) \times 10^{14.0}$  & $4.4 \pm 0.2$  \\ 
CH$_3$OH & $(4.5 \pm 0.5) \times 10^{14.0}$  & $8.3 \pm 0.5$  \\ 
H$_2$CO & $(1.0 \pm 0.2) \times 10^{13.0}$  & $20.0 \pm ...$  \\ 
NNH$^+$ & $(7.2 \pm 1.0) \times 10^{12.0}$  & $10.0 \pm ...$  \\ 
HCCCN & $<1.6\times 10^{13.0}$  & $...$  \\ 
CH3CCH & $(1.2 \pm 0.4) \times 10^{14.0}$  & $24.8 \pm 5.2$  \\ 
CH3CN & $(6.2 \pm 0.8) \times 10^{12.0}$  & $10.0 \pm ...$  \\ 
SiO & $<5.0\times 10^{12.0}$  & $...$  \\ 
H$^{13}$CO$^+$ & $<2.5\times 10^{12.0}$  & $...$  \\ 
HOC$^+$ & $<2.0\times 10^{12.0}$  & $...$  \\ 
c-C$_3$H$_2$ & $<2.0\times 10^{13.0}$  & $...$  \\ 
SO & $(3.6 \pm 0.7) \times 10^{13.0}$  & $10.0 \pm ...$  \\ 
C$^{34}$S & $<7.9\times 10^{12.0}$  & $...$  \\ 
NO & $<1.0\times 10^{15.0}$  & $...$  \\ 
CO$^+$ & $<1.6\times 10^{13.0}$  & $...$  \\ 
HNCO & $<2.0\times 10^{13.0}$  & $...$  \\ 
\enddata 
\end{deluxetable} 
\begin{deluxetable}{ccc} 
\tablecolumns{3} 
\tablewidth{0pc} 
\tablecaption{Column densities at TSW\label{tab:colTSW}} 
\tablehead{\colhead{Molecule} & \colhead{N (cm$^{-2}$)} & \colhead{T$_{\rm ex}$ (K)} } 
\startdata 
$^{13}$CO & $(3.0 \pm 0.3) \times 10^{16.0}$  & $5.6 \pm 0.3$  \\ 
C$^{18}$O & $(6.9 \pm 1.5) \times 10^{15.0}$  & $4.9 \pm 0.4$  \\ 
C$^{17}$O & $<1.3\times 10^{15.0}$  & $...$  \\ 
HCN & $(1.2 \pm 0.1) \times 10^{14.0}$  & $4.4 \pm 0.1$  \\ 
HCO$^+$ & $(9.9 \pm 1.8) \times 10^{13.0}$  & $4.8 \pm 0.2$  \\ 
HNC & $(3.1 \pm 0.9) \times 10^{13.0}$  & $4.3 \pm 0.3$  \\ 
CS & $(7.3 \pm 1.5) \times 10^{13.0}$  & $6.1 \pm 0.5$  \\ 
CN & $(4.4 \pm 0.7) \times 10^{14.0}$  & $3.4 \pm 0.1$  \\ 
CCH & $(9.3 \pm 2.5) \times 10^{14.0}$  & $3.9 \pm 0.2$  \\ 
CH$_3$OH & $(5.4 \pm 0.4) \times 10^{14.0}$  & $7.6 \pm 0.3$  \\ 
H$_2$CO & $(1.0 \pm 0.3) \times 10^{13.0}$  & $10.0 \pm ...$  \\ 
NNH$^+$ & $(7.8 \pm 0.8) \times 10^{12.0}$  & $10.0 \pm ...$  \\ 
HCCCN & $(1.8 \pm 0.9) \times 10^{13.0}$  & $10.9 \pm 2.9$  \\ 
CH3CCH & $(1.1 \pm 0.2) \times 10^{14.0}$  & $24.9 \pm 2.8$  \\ 
CH3CN & $<4.0\times 10^{13.0}$  & $...$  \\ 
SiO & $(8.3 \pm 1.5) \times 10^{12.0}$  & $10.0 \pm ...$  \\ 
H$^{13}$CO$^+$ & $<1.6\times 10^{12.0}$  & $...$  \\ 
HOC$^+$ & $<1.6\times 10^{12.0}$  & $...$  \\ 
c-C$_3$H$_2$ & $<7.9\times 10^{12.0}$  & $...$  \\ 
SO & $(4.9 \pm 0.5) \times 10^{13.0}$  & $7.7 \pm 3.1$  \\ 
C$^{34}$S & $(9.2 \pm 2.4) \times 10^{12.0}$  & $6.1 \pm ...$  \\ 
NO & $<7.9\times 10^{14.0}$  & $...$  \\ 
CO$^+$ & $<1.3\times 10^{13.0}$  & $...$  \\ 
HNCO & $(6.3 \pm 0.6) \times 10^{13.0}$  & $6.9 \pm 0.6$  \\ 
\enddata 
\end{deluxetable} 
\begin{deluxetable}{ccc} 
\tablecolumns{3} 
\tablewidth{0pc} 
\tablecaption{Column densities at OS\label{tab:colOS}} 
\tablehead{\colhead{Molecule} & \colhead{N (cm$^{-2}$)} & \colhead{T$_{\rm ex}$ (K)} } 
\startdata 
$^{13}$CO & $(1.0 \pm 0.4) \times 10^{16.0}$  & $5.2 \pm 0.9$  \\ 
 & $< 6.4 \times 10^{15.0}$  & $...$  \\ 
C$^{18}$O & $(3.1 \pm 2.2) \times 10^{15.0}$  & $4.2 \pm 0.9$  \\ 
 & $(8.8 \pm 2.2) \times 10^{14.0}$  & $5.0 \pm ...$  \\ 
C$^{17}$O & $<1.1\times 10^{15.0}$  & $...$  \\ 
 & $<1.9 \times 10^{15.0}$  & $...$  \\ 
HCN & $(4.2 \pm 0.3) \times 10^{13.0}$  & $4.9 \pm 0.1$  \\ 
 & $(2.7 \pm 0.3) \times 10^{13.0}$  & $5.6 \pm 0.1$  \\ 
HCO$^+$ & $(4.3 \pm 0.5) \times 10^{13.0}$  & $5.0 \pm 0.1$  \\ 
 & $(2.4 \pm 0.4) \times 10^{13.0}$  & $5.3 \pm 0.2$  \\ 
HNC & $(1.3 \pm 0.1) \times 10^{13.0}$  & $5.0 \pm ...$  \\ 
 & $(7.1 \pm 0.7) \times 10^{12.0}$  & $5.0 \pm ...$  \\ 
CS & $(3.1 \pm 0.8) \times 10^{13.0}$  & $7.2 \pm 0.6$  \\ 
 & $(1.5 \pm 0.7) \times 10^{13.0}$  & $7.2 \pm ...$  \\ 
CN & $(1.1 \pm 0.1) \times 10^{14.0}$  & $3.9 \pm 0.1$  \\ 
 & $(1.2 \pm 0.1) \times 10^{14.0}$  & $3.9 \pm 0.1$  \\ 
CCH & $(2.7 \pm 0.2) \times 10^{14.0}$  & $5.0 \pm ...$  \\ 
 & $(2.1 \pm 1.8) \times 10^{13.0}$  & $5.0 \pm ...$  \\ 
CH$_3$OH & $(1.6 \pm 0.1) \times 10^{14.0}$  & $11.5 \pm 0.6$  \\ 
 & $(6.4 \pm 1.1) \times 10^{13.0}$  & $9.6 \pm 1.2$  \\ 
H$_2$CO & $<5.8\times 10^{12.0}$  & $...$  \\ 
 & $<5.8 \times 10^{12.0}$  & $...$  \\ 
NNH$^+$ & $<4.4\times 10^{12.0}$  & $...$  \\ 
 & $<4.4 \times 10^{12.0}$  & $...$  \\ 
HCCCN & $<1.3\times 10^{13.0}$  & $...$  \\ 
 & $<1.3 \times 10^{13.0}$  & $...$  \\ 
CH3CCH & $<1.2\times 10^{14.0}$  & $...$  \\ 
 & $<1.2 \times 10^{14.0}$  & $...$  \\ 
CH3CN & $<3.3\times 10^{12.0}$  & $...$  \\ 
 & $<3.3 \times 10^{12.0}$  & $...$  \\ 
SiO & $(4.9 \pm 0.7) \times 10^{12.0}$  & $10.0 \pm ...$  \\ 
 & $(2.5 \pm 0.7) \times 10^{12.0}$  & $10.0 \pm ...$  \\ 
H$^{13}$CO$^+$ & $<1.9\times 10^{12.0}$  & $...$  \\ 
 & $<1.9 \times 10^{12.0}$  & $...$  \\ 
HOC$^+$ & $(1.7 \pm 0.3) \times 10^{12.0}$  & $7.0 \pm ...$  \\ 
 & $(7.7 \pm 3.5) \times 10^{11.0}$  & $7.0 \pm ...$  \\ 
c-C$_3$H$_2$ & $<4.4\times 10^{12.0}$  & $...$  \\ 
 & $<4.4 \times 10^{12.0}$  & $...$  \\ 
SO & $(1.9 \pm 0.3) \times 10^{13.0}$  & $10.0 \pm ...$  \\ 
 & $(5.3 \pm 3.3) \times 10^{12.0}$  & $10.0 \pm ...$  \\ 
C$^{34}$S & $<5.8\times 10^{12.0}$  & $...$  \\ 
 & $<5.8 \times 10^{12.0}$  & $...$  \\ 
NO & $<8.3\times 10^{14.0}$  & $...$  \\ 
 & $<8.3 \times 10^{14.0}$  & $...$  \\ 
CO$^+$ & $<1.3\times 10^{13.0}$  & $...$  \\ 
 & $<1.3 \times 10^{13.0}$  & $...$  \\ 
HNCO & $<1.7\times 10^{13.0}$  & $...$  \\ 
 & $<1.7 \times 10^{13.0}$  & $...$  \\ 
\enddata 
\end{deluxetable} 
\begin{deluxetable}{ccc} 
\tablecolumns{3} 
\tablewidth{0pc} 
\tablecaption{Column densities at SW\label{tab:colSW}} 
\tablehead{\colhead{Molecule} & \colhead{N (cm$^{-2}$)} & \colhead{T$_{\rm ex}$ (K)} } 
\startdata 
$^{13}$CO & $(6.7 \pm 0.4) \times 10^{16.0}$  & $7.8 \pm 0.3$  \\ 
C$^{18}$O & $(9.8 \pm 0.7) \times 10^{15.0}$  & $7.6 \pm 0.4$  \\ 
C$^{17}$O & $(2.1 \pm 0.3) \times 10^{15.0}$  & $7.8 \pm ...$  \\ 
HCN & $(1.6 \pm 0.1) \times 10^{14.0}$  & $4.9 \pm 0.1$  \\ 
HCO$^+$ & $(1.4 \pm 0.1) \times 10^{14.0}$  & $6.0 \pm 0.2$  \\ 
HNC & $(6.2 \pm 0.6) \times 10^{13.0}$  & $5.0 \pm 0.1$  \\ 
CS & $(1.3 \pm 0.1) \times 10^{14.0}$  & $7.3 \pm 0.2$  \\ 
CN & $(6.0 \pm 0.3) \times 10^{14.0}$  & $4.1 \pm 0.1$  \\ 
CCH & $(1.3 \pm 0.1) \times 10^{15.0}$  & $5.4 \pm 0.1$  \\ 
CH$_3$OH & $(4.3 \pm 0.1) \times 10^{14.0}$  & $10.1 \pm 0.2$  \\ 
H$_2$CO & $(2.3 \pm 0.5) \times 10^{13.0}$  & $34.6 \pm 8.9$  \\ 
NNH$^+$ & $(1.0 \pm 0.1) \times 10^{13.0}$  & $10.0 \pm ...$  \\ 
HCCCN & $(3.7 \pm 1.0) \times 10^{13.0}$  & $11.1 \pm 1.7$  \\ 
CH3CCH & $(4.1 \pm 0.3) \times 10^{14.0}$  & $25.1 \pm 0.9$  \\ 
CH3CN & $<5.0\times 10^{12.0}$  & $...$  \\ 
SiO & $(8.7 \pm 1.2) \times 10^{12.0}$  & $10.0 \pm ...$  \\ 
H$^{13}$CO$^+$ & $(4.6 \pm 0.5) \times 10^{12.0}$  & $6.0 \pm ...$  \\ 
HOC$^+$ & $(4.9 \pm 0.5) \times 10^{12.0}$  & $10.0 \pm ...$  \\ 
c-C$_3$H$_2$ & $(9.2 \pm 2.5) \times 10^{12.0}$  & $13.5 \pm 6.3$  \\ 
SO & $(8.5 \pm 0.8) \times 10^{13.0}$  & $10.0 \pm ...$  \\ 
C$^{34}$S & $(9.6 \pm 2.6) \times 10^{12.0}$  & $7.3 \pm ...$  \\ 
NO & $(9.4 \pm 0.7) \times 10^{14.0}$  & $10.0 \pm ...$  \\ 
CO$^+$ & $<1.3\times 10^{13.0}$  & $...$  \\ 
HNCO & $(4.6 \pm 0.4) \times 10^{13.0}$  & $9.4 \pm 1.5$  \\ 
\enddata 
\end{deluxetable} 
\clearpage

\clearpage

\startlongtable
\begin{deluxetable}{cccccccc} 
\tablecolumns{5} 
\tablewidth{0pc} 
\tablecaption{Observed peak intensities and velocity-integrated intensities for detected lines. \label{tab:intens}} 
\tablehead{\colhead{Position} &\colhead{Species} & \colhead{Transitions} & \colhead{Restfreq} & \colhead{I$_{\rm peak}$} &\colhead{$\Delta$I$_{\rm peak}$} & \colhead{I $\Delta$V} & \colhead{$\Delta($I $\Delta$V)}\\
\colhead{} &\colhead{} &\colhead{} & \colhead{(GHz)} & \colhead{(mK)}& \colhead{(mK)} & \colhead{(K km/s)} & \colhead{(K km/s)} } 
\startdata 
N &H$^{13}$CO$^+$ & $1-0$ & 86.754 & 56.3 & 10.41 & 9.4 & 0.95 \\ 
N &SiO & $2-1$ & 86.847 & 40.85 & 10.39 & 6.41 & 0.95 \\ 
N &CCH & $1_{3/2}-0_{1/2}$ & 87.317 & 373.18 & 10.28 & 76.97 & 0.94 \\ 
N &CCH & $1_{1/2}-0_{1/2}$ & 87.402 & 201.53 & 10.26 & 41.4 & 0.94 \\ 
N &HCN & $1-0$ & 88.634 & 1111.22 & 10.64 & 197.23 & 0.96 \\ 
N &HCO$^+$ & $1-0$ & 89.189 & 1173.21 & 10.51 & 205.25 & 0.95 \\ 
N &HOC$^+$ & $1-0$ & 89.487 & 26.83 & 10.44 & 3.88 & 0.94 \\ 
N &HNC & $1-0$ & 90.664 & 629.27 & 9.54 & 110.23 & 0.85 \\ 
N &HC$_3$N & $10-9$ & 90.979 & 80.17 & 9.47 & 11.47 & 0.85 \\ 
N &N$_2$H$^+$ & $1-0$ & 93.176 & 100.99 & 24.15 & 14.96 & 2.13 \\ 
N &C$^{34}$S & $1-0$ & 96.413 & 38.25 & 7.69 & 6.59 & 0.67 \\ 
N &CH$_3$OH & $2_k-1_k$ & 96.741 & 26.23 & 7.64 & 3.09 & 0.66 \\ 
N &CS & $2-1$ & 97.981 & 279.51 & 7.45 & 48.99 & 0.64 \\ 
N &H40$\alpha$    & ---     &99.023  &35.47 &9.50     &5.68&0.81\\ 
N &SO & $2_1-3_2$ & 99.3 & 38.95 & 9.45 & 4.99 & 0.81 \\ 
N &HC$_3$N & $11-10$ & 100.076 & 70.48 & 9.31 & 10.51 & 0.79 \\ 
N &CH$_3$CCH & $6_k-5_k$ & 102.548 & 54.49 & 8.25 & 9.4 & 0.69 \\ 
N &HC$_3$N & $12-11$ & 109.174 & 74.57 & 6.99 & 9.63 & 0.57 \\ 
N &C$^{18}$O & $1-0$ & 109.782 & 135.08 & 6.91 & 21.37 & 0.56 \\ 
N &HNCO & $5_{0,5}-4_{0,4}$ & 109.906 & 22.57 & 6.9 & 3.5 & 0.56 \\ 
N &$^{13}$CO & $1-0$ & 110.201 & 349.93 & 6.86 & 51.13 & 0.56 \\ 
N &CH$_3$CN & $6_k-5_k$ & 110.383 & 23.73 & 6.84 & 4.75 & 0.55 \\ 
N &CN & $1_{3/2}-0_{1/2}$ & 113.191 & 485.15 & 23.72 & 101.7 & 2.12 \\ 
N &CN & $1_{1/2}-0_{1/2}$ & 113.491 & 954.47 & 23.59 & 178.52 & 2.11 \\ 
N &c-C$_3$H$_2$ & $3_{3,0}-2_{2,1}$ & 216.279 & 18.89 & 5.97 & 2.08 & 0.55 \\ 
N &c-C$_3$H$_2$ & $6_k-5_k$ & 217.822 & 28.73 & 5.89 & 4.8 & 0.54 \\ 
N &c-C$_3$H$_2$ & $5_{1,4}-4_{2,3}$ & 217.94 & 28.7 & 5.88 & 4.04 & 0.54 \\ 
N &c-C$_3$H$_2$ & $5_{2,4}-4_{1,3}$ & 218.16 & 22.92 & 5.87 & 4.25 & 0.54 \\ 
N &H$_2$CO & $3_{0,3}-2_{0,2}$ & 218.222 & 22.91 & 5.87 & 4.49 & 0.53 \\ 
N &H$_2$CO & $3_{2,2}-2_{2,1}$ & 218.476 & 12.87 & 5.85 & 1.83 & 0.53 \\ 
N &C$^{18}$O & $2-1$ & 219.56 & 232.37 & 5.8 & 35.09 & 0.53 \\ 
N &HNCO & $10_{0,10}-9_{0,9}$ & 219.798 & 28.75 & 8.2 & 3.69 & 0.75 \\ 
N &SO & $6_5-5_4$ & 219.949 & 28.71 & 8.19 & 2.42 & 0.74 \\ 
N &$^{13}$CO & $2-1$ & 220.399 & 606.9 & 8.16 & 85.29 & 0.74 \\ 
N &CH$_3$CCH & $13_k-12_k$ & 222.167 & 54.25 & 8.03 & 7.61 & 0.73 \\ 
N &C$^{17}$O                &$2-1$                    &224.714     &20.81& 6.94    & 2.71&0.62\\ 
N &H$_2$CO & $3_{1,2}-2_{1,1}$ & 225.698 & 42.28 & 6.88 & 4.84 & 0.62 \\ 
N &CN & $2_{3/2}-1_{3/2}$ & 226.36 & 139.81 & 6.84 & 26.33 & 0.58 \\ 
N &CN & $2_{3/2}-1_{1/2}$ & 226.66 & 331.56 & 6.82 & 50.92 & 0.46 \\ 
N &CN & $2_{5/2}-1_{3/2}$ & 226.875 & 518.34 & 6.81 & 91.61 & 0.58 \\ 
N &c-C$_3$H$_2$ & $4_{3,2}-3_{2,1}$ & 227.169 & 16.2 & 7.43 & 1.97 & 0.66 \\ 
N &CO$^+$ & $2_{3/2}-1_{1/2}$ & 235.79 & 12.47 & 5.51 & 1.26 & 0.48 \\ 
N &CO$^+$ & $2_{5/2}-1_{3/2}$ & 236.063 & 22.48 & 7.41 & 1.61 & 0.65 \\ 
N &CH$_3$CCH & $14_k-13_k$ & 239.252 & 47.7 & 7.22 & 8.8 & 0.63 \\ 
N &C$^{34}$S & $5-4$ & 241.016 & 24.5 & 6.56 & 2.12 & 0.57 \\ 
N &CH$_3$OH & $5_k-4_k$ & 241.791 & 37.99 & 6.52 & 5.84 & 0.57 \\ 
N &CS & $5-4$ & 244.936 & 108.3 & 7.35 & 15.45 & 0.63 \\ 
N &NO & $5/2+-3/2-$ & 250.483 & 26.75 & 6.0 & 3.18 & 0.51 \\ 
N &NO & $5/2--3/2+$ & 250.817 & 14.28 & 7.08 & 1.54 & 0.6 \\ 
N &CH$_3$CCH & $15_k-14_k$ & 256.337 & 52.6 & 5.89 & 7.66 & 0.5 \\ 
N &H$^{13}$CO$^+$ & $3-2$ & 260.255 & 19.05 & 5.84 & 2.6 & 0.49 \\ 
N &CH$_3$OH & $2_{1,1}-1_{0,1}$ & 261.806 & 208.08 & 6.05 & 13.46 & 0.5 \\ 
N &CCH & $3-2$ & 262.004 & 286.7 & 6.04 & 53.6 & 0.5 \\ 
N &HCN & $3-2$ & 265.886 & 402.69 & 7.58 & 71.38 & 0.63 \\ 
N &HCO$^+$ & $3-2$ & 267.558 & 531.35 & 7.49 & 91.56 & 0.62 \\ 
N &HOC$^+$ & $3-2$ & 268.451 & 22.04 & 7.44 & 2.8 & 0.61 \\ 
N &HNC & $3-2$ & 271.981 & 241.95 & 6.3 & 37.97 & 0.51 \\ 
N &CH$_3$CCH & $16_k-15_k$ & 273.42 & 32.6 & 6.26 & 5.6 & 0.51 \\ 
S &SiO & $2-1$ & 86.847 & 29.12 & 10.39 & 2.75 & 0.95 \\ 
S &CCH & $1_{3/2}-0_{1/2}$ & 87.317 & 243.63 & 10.28 & 37.41 & 0.94 \\ 
S &CCH & $1_{1/2}-0_{1/2}$ & 87.402 & 137.71 & 10.26 & 20.28 & 0.94 \\ 
S &HCN & $1-0$ & 88.634 & 510.57 & 10.64 & 56.68 & 0.96 \\ 
S &HCO$^+$ & $1-0$ & 89.189 & 773.12 & 10.51 & 87.87 & 0.95 \\ 
S &HOC$^+$ & $1-0$ & 89.487 & 27.53 & 10.44 & 3.02 & 0.94 \\ 
S &HNC & $1-0$ & 90.664 & 275.69 & 9.54 & 28.08 & 0.85 \\ 
S &HC$_3$N & $10-9$ & 90.979 & 40.77 & 9.48 & 5.06 & 0.85 \\ 
S &C$^{34}$S & $1-0$ & 96.413 & 17.5 & 7.7 & 1.07 & 0.67 \\ 
S &CH$_3$OH & $2_k-1_k$ & 96.741 & 48.45 & 7.64 & 6.22 & 0.66 \\ 
S &CS & $2-1$ & 97.981 & 150.92 & 7.45 & 22.92 & 0.64 \\ 
S &H40$\alpha$              &---                    &99.023     &34.76&9.51     &5.50&0.81\\ 
S &SO & $2_1-3_2$ & 99.3 & 57.63 & 9.46 & 6.08 & 0.81 \\ 
S &HC$_3$N & $11-10$ & 100.076 & 34.12 & 9.31 & 3.49 & 0.79 \\ 
S &CH$_3$CCH & $6_k-5_k$ & 102.548 & 78.86 & 8.25 & 10.22 & 0.69 \\ 
S &HC$_3$N & $12-11$ & 109.174 & 26.9 & 6.99 & 2.82 & 0.57 \\ 
S &C$^{18}$O & $1-0$ & 109.782 & 73.14 & 6.92 & 9.09 & 0.56 \\ 
S &$^{13}$CO & $1-0$ & 110.201 & 419.11 & 6.86 & 53.22 & 0.56 \\ 
S &CN & $1_{3/2}-0_{1/2}$ & 113.191 & 133.03 & 23.73 & 24.62 & 2.12 \\ 
S &CN & $1_{1/2}-0_{1/2}$ & 113.491 & 375.3 & 23.6 & 48.44 & 2.11 \\ 
S &H$_2$CO & $3_{0,3}-2_{0,2}$ & 218.222 & 21.13 & 5.88 & 1.64 & 0.54 \\ 
S &C$^{18}$O & $2-1$ & 219.56 & 87.1 & 5.81 & 10.09 & 0.53 \\ 
S &$^{13}$CO & $2-1$ & 220.399 & 549.1 & 8.18 & 66.45 & 0.74 \\ 
S &CH$_3$CCH & $13_k-12_k$ & 222.167 & 55.88 & 8.05 & 7.78 & 0.73 \\ 
S &C$^{17}$O                &$2-1$                    &224.714     &20.62&6.95     &2.38&0.62\\ 
S &H$_2$CO & $3_{1,2}-2_{1,1}$ & 225.698 & 33.19 & 6.89 & 3.31 & 0.62 \\ 
S &CN & $2_{3/2}-1_{3/2}$ & 226.36 & 29.88 & 6.85 & 4.86 & 0.61 \\ 
S &CN & $2_{3/2}-1_{1/2}$ & 226.66 & 106.69 & 6.84 & 14.88 & 0.61 \\ 
S &CN & $2_{5/2}-1_{3/2}$ & 226.875 & 198.24 & 6.82 & 25.84 & 0.61 \\ 
S &CH$_3$CCH & $14_k-13_k$ & 239.252 & 55.34 & 7.23 & 6.75 & 0.63 \\ 
S &CH$_3$OH & $5_k-4_k$ & 241.791 & 26.22 & 6.54 & 4.35 & 0.57 \\ 
S &CS & $5-4$ & 244.936 & 41.85 & 7.36 & 4.62 & 0.63 \\ 
S &NO & $5/2+-3/2-$ & 250.483 & 17.18 & 6.01 & 2.06 & 0.51 \\ 
S &NO & $5/2--3/2+$ & 250.817 & 14.49 & 7.1 & 2.01 & 0.6 \\ 
S &CH$_3$CCH & $15_k-14_k$ & 256.337 & 39.52 & 5.9 & 5.59 & 0.5 \\ 
S &CCH & $3-2$ & 262.004 & 165.19 & 6.06 & 28.91 & 0.5 \\ 
S &HCN & $3-2$ & 265.886 & 139.89 & 7.6 & 17.72 & 0.63 \\ 
S &HCO$^+$ & $3-2$ & 267.558 & 371.92 & 7.51 & 43.69 & 0.62 \\ 
S &HNC & $3-2$ & 271.981 & 92.91 & 6.32 & 9.16 & 0.52 \\ 
S &CH$_3$CCH & $16_k-15_k$ & 273.42 & 31.35 & 6.28 & 4.38 & 0.51 \\ 
C &H$^{13}$CO$^+$ & $1-0$ & 86.754 & 28.84 & 10.38 & 3.44 & 0.95 \\ 
C &SiO & $2-1$ & 86.847 & 57.26 & 10.35 & 5.26 & 0.95 \\ 
C &CCH & $1_{3/2}-0_{1/2}$ & 87.317 & 281.38 & 10.24 & 38.37 & 0.93 \\ 
C &CCH & $1_{1/2}-0_{1/2}$ & 87.402 & 144.99 & 10.22 & 18.36 & 0.93 \\ 
C &HNCO & $4_{0,4}-3_{0,3}$ & 87.926 & 55.98 & 10.1 & 3.96 & 0.92 \\ 
C &HCN & $1-0$ & 88.634 & 797.6 & 10.6 & 93.15 & 0.96 \\ 
C &HCO$^+$ & $1-0$ & 89.189 & 1112.08 & 10.47 & 139.46 & 0.94 \\ 
C &HOC$^+$ & $1-0$ & 89.487 & 50.54 & 10.4 & 5.6 & 0.94 \\ 
C &HNC & $1-0$ & 90.664 & 384.4 & 9.5 & 38.95 & 0.85 \\ 
C &HC$_3$N & $10-9$ & 90.979 & 73.71 & 9.44 & 8.03 & 0.84 \\ 
C &N$_2$H$^+$ & $1-0$ & 93.176 & 101.29 & 24.06 & 6.46 & 2.12 \\ 
C &C$^{34}$S & $1-0$ & 96.413 & 30.82 & 7.66 & 2.04 & 0.66 \\ 
C &CH$_3$OH & $2_k-1_k$ & 96.741 & 202.4 & 7.61 & 15.26 & 0.66 \\ 
C &CS & $2-1$ & 97.981 & 316.68 & 7.42 & 29.89 & 0.64 \\ 
C &H40$\alpha$             &---                    &99.023    & 21.03&9.46     &0.88&0.81\\ 
C &SO & $2_1-3_2$ & 99.3 & 51.59 & 9.41 & 5.6 & 0.8 \\ 
C &HC$_3$N & $11-10$ & 100.076 & 72.21 & 9.26 & 7.15 & 0.79 \\ 
C &CH$_3$CCH & $6_k-5_k$ & 102.548 & 46.12 & 8.21 & 4.92 & 0.69 \\ 
C &CH$_3$OH & $0_{0,0}-1_{1,1}$ & 108.894 & 42.87 & 6.99 & 2.55 & 0.57 \\ 
C &HC$_3$N & $12-11$ & 109.174 & 47.64 & 6.95 & 4.42 & 0.57 \\ 
C &C$^{18}$O & $1-0$ & 109.782 & 158.38 & 6.87 & 12.53 & 0.56 \\ 
C &HNCO & $5_{0,5}-4_{0,4}$ & 109.906 & 67.01 & 6.86 & 6.18 & 0.56 \\ 
C &$^{13}$CO & $1-0$ & 110.201 & 541.15 & 6.82 & 41.32 & 0.55 \\ 
C &CH$_3$CN & $6_k-5_k$ & 110.383 & 21.96 & 6.8 & 2.69 & 0.55 \\ 
C &CN & $1_{3/2}-0_{1/2}$ & 113.191 & 196.54 & 23.58 & 33.0 & 2.11 \\ 
C &CN & $1_{1/2}-0_{1/2}$ & 113.491 & 441.29 & 23.45 & 62.01 & 2.1 \\ 
C &H$_2$CO & $3_{0,3}-2_{0,2}$ & 218.222 & 26.6 & 5.74 & 2.39 & 0.52 \\ 
C &H$_2$CO & $3_{2,2}-2_{2,1}$ & 218.476 & 16.17 & 5.72 & 1.67 & 0.52 \\ 
C &C$^{18}$O & $2-1$ & 219.56 & 197.0 & 5.67 & 14.75 & 0.52 \\ 
C &$^{13}$CO & $2-1$ & 220.399 & 576.38 & 7.98 & 47.41 & 0.72 \\ 
C &CH$_3$CCH & $13_k-12_k$ & 222.167 & 23.7 & 7.85 & 2.59 & 0.71 \\ 
C &C$^{17}$O                &$2-1$                   & 224.714     &21.54&6.78     &1.09&0.61\\ 
C &H$_2$CO & $3_{1,2}-2_{1,1}$ & 225.698 & 39.06 & 6.72 & 1.28 & 0.6 \\ 
C &CN & $2_{3/2}-1_{3/2}$ & 226.36 & 28.52 & 6.68 & 4.11 & 0.6 \\ 
C &CN & $2_{3/2}-1_{1/2}$ & 226.66 & 98.48 & 6.66 & 14.13 & 0.6 \\ 
C &CN & $2_{5/2}-1_{3/2}$ & 226.875 & 187.62 & 6.65 & 21.08 & 0.59 \\ 
C &CH$_3$CCH & $14_k-13_k$ & 239.252 & 22.52 & 7.03 & 0.23 & 0.61 \\ 
C &CH$_3$OH & $5_k-4_k$ & 241.791 & 63.94 & 6.35 & 5.95 & 0.55 \\ 
C &CS & $5-4$ & 244.936 & 56.25 & 7.14 & 5.37 & 0.62 \\ 
C &NO & $5/2+-3/2-$ & 250.483 & 25.51 & 5.83 & 2.1 & 0.5 \\ 
C &NO & $5/2--3/2+$ & 250.817 & 18.56 & 6.87 & 0.37 & 0.58 \\ 
C &CH$_3$CCH & $15_k-14_k$ & 256.337 & 17.95 & 5.71 & 1.29 & 0.48 \\ 
C &CH$_3$OH & $2_{1,1}-1_{0,1}$ & 261.806 & 35.11 & 5.86 & 3.05 & 0.49 \\ 
C &CCH & $3-2$ & 262.004 & 86.62 & 5.85 & 11.52 & 0.49 \\ 
C &HCN & $3-2$ & 265.886 & 212.11 & 7.34 & 19.24 & 0.61 \\ 
C &HCO$^+$ & $3-2$ & 267.558 & 439.54 & 7.25 & 44.61 & 0.6 \\ 
C &HNC & $3-2$ & 271.981 & 60.43 & 6.09 & 4.71 & 0.5 \\ 
C &CH$_3$CCH & $16_k-15_k$ & 273.42 & 13.82 & 6.05 & 0.76 & 0.49 \\ 
TNE &H$^{13}$CO$^+$ & $1-0$ & 86.754 & 28.32 & 10.67 & 3.21 & 0.98 \\ 
TNE &CCH & $1_{3/2}-0_{1/2}$ & 87.317 & 162.65 & 10.54 & 17.15 & 0.96 \\ 
TNE &CCH & $1_{1/2}-0_{1/2}$ & 87.402 & 87.83 & 10.52 & 9.6 & 0.96 \\ 
TNE &HNCO & $4_{0,4}-3_{0,3}$ & 87.926 & 24.61 & 10.39 & 1.06 & 0.94 \\ 
TNE &HCN & $1-0$ & 88.634 & 459.05 & 10.91 & 37.29 & 0.99 \\ 
TNE &HCO$^+$ & $1-0$ & 89.189 & 844.66 & 10.78 & 71.87 & 0.97 \\ 
TNE &HNC & $1-0$ & 90.664 & 219.31 & 9.79 & 16.38 & 0.88 \\ 
TNE &HC$_3$N & $10-9$ & 90.979 & 25.18 & 9.73 & 1.3 & 0.87 \\ 
TNE &CH$_3$OH          &$2_k-1_k$                   & 96.741     &68.40&7.87    & 4.33& 0.68\\ 
TNE &CS & $2-1$ & 97.981 & 150.52 & 7.68 & 12.45 & 0.66 \\ 
TNE &H40$\alpha$            &---                   & 99.023     &25.57&9.81    & 2.77&0.84\\ 
TNE &SO & $2_1-3_2$ & 99.3 & 39.38 & 9.76 & 2.49 & 0.83 \\ 
TNE &HC$_3$N & $11-10$ & 100.076 & 41.96 & 9.61 & 2.93 & 0.82 \\ 
TNE &CH$_3$CCH & $6_k-5_k$ & 102.548 & 74.84 & 8.53 & 4.68 & 0.72 \\ 
TNE &CH$_3$OH                & $0_{0,0}-1_{1,1}$        &108.894     &26.31&7.30    & 2.07&0.60\\ 
TNE &HC$_3$N & $12-11$ & 109.174 & 16.43 & 7.26 & 0.91 & 0.59 \\ 
TNE &SO & $2_3-1_2$ & 109.252 & 17.71 & 7.25 & 0.87 & 0.59 \\ 
TNE &C$^{18}$O & $1-0$ & 109.782 & 78.58 & 7.19 & 5.69 & 0.58 \\ 
TNE &HNCO & $5_{0,5}-4_{0,4}$ & 109.906 & 23.25 & 7.17 & 1.61 & 0.58 \\ 
TNE &$^{13}$CO & $1-0$ & 110.201 & 449.06 & 7.13 & 35.71 & 0.58 \\ 
TNE &CN & $1_{3/2}-0_{1/2}$ & 113.191 & 96.05 & 24.71 & 9.16 & 2.21 \\ 
TNE &CN & $1_{1/2}-0_{1/2}$ & 113.491 & 205.02 & 24.59 & 15.34 & 2.2 \\ 
TNE &c-C$_3$H$_2$ & $6_k-5_k$ & 217.822 & 17.86 & 6.86 & 0.78 & 0.63 \\ 
TNE &c-C$_3$H$_2$ & $5_{2,4}-4_{1,3}$ & 218.16 & 18.62 & 6.84 & 1.71 & 0.62 \\ 
TNE &H$_2$CO & $3_{0,3}-2_{0,2}$ & 218.222 & 18.61 & 6.84 & 1.55 & 0.62 \\ 
TNE &C$^{18}$O & $2-1$ & 219.56 & 108.09 & 6.77 & 6.92 & 0.62 \\ 
TNE &$^{13}$CO & $2-1$ & 220.399 & 550.16 & 9.54 & 42.02 & 0.87 \\ 
TNE &CH$_3$CCH & $13_k-12_k$ & 222.167 & 42.72 & 9.42 & 2.38 & 0.85 \\ 
TNE &H$_2$CO & $3_{1,2}-2_{1,1}$ & 225.698 & 29.55 & 8.1 & 2.4 & 0.73 \\ 
TNE &CN & $2_{3/2}-1_{3/2}$ & 226.36 & 23.01 & 8.07 & 0.21 & 0.72 \\ 
TNE &CN & $2_{3/2}-1_{1/2}$ & 226.66 & 34.19 & 8.05 & 1.24 & 0.72 \\ 
TNE &CN & $2_{5/2}-1_{3/2}$ & 226.875 & 123.5 & 8.04 & 6.89 & 0.72 \\ 
TNE &CH$_3$CCH & $14_k-13_k$ & 239.252 & 37.16 & 8.68 & 2.3 & 0.76 \\ 
TNE &CH$_3$OH & $5_k-4_k$ & 241.791 & 35.34 & 7.87 & 2.11 & 0.68 \\ 
TNE &CS & $5-4$ & 244.936 & 28.97 & 8.91 & 2.05 & 0.77 \\ 
TNE &NO & $5/2+-3/2-$ & 250.483 & 25.95 & 7.34 & -0.01 & 0.62 \\ 
TNE &NO & $5/2--3/2+$ & 250.817 & 20.85 & 8.67 & 0.37 & 0.74 \\ 
TNE &CH$_3$OH & $2_{0,2}-1_{-1,1}$ & 254.015 & 15.59 & 8.5 & 0.46 & 0.72 \\ 
TNE &CH$_3$CCH & $15_k-14_k$ & 256.337 & 33.82 & 7.27 & 2.3 & 0.61 \\ 
TNE &H$^{13}$CO$^+$ & $3-2$ & 260.255 & 21.65 & 7.26 & 1.03 & 0.61 \\ 
TNE &CH$_3$OH & $2_{1,1}-1_{0,1}$ & 261.806 & 20.93 & 7.55 & 1.34 & 0.63 \\ 
TNE &CCH & $3-2$ & 262.004 & 68.71 & 7.54 & 6.15 & 0.63 \\ 
TNE &HCN & $3-2$ & 265.886 & 90.5 & 9.52 & 5.88 & 0.79 \\ 
TNE &HCO$^+$ & $3-2$ & 267.558 & 360.03 & 9.43 & 23.59 & 0.78 \\ 
TNE &HNC & $3-2$ & 271.981 & 44.21 & 7.99 & 3.07 & 0.65 \\ 
TNE &CH$_3$CCH & $16_k-15_k$ & 273.42 & 14.3 & 7.96 & 0.97 & 0.65 \\ 
TSE &CCH & $1_{3/2}-0_{1/2}$ & 87.317 & 95.59 & 10.42 & 10.02 & 0.95 \\ 
TSE &CCH & $1_{1/2}-0_{1/2}$ & 87.402 & 44.34 & 10.4 & 6.83 & 0.95 \\ 
TSE &HCN & $1-0$ & 88.634 & 330.75 & 10.78 & 36.63 & 0.98 \\ 
TSE &HCO$^+$ & $1-0$ & 89.189 & 560.87 & 10.65 & 70.52 & 0.96 \\ 
TSE &HNC & $1-0$ & 90.664 & 123.88 & 9.67 & 13.8 & 0.87 \\ 
TSE &N$_2$H$^+$ & $1-0$ & 93.176 & 65.16 & 24.52 & 4.32 & 2.16 \\ 
TSE &CH$_3$OH & $2_k-1_k$ & 96.741 & 82.94 & 7.76 & 7.93 & 0.67 \\ 
TSE &CS & $2-1$ & 97.981 & 143.78 & 7.57 & 15.96 & 0.65 \\ 
TSE &SO & $2_1-3_2$ & 99.3 & 44.24 & 9.61 & 3.24 & 0.82 \\ 
TSE &CH$_3$CCH & $6_k-5_k$ & 102.548 & 27.33 & 8.4 & 0.45 & 0.71 \\ 
TSE &C$^{18}$O & $1-0$ & 109.782 & 45.97 & 7.06 & 5.73 & 0.57 \\ 
TSE &$^{13}$CO & $1-0$ & 110.201 & 292.94 & 7.01 & 31.98 & 0.57 \\ 
TSE &CH$_3$CN & $6_k-5_k$ & 110.383 & 22.06 & 6.98 & 2.05 & 0.57 \\ 
TSE &CN & $1_{3/2}-0_{1/2}$ & 113.191 & 71.36 & 24.25 & 10.04 & 2.17 \\ 
TSE &CN & $1_{1/2}-0_{1/2}$ & 113.491 & 136.68 & 24.12 & 14.94 & 2.16 \\ 
TSE &H$_2$CO & $3_{0,3}-2_{0,2}$ & 218.222 & 12.03 & 6.37 & 0.96 & 0.58 \\ 
TSE &C$^{18}$O & $2-1$ & 219.56 & 30.01 & 6.3 & 3.21 & 0.57 \\ 
TSE &$^{13}$CO & $2-1$ & 220.399 & 297.26 & 8.87 & 29.8 & 0.81 \\ 
TSE &CH$_3$CCH & $13_k-12_k$ & 222.167 & 20.33 & 8.75 & 0.21 & 0.79 \\ 
TSE &H$_2$CO & $3_{1,2}-2_{1,1}$ & 225.698 & 26.95 & 7.51 & 1.3 & 0.67 \\ 
TSE &CN & $2_{3/2}-1_{1/2}$ & 226.66 & 27.88 & 7.45 & 0.85 & 0.67 \\ 
TSE &CN & $2_{5/2}-1_{3/2}$ & 226.875 & 48.68 & 7.44 & 3.32 & 0.67 \\ 
TSE &CH$_3$OH & $5_k-4_k$ & 241.791 & 35.04 & 7.22 & 4.07 & 0.63 \\ 
TSE &CS & $5-4$ & 244.936 & 30.26 & 8.15 & 0.68 & 0.7 \\ 
TSE &CH$_3$OH & $2_{1,1}-1_{0,1}$ & 261.806 & 29.02 & 6.81 & 2.49 & 0.57 \\ 
TSE &CCH & $3-2$ & 262.004 & 24.84 & 6.8 & 0.23 & 0.57 \\ 
TSE &HCN & $3-2$ & 265.886 & 69.33 & 8.57 & 6.13 & 0.71 \\ 
TSE &HCO$^+$ & $3-2$ & 267.558 & 178.16 & 8.48 & 17.21 & 0.7 \\ 
TSE &HNC & $3-2$ & 271.981 & 20.38 & 7.16 & 2.22 & 0.58 \\ 
TSW &SiO & $2-1$ & 86.847 & 31.41 & 10.6 & 3.95 & 0.97 \\ 
TSW &CCH & $1_{3/2}-0_{1/2}$ & 87.317 & 148.27 & 10.49 & 17.04 & 0.96 \\ 
TSW &CCH & $1_{1/2}-0_{1/2}$ & 87.402 & 95.46 & 10.47 & 8.64 & 0.95 \\ 
TSW &HNCO & $4_{0,4}-3_{0,3}$ & 87.926 & 69.09 & 10.35 & 5.53 & 0.94 \\ 
TSW &HCN & $1-0$ & 88.634 & 496.89 & 10.86 & 51.9 & 0.98 \\ 
TSW &HCO$^+$ & $1-0$ & 89.189 & 688.99 & 10.73 & 84.51 & 0.97 \\ 
TSW &HNC & $1-0$ & 90.664 & 202.15 & 9.74 & 17.53 & 0.87 \\ 
TSW &HC$_3$N & $10-9$ & 90.979 & 27.63 & 9.68 & 1.77 & 0.86 \\ 
TSW &N$_2$H$^+$ & $1-0$ & 93.176 & 83.15 & 24.71 & 6.26 & 2.18 \\ 
TSW &C$^{34}$S & $1-0$ & 96.413 & 29.57 & 7.88 & 2.09 & 0.68 \\ 
TSW &CH$_3$OH & $2_k-1_k$ & 96.741 & 222.31 & 7.83 & 15.4 & 0.68 \\ 
TSW &CS & $2-1$ & 97.981 & 187.28 & 7.64 & 15.6 & 0.66 \\ 
TSW &SO & $2_1-3_2$ & 99.3 & 46.56 & 9.7 & 4.29 & 0.83 \\ 
TSW &HC$_3$N & $11-10$ & 100.076 & 28.13 & 9.55 & 2.46 & 0.81 \\ 
TSW &CH$_3$CCH & $6_k-5_k$ & 102.548 & 28.19 & 8.48 & 2.63 & 0.71 \\ 
TSW &HC$_3$N & $12-11$ & 109.174 & 23.39 & 7.21 & 1.2 & 0.59 \\ 
TSW &C$^{18}$O & $1-0$ & 109.782 & 95.16 & 7.13 & 6.26 & 0.58 \\ 
TSW &HNCO & $5_{0,5}-4_{0,4}$ & 109.906 & 54.65 & 7.12 & 2.54 & 0.58 \\ 
TSW &$^{13}$CO & $1-0$ & 110.201 & 381.66 & 7.08 & 29.99 & 0.57 \\ 
TSW &CN & $1_{3/2}-0_{1/2}$ & 113.191 & 76.31 & 24.52 & 4.41 & 2.2 \\ 
TSW &CN & $1_{1/2}-0_{1/2}$ & 113.491 & 201.53 & 24.4 & 18.96 & 2.18 \\ 
TSW &C$^{18}$O & $2-1$ & 219.56 & 48.71 & 6.57 & 2.99 & 0.6 \\ 
TSW &$^{13}$CO & $2-1$ & 220.399 & 304.86 & 9.26 & 17.61 & 0.84 \\ 
TSW &H$_2$CO & $3_{1,2}-2_{1,1}$ & 225.698 & 17.04 & 7.86 & 1.18 & 0.7 \\ 
TSW &CN & $2_{3/2}-1_{3/2}$ & 226.36 & 13.95 & 7.82 & 0.27 & 0.7 \\ 
TSW &CN & $2_{3/2}-1_{1/2}$ & 226.66 & 26.37 & 7.8 & 1.84 & 0.7 \\ 
TSW &CN & $2_{5/2}-1_{3/2}$ & 226.875 & 67.62 & 7.79 & 3.85 & 0.7 \\ 
TSW &CH$_3$OH & $5_k-4_k$ & 241.791 & 50.67 & 7.6 & 2.5 & 0.66 \\ 
TSW &CS & $5-4$ & 244.936 & 19.23 & 8.59 & -0.21 & 0.74 \\ 
TSW &CH$_3$OH & $2_{0,2}-1_{-1,1}$ & 254.015 & 23.75 & 8.17 & 0.57 & 0.69 \\ 
TSW &CCH & $3-2$ & 262.004 & 19.45 & 7.23 & 0.23 & 0.6 \\ 
TSW &HCN & $3-2$ & 265.886 & 81.2 & 9.12 & 4.14 & 0.75 \\ 
TSW &HCO$^+$ & $3-2$ & 267.558 & 155.94 & 9.03 & 10.14 & 0.74 \\ 
TSW &HNC & $3-2$ & 271.981 & 23.86 & 7.64 & 2.05 & 0.62 \\ 
OS &SiO & $2-1$ & 86.847 & 29.29 & 10.57 & 3.83 & 0.97 \\ 
OS &CCH & $1_{3/2}-0_{1/2}$ & 87.317 & 52.69 & 10.46 & 9.79 & 0.95 \\ 
OS &CCH & $1_{1/2}-0_{1/2}$ & 87.402 & 39.54 & 10.44 & 4.01 & 0.95 \\ 
OS &HCN & $1-0$ & 88.634 & 228.92 & 10.83 & 40.42 & 0.98 \\ 
OS &HCO$^+$ & $1-0$ & 89.189 & 369.77 & 10.7 & 64.6 & 0.97 \\ 
OS &HOC$^+$ & $1-0$ & 89.487 & 21.55 & 10.63 & 0.92 & 0.96 \\ 
OS &HNC & $1-0$ & 90.664 & 90.76 & 9.71 & 12.34 & 0.87 \\ 
OS &CH$_3$OH & $2_k-1_k$ & 96.741 & 57.49 & 7.8 & 8.35 & 0.68 \\ 
OS &CS & $2-1$ & 97.981 & 83.96 & 7.61 & 10.34 & 0.66 \\ 
OS &SO & $2_1-3_2$ & 99.3 & 25.86 & 9.66 & 2.1 & 0.83 \\ 
OS &C$^{18}$O & $1-0$ & 109.782 & 34.8 & 7.1 & 3.44 & 0.58 \\ 
OS &$^{13}$CO & $1-0$ & 110.201 & 117.67 & 7.05 & 10.91 & 0.57 \\ 
OS &CN & $1_{3/2}-0_{1/2}$ & 113.191 & 68.06 & 24.41 & 13.07 & 2.19 \\ 
OS &CN & $1_{1/2}-0_{1/2}$ & 113.491 & 118.29 & 24.28 & 19.56 & 2.17 \\ 
OS &C$^{18}$O & $2-1$ & 219.56 & 17.85 & 6.46 & 0.77 & 0.59 \\ 
OS &$^{13}$CO & $2-1$ & 220.399 & 113.47 & 9.1 & 7.07 & 0.83 \\ 
OS &CN & $2_{3/2}-1_{1/2}$ & 226.66 & 27.86 & 7.65 & 2.21 & 0.69 \\ 
OS &CN & $2_{5/2}-1_{3/2}$ & 226.875 & 38.21 & 7.64 & 6.55 & 0.68 \\ 
OS &CH$_3$OH & $5_k-4_k$ & 241.791 & 37.05 & 7.44 & 5.48 & 0.64 \\ 
OS &CS & $5-4$ & 244.936 & 22.38 & 8.4 & -0.13 & 0.72 \\ 
OS &CCH & $3-2$ & 262.004 & 23.6 & 7.05 & 1.99 & 0.59 \\ 
OS &HCN & $3-2$ & 265.886 & 54.12 & 8.88 & 7.3 & 0.73 \\ 
OS &HCO$^+$ & $3-2$ & 267.558 & 95.61 & 8.8 & 11.99 & 0.72 \\ 
OS &HOC$^+$ & $3-2$ & 268.451 & 19.6 & 8.75 & 1.41 & 0.72 \\ 
SW &H$^{13}$CO$^+$ & $1-0$ & 86.754 & 45.52 & 10.44 & 3.78 & 0.96 \\ 
SW &SiO & $2-1$ & 86.847 & 54.12 & 10.42 & 4.69 & 0.95 \\ 
SW &CCH & $1_{3/2}-0_{1/2}$ & 87.317 & 314.32 & 10.31 & 38.99 & 0.94 \\ 
SW &CCH & $1_{1/2}-0_{1/2}$ & 87.402 & 158.32 & 10.29 & 17.79 & 0.94 \\ 
SW &HNCO & $4_{0,4}-3_{0,3}$ & 87.926 & 56.2 & 10.17 & 3.52 & 0.92 \\ 
SW &HCN & $1-0$ & 88.634 & 745.59 & 10.67 & 76.73 & 0.97 \\ 
SW &HCO$^+$ & $1-0$ & 89.189 & 1106.6 & 10.54 & 124.11 & 0.95 \\ 
SW &HOC$^+$ & $1-0$ & 89.487 & 47.47 & 10.47 & 5.01 & 0.94 \\ 
SW &HNC & $1-0$ & 90.664 & 371.75 & 9.57 & 33.33 & 0.86 \\ 
SW &HC$_3$N & $10-9$ & 90.979 & 69.12 & 9.5 & 5.0 & 0.85 \\ 
SW &N$_2$H$^+$ & $1-0$ & 93.176 & 117.54 & 24.24 & 6.76 & 2.14 \\ 
SW &C$^{34}$S & $1-0$ & 96.413 & 22.23 & 7.72 & 2.55 & 0.67 \\ 
SW &CH$_3$OH & $2_k-1_k$ & 96.741 & 226.19 & 7.67 & 16.09 & 0.66 \\ 
SW &CS & $2-1$ & 97.981 & 362.43 & 7.48 & 31.55 & 0.64 \\ 
SW &H40$\alpha$         &---         &99.023     &19.93&9.54     &0.35&0.82\\ 
SW &SO & $2_1-3_2$ & 99.3 & 96.97 & 9.49 & 7.96 & 0.81 \\ 
SW &HC$_3$N & $11-10$ & 100.076 & 49.29 & 9.34 & 5.13 & 0.8 \\ 
SW &CH$_3$CCH & $6_k-5_k$ & 102.548 & 108.85 & 8.28 & 7.71 & 0.7 \\ 
SW &CH$_3$OH & $0_{0,0}-1_{1,1}$ & 108.894 & 40.65 & 7.06 & 2.88 & 0.58 \\ 
SW &HC$_3$N & $12-11$ & 109.174 & 39.1 & 7.02 & 3.84 & 0.57 \\ 
SW &C$^{18}$O & $1-0$ & 109.782 & 145.68 & 6.95 & 11.28 & 0.56 \\ 
SW &HNCO & $5_{0,5}-4_{0,4}$ & 109.906 & 42.04 & 6.93 & 3.43 & 0.56 \\ 
SW &$^{13}$CO & $1-0$ & 110.201 & 942.81 & 6.89 & 71.38 & 0.56 \\ 
SW &CN & $1_{3/2}-0_{1/2}$ & 113.191 & 179.74 & 23.84 & 24.35 & 2.13 \\ 
SW &CN & $1_{1/2}-0_{1/2}$ & 113.491 & 374.89 & 23.71 & 42.88 & 2.12 \\ 
SW &c-C$_3$H$_2$ & $3_{3,0}-2_{2,1}$ & 216.279 & 22.71 & 6.08 & 1.14 & 0.56 \\ 
SW &H$_2$CO & $3_{0,3}-2_{0,2}$ & 218.222 & 26.21 & 5.98 & 2.93 & 0.55 \\ 
SW &H$_2$CO & $3_{2,2}-2_{2,1}$ & 218.476 & 21.41 & 5.96 & 1.46 & 0.54 \\ 
SW &C$^{18}$O & $2-1$ & 219.56 & 168.14 & 5.91 & 11.45 & 0.54 \\ 
SW &$^{13}$CO & $2-1$ & 220.399 & 1009.89 & 8.32 & 75.17 & 0.76 \\ 
SW &CH$_3$CCH & $13_k-12_k$ & 222.167 & 39.65 & 8.19 & 2.86 & 0.74 \\ 
SW &C$^{17}$O                &$2-1$                    &224.714    & 37.07&7.08     &3.26&0.64\\ 
SW &H$_2$CO & $3_{1,2}-2_{1,1}$ & 225.698 & 54.69 & 7.02 & 4.33 & 0.63 \\ 
SW &CN & $2_{3/2}-1_{3/2}$ & 226.36 & 21.52 & 6.98 & 3.04 & 0.63 \\ 
SW &CN & $2_{3/2}-1_{1/2}$ & 226.66 & 85.01 & 6.96 & 6.79 & 0.62 \\ 
SW &CN & $2_{5/2}-1_{3/2}$ & 226.875 & 157.89 & 6.95 & 15.05 & 0.62 \\ 
SW &CH$_3$CCH & $14_k-13_k$ & 239.252 & 36.7 & 7.38 & 2.91 & 0.64 \\ 
SW &CH$_3$OH & $5_k-4_k$ & 241.791 & 79.12 & 6.68 & 7.88 & 0.58 \\ 
SW &CS & $5-4$ & 244.936 & 70.96 & 7.52 & 2.75 & 0.65 \\ 
SW &NO & $5/2+-3/2-$ & 250.483 & 17.52 & 6.15 & 1.75 & 0.52 \\ 
SW &NO & $5/2--3/2+$ & 250.817 & 16.94 & 7.26 & 0.89 & 0.62 \\ 
SW &CH$_3$CCH & $15_k-14_k$ & 256.337 & 24.03 & 6.04 & 1.14 & 0.51 \\ 
SW &CH$_3$OH & $2_{1,1}-1_{0,1}$ & 261.806 & 38.48 & 6.22 & 2.64 & 0.52 \\ 
SW &CCH & $3-2$ & 262.004 & 96.45 & 6.21 & 12.31 & 0.52 \\ 
SW &HCN & $3-2$ & 265.886 & 163.25 & 7.8 & 12.26 & 0.64 \\ 
SW &HCO$^+$ & $3-2$ & 267.558 & 403.21 & 7.71 & 35.26 & 0.64 \\ 
SW &HNC & $3-2$ & 271.981 & 71.28 & 6.49 & 4.75 & 0.53 \\ 
SW &CH$_3$CCH & $16_k-15_k$ & 273.42 & 21.45 & 6.45 & 2.79 & 0.53 \\ 
\enddata 
\end{deluxetable} 



\end{document}